\documentclass[aps,prd,showpacs,twocolumn,notitlepage,eqsecnum,
nofootinbib]{revtex4-1}

\usepackage[centertags]{amsmath}
\usepackage{amssymb}
\usepackage{latexsym}
\usepackage{enumerate}
\usepackage{graphicx}
\usepackage{mathrsfs}
\usepackage{hyperref}
\usepackage{stmaryrd}
\usepackage{import}
\usepackage{tensor}
\usepackage{color}
\usepackage{bm}
\usepackage{multirow}
\usepackage{breakurl}
\usepackage{float}

\newcommand{\bi}{\begin{itemize}}
\newcommand{\ei}{\end{itemize}}
\newcommand{\be}{\begin{equation}}
\newcommand{\ee}{\end{equation}}

\renewcommand{\l}{\left(}
\renewcommand{\r}{\right)}
\renewcommand{\a}{\alpha}
\renewcommand{\b}{\beta}
\newcommand{\g}{\gamma}
\newcommand{\G}{\Gamma}
\renewcommand{\d}{\delta}

\newcommand{\ve}{\varepsilon}
\newcommand{\La}{\Lambda}
\newcommand{\la}{\lambda}
\renewcommand{\O}{\Omega}
\renewcommand{\o}{\omega}
\renewcommand{\th}{\theta}

\newcommand{\q}{\quad}
\newcommand{\s}{\sigma}
\newcommand{\vp}{\varphi}
\newcommand{\ti}{\tilde}

\newcommand{\pa}{\partial}
\newcommand{\bs}{\boldsymbol}
\newcommand{\bscal}[1]{\boldsymbol{\mathcal{#1}}}
\renewcommand{\Bar}[1]{\makebox{$\bar{#1}$}}

\begin{document}

\title{Lorenz gauge gravitational self-force calculations of eccentric 
binaries using a frequency domain procedure}

\author{Thomas Osburn}
\author{Erik Forseth}
\author{Charles R. Evans}
\affiliation{Department of Physics and Astronomy, \\ 
University of North Carolina, Chapel Hill, North Carolina 27599, USA}
\author{Seth Hopper}
\affiliation{Max-Planck-Institut f\"ur Gravitationsphysik, 
Albert-Einstein-Institut, \\ 
Am M\"uhlenberg 1, D-14476 Golm, Germany}

\begin{abstract}
We present an algorithm for calculating the metric perturbations and 
gravitational self-force for extreme-mass-ratio inspirals (EMRIs) with 
eccentric orbits.  The massive black hole is taken to be Schwarzschild and 
metric perturbations are computed in Lorenz gauge.  The perturbation 
equations are solved as coupled systems of ordinary differential equations 
in the frequency domain.  Accurate local behavior of the metric is attained 
through use of the method of extended homogeneous solutions and mode-sum 
regularization is used to find the self-force.  We focus on calculating the 
self-force with sufficient accuracy to ensure its error contributions to the 
phase in a long term orbital evolution will be $\delta\Phi \lesssim 10^{-2}$ 
radians.  This requires the orbit-averaged 
force to have fractional errors $\lesssim 10^{-8}$ and the oscillatory part 
of the self-force to have errors $\lesssim 10^{-3}$ (a level frequently 
easily exceeded).  Our code meets this error requirement in the oscillatory 
part, extending the reach to EMRIs with eccentricities of $e \lesssim 0.8$, 
if augmented by use of fluxes for the orbit-averaged force, or to 
eccentricities of $e \lesssim 0.5$ when used as a stand-alone code.  Further, 
we demonstrate accurate calculations up to orbital separations of 
$a \simeq 100 M$, beyond that required for EMRI models and useful for 
comparison with post-Newtonian theory.  Our principal developments include 
(1) use of fully constrained field equations, (2) discovery of analytic 
solutions for even-parity static modes, (3) finding a pre-conditioning 
technique for outer homogeneous solutions, (4) adaptive use of 
quad-precision and (5) jump conditions to handle near-static modes, and (6) 
a hybrid scheme for high eccentricities.
\end{abstract}


\maketitle

\section{Introduction}
\label{sec:intro}

Merging compact binaries are thought to be a promising source of gravitational 
waves that may be found by ground-based or future space-based detectors.  
Theoretical models play a role in the experimental efforts, both in assisting 
detection and in allowing binary parameter estimation.  Three principal 
theoretical approaches exist, numerical relativity \cite{BaumShap10,
LehnPret14}, post-Newtonian (PN) theory~\cite{Blan14}, and gravitational 
self-force (GSF) calculations \cite{Bara09,PoisPounVega11,Thor11}, with 
the effective-one-body formalism providing a synthesis of the three 
\cite{BuonDamo99,BuonETC09,Damo10}.  The GSF approach is relevant when the 
binary mass ratio $\ve$ is sufficiently small that the motion and field of 
the smaller mass can be treated in a perturbation expansion.  In this black 
hole perturbation theory, the background field is that of the heavier 
stationary black hole and the zeroth-order motion of the small mass is a 
geodesic in this background.  Then the perturbation in the metric is 
calculated to first order in the mass ratio and the action of the field of 
the small body back on its own motion is computed (i.e., the first-order 
GSF) \cite{MinoSasaTana97,QuinWald97}.  In principle the calculation proceeds 
to second order \cite{Gral12, Poun12a} and beyond.  Over the past fifteen 
years a number of key formal developments have been established 
\cite{MinoSasaTana97,QuinWald97,BaraOri00,DetwWhit03,GralWald08,Poun10}.

Work on the GSF approach has been motivated in part by prospects of 
detecting extreme-mass-ratio inspirals (EMRIs) using a space-based 
gravitational wave detector like LISA or eLISA \cite{AmarETC13,NASA11,ESA12}.  
For a LISA-like detector with $f_{\rm min} \simeq 10^{-4}$ Hz, an EMRI 
consists of a small compact object of mass $\mu \simeq 1-10 M_{\odot}$ 
(neutron star or black hole) in orbit about a supermassive black hole (SMBH) 
of mass $M \sim 10^5 - 10^7 M_{\odot}$.  The mass ratio would lie in the range 
$\ve = \mu/M \simeq 10^{-7}-10^{-4}$, small enough to allow a gradual, 
adiabatic inspiral and provide a natural application of perturbation theory.  
As the EMRI crosses the detector passband prior to merger its orbital motion 
accumulates a total change in phase of order $\ve^{-1} \sim 10^{4}-10^{7}$ 
radians.  

Less extreme mass ratios may also be important.  A class of intermediate 
mass black holes (IMBHs) may exist with masses $M \sim 10^2 - 10^4 M_{\odot}$. 
These are suggested~\cite{MillColb04} by observations of ultraluminous X-ray 
sources and by theoretical simulations of globular cluster dynamical 
evolution.  Stellar mass black holes or neutron stars spiralling into IMBHs 
with masses $M \sim 50-350 M_{\odot}$, referred to as intermediate-mass-ratio 
inspirals (IMRIs), would lie in the passband of Advanced LIGO and are 
potentially promising sources~\cite{BrowETC07,AmarETC07}.  An IMRI might also 
result from binaries composed of an IMBH and a SMBH~\cite{AmarETC07},
which would appear as an eLISA source.  While IMRIs execute fewer total orbits 
(i.e., $\ve^{-1} \sim 10^2 - 10^3$) than EMRIs in making, say, a decade of 
frequency change, the theoretical approach is nearly the same.  Detection of 
E/IMRIs would represent a strong field test of general relativity and 
measurement of the primary's multipole structure would confirm or not the 
presence of a Kerr black hole~\cite{VigeHugh10,BaraCutl07,BrowETC07}.

In tandem with the more formal GSF developments have come a host of practical 
numerical calculations.  The dominant approach to date takes the small body 
to be a point mass \cite{PoisPounVega11}, computes the metric perturbation (MP)
in the time domain (TD) \cite{Mart04,BaraSago07,FielHestLau09,CaniSopu14} or 
frequency domain (FD) \cite{Detw08,HoppEvan10,Akca11}, and obtains a finite 
self-force from the divergent retarded field by mode-sum regularization 
\cite{BaraOri00,BaraSago07,Detw08,BaraSago09,BaraSago10,BaraSago11,Akca11,
WarbETC12,AkcaWarbBara13}.  Work on the gauge dependent GSF has benefitted 
from analogous scalar field models \cite{DetwMessWhit03,Haas07,BaraOriSago08}.
Applications to Kerr EMRIs, both with scalar and gravitational self-force, 
have been made \cite{BaraGolbSago07,WarbBara10,KeidETC10,DolaBaraWard11,
ShahFrieKeid12,DolaBara13,IsoyETC14,Warb14c}.  Availability of analytic 
mode-sum regularization parameters \cite{HeffOtteWard12a,HeffOtteWard12b} has 
been beneficial.  Calculations of perturbations and the GSF have now been 
made with very high accuracy, arbitrary precision arithmetic \cite{Fuji12,
ShahFrieWhit14,Shah14,ForsEvanHopp14}, allowing detailed comparison with PN 
theory (see also \cite{Detw08,BlanETC10}).  Finally, alternative means of 
calculating the self-force, both effective source calculations 
\cite{VegaDetw08,VegaWardDien11,WardETC12} and direct Green function 
calculations \cite{CasaETC09,CasaETC13,WardETC14}, are being developed.  

This paper reports the development of a method and computer code for 
accurately calculating the GSF of Schwarzschild EMRIs with eccentric orbits.  
We use a point mass description for the stress-energy tensor of the small body 
and work in Lorenz gauge.  Tensor spherical harmonic and Fourier decomposition 
are used and the MP amplitudes are computed initially in the FD.  These 
amplitudes are then transferred to the TD using a generalization of the method 
of extended homogeneous solutions (EHS) \cite{BaraOriSago08,HoppEvan10} for 
systems of equations \cite{Golb09,BaraSago10,Akca11,EvanOsbuFors12,
AkcaWarbBara13}.  The GSF is then calculated using standard mode-sum 
regularization.  Our code was developed over the past several years and was 
reported in a series of talks at the 15th, 16th, and 17th Capra meetings 
\cite{EvanOsbuFors12,*Osbu13,*Fors13,*Osbu14}.  A similar effort by a group 
in Southampton, initiated earlier \cite{Akca11} but developed in part 
concurrently with ours, has been reported in full elsewhere 
\cite{AkcaWarbBara13}.

Our use of Lorenz gauge in the FD and generalization of EHS is in common 
with \cite{AkcaWarbBara13}.  The FD is used to achieve high accuracy and the 
method of EHS circumvents the Gibbs phenomenon in returning to the TD.  We 
calculate also the ``geodesic self-force,'' that is the GSF as a function of 
time along an undisturbed geodesic orbit.  The intent is to provide GSF 
curves at points that densely cover a region of orbital parameter space 
(parameterized by eccentricity $e$ and dimensionless 
semi-latus rectum $p$).  As shown in \cite{WarbETC12} these data can then 
serve as an interpolated input to an osculating orbits evolution code.  

\begin{figure}
\includegraphics[scale=1.05]{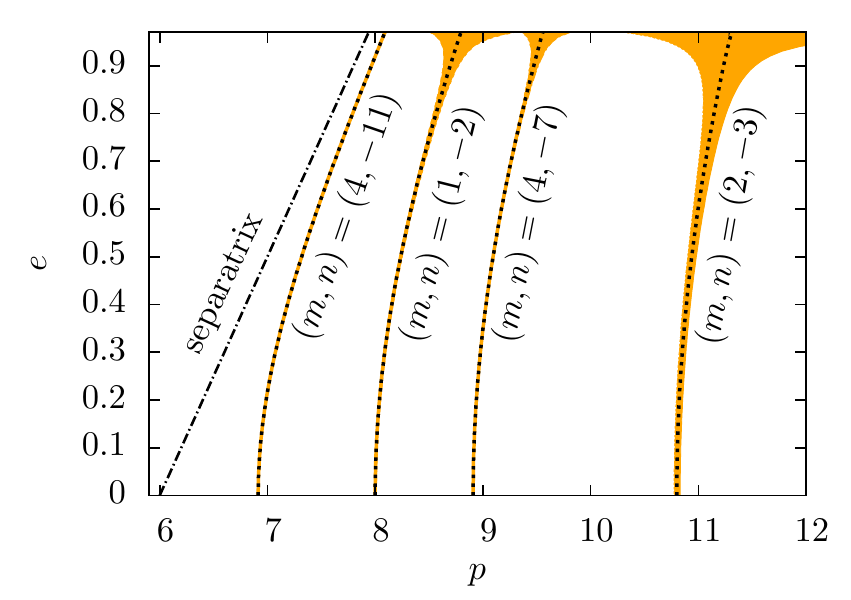}
\caption{Orbital parameter space, resonances, and regions with near-static 
modes.  Relativistic definitions of semi-latus rectum $p$ and eccentricity 
$e$ are adopted [Eqn.~\eqref{eqn:defeandp}]. Dotted curves indicate, as in 
\cite{AkcaWarbBara13}, a closed orbit with the ratio $\Omega_{\vp}/\Omega_r$ 
being a rational number.  On any such curve there exists a static mode 
$\omega_{mn} = m \Omega_{\vp} + n \Omega_r = 0$ for indicated $m$ and $n$.
Within the vicinity of these curves these modes will be nearly static.  For 
near-static modes with frequencies below $|\o|< 10^{-4} M^{-1}$ (shaded 
region) we use 128-bit floating point arithmetic for part of the mode 
calculation.  Our calculations are extended to frequencies as small as 
$|\o|< 10^{-6} M^{-1}$, which exist in regions narrower than the dotted 
curves.  
\label{fig:near_static}} 
\end{figure}

Our approach is distinguished, however, in several respects.  We devise and 
use here a fully constrained system of equations for even parity, as well 
as use the comparable system \cite{AkcaWarbBara13} for odd parity.  We 
have found and use a set of analytic solutions for even-parity static modes, 
which complement published solutions \cite{BaraLous05} for odd parity.  
Particular attention is paid to accurately calculating near-static modes that 
occur for certain orbital parameters that produce a near resonance between 
the radial $\Omega_r$ and azimuthal $\Omega_{\vp}$ orbital frequencies 
(see Fig.~\ref{fig:near_static}).  To compute this subset of modes accurately 
we resort to occasional (more expensive) use of 128-bit arithmetic (i.e., 
quad precision).  This has two effects.  Firstly, we are able to trade some 
computational speed for more uniform accuracy across $e$ and $p$ space.  
Secondly, the technique significantly expands the region of $e$ and $p$ space 
within which the GSF can be computed accurately.  For a given $l$ and $m$ mode 
there will exist a harmonic $n$ that produces the lowest magnitude frequency, 
$\omega_{mn} = m \Omega_{\vp} + n \Omega_r$.  When a mode exists with 
frequency at or below $|\o|< 10^{-4} M^{-1}$ we switch the critical parts of 
the computation over to quad precision.  Furthermore, there is an added 
device that can be used for this single $(l,m,n)$ mode--we can eliminate 
part of the integration by using the jump conditions 
to normalize the mode.  This procedure increases accuracy and restores some 
computational speed.  With these techniques we are able to extend the reach 
of the code in computing the GSF to wider orbital separations, out to 
$p\lesssim 100$, and to higher eccentricities, reaching as high as 
$e\lesssim 0.8$ with acceptable errors when all available techniques are 
used.

The accuracy criteria we adopt in this paper stem from envisioned use of 
computed inspirals and resulting waveforms in the matched filtering 
applications of gravitational wave detectors.  A detector like 
eLISA~\cite{AmarETC07,Bara09} would employ template matching to separate 
individual sources and extract physical parameters from events buried in 
detector noise.  To take full advantage of a signal when doing parameter 
matching~\cite{BaraCutl07,BrowETC07, VigeHugh10}, theoretical waveform phases 
must be sufficiently accurate that they not contribute dephasing errors and 
thus degrade available signal-to-noise ratios 
\cite{BaraCutl07,HindFlan08,Thor11}.  The oscillations within the 
gravitational waveform will depend upon the orbital 
motion.  For Schwarzschild EMRIs there are cumulative radial 
$\Phi_r = \chi_p(T)$ and azimuthal $\Phi_{\varphi} = \varphi_p(T)$ orbital 
phases (here $T \sim M^2/\mu$ is the cumulative time in the inspiral and see
Sec.~\ref{sec:orbits} for discussion of eccentric orbital motion).  For 
schematic purposes, we simply take here the radial phase as a proxy for the 
waveform phase.  Further, we assume that theoretical orbital phase 
uncertainties should be no larger than $\delta\Phi_r \simeq 0.01$ radians 
over a cumulative phase in the inspiral of as much as $\Phi_r \sim 10^6$ (for 
an EMRI) (see discussion in \cite{Thor11}).  Thus the GSF and inspiral 
calculation should have fractional errors in the phase of order $10^{-8}$. 

The GSF is often split into dissipative and conservative parts
\cite{HindFlan08}.  It is useful to also split the dissipative part into 
orbit-averaged and oscillatory parts.  The orbit-averaged, dissipative GSF 
(i.e., energy and angular momentum fluxes to infinity and down the horizon) 
produces secular changes that drive the adiabatic inspiral.  For a small mass 
ratio $\ve$ the inspiral will schematically accumulate an orbital phase of
\be
\label{eqn:phase}
\Phi_r = \kappa_0(e,p,\eta) \, \frac{1}{\ve} 
+ \kappa_1(e,p,\eta) 
+ \kappa_2(e,p,\eta) \, \ve + \cdots ,
\ee
where $e$ and $p$ are orbital parameters when the EMRI enters a detector 
passband and $\eta$ is the ratio between ingress frequency $f_i$ and egress 
(or merger) frequency $f_e$.  The $\kappa$'s are dimensionless functions of 
order unity that do not depend on $\ve$.  We are here assuming a Schwarzschild 
E/IMRI and absence of Kerr transient resonances~\cite{FlanHind10}.  Also 
beyond our present concern are the recently recognized effects of resonances 
in Schwarzschild EMRIs \cite{vdMe14}, which appear to come in at order $\ve$ 
(i.e., produce contributions to $\kappa_2$).  The orbit-averaged, dissipative 
part of the first-order GSF will determine $\kappa_0$.  The rest of the 
first-order GSF, the oscillatory part of the dissipative piece and the 
(oscillatory) conservative part, contribute to $\kappa_1$.  This term in 
$\Phi_r$ is of order unity and represents the post-1-adiabatic correction 
\cite{HindFlan08}.  The implications for our work are this: if we require 
$\delta\Phi_r \simeq 10^{-2}$, we must compute the orbit-averaged first-order 
GSF with fractional errors at or below 
$\epsilon_0 \simeq 10^{-8} \lesssim \ve\delta\Phi_r$ and compute 
the oscillatory parts with fractional errors of order 
$\epsilon_1 \simeq 10^{-3}\lesssim \delta\Phi_r$ or less.  The retarded MPs 
themselves must be known even more accurately, since mode-sum regularization 
is a numerically subtractive procedure.

Ultimately these contributions to $\kappa_1$ are necessary but not
sufficient.  It has long been understood that $\kappa_1$ also depends on the 
orbit-averaged part of the second-order GSF \cite{Pois02,Rose06a,Rose06b,
HindFlan08,Poun10,Gral12}, which our code (and the one described in 
\cite{AkcaWarbBara13}) does not calculate.  Moreover, there is expected to be 
an error in computing $\kappa_1$ by using FD methods and the ``geodesic'' 
GSF.  In curved space, the real GSF will depend upon the entire past history 
of the particle's motion and the self-consistently evolved retarded field.  
In the geodesic approximation there is no encoding of the prior history of 
an inspiral.  For adiabatic inspiral the discrepancy is expected to appear at 
a relative order of $\ve$ (thus in $\kappa_1$) \cite{BurkKhan13}.  It was 
stressed in \cite{DienETC12} that this discrepancy could be assessed by 
comparing a self-consistent TD self-force calculation with an osculating 
orbits evolution using a FD-derived geodesic self-force.  Such calculations 
are now in progress \cite{DienETC14,Warb14a}, pitting a scalar field 
self-consistent TD evolution against an osculating orbits inspiral driven 
by a geodesic scalar self-force calculation.  Preliminary results 
\cite{Warb14b} show small differences that are (so far) nearly 
indistinguishable from errors in the TD evolution.

Achievable GSF accuracy will depend on orbital parameters, particularly the
eccentricity.  Theoretical studies suggest that EMRIs may form via several 
mechanisms~\cite{AmarETC07}.  The standard channel involves weak two-body 
relaxation within the nuclear star cluster that scatters a compact object into 
a high eccentricity orbit about a SMBH.  It is then captured by the SMBH 
through successive bursts of GW emission near pericenter, a process referred 
to as one-body 
inspiral \cite{HopmAlex05}.  These stars are captured initially in very high 
eccentricity orbits, which then proceed to circularize as the orbit shrinks.  
For $M \simeq 3 \times 10^6 M_{\odot}$, EMRIs formed in this way will have a 
distribution of eccentricities peaked about $e \simeq 0.7$ (and a maximum of 
$e \simeq 0.81$) as they enter the eLISA passband (see~\cite{HopmAlex05} and 
their Figure 4).  Because of the likelihood that EMRIs will have high 
eccentricities, we have focused on extending the ability of our code to 
calculate up to $e \simeq 0.8$. 

An alternative EMRI formation channel posits that compact \emph{binaries} may 
scatter into high eccentricity orbits about the SMBH, with the binary being
subsequently tidally disrupted.  The dissolution of the binary may then inject 
a compact object into orbit, which will typically be less eccentric, about the 
SMBH.  These EMRIs will subsequently have nearly circular orbits by the time 
they enter the eLISA passband~\cite{AmarETC07}.  As Fig.~\ref{fig:near_static} 
makes clear, there is less likelihood of encountering troublesome near-static 
modes at low eccentricity, and our code correspondingly has higher accuracy 
and speed in these cases.

This paper is organized as follows.  In Sec.~\ref{sec:formalism} we review 
the formalism for calculating the first-order MPs and the GSF for bound 
eccentric orbits on Schwarzschild.  There we establish our notation for 
bound geodesic motion, our convention for spherical harmonic decomposition 
and definition of MP amplitudes, and give the coupled MP equations in Lorenz 
gauge.  We also show in Sec.~\ref{sec:formalism} how the size of these systems 
of coupled equations can be reduced, from seven down to four equations for 
even parity and from three down to two equations for odd parity, using the 
gauge conditions.  These fully constrained equations are the ones we solve 
numerically, deriving the remaining MP amplitudes from the gauge conditions.  
In Sec.~\ref{sec:FD_Methods} we outline how we apply the method of EHS to 
coupled systems of equations.  In Sec.~\ref{sec:numerical}, where the heart 
of our numerical method is presented, we provide a roadmap and details on how 
various classes of Fourier-harmonic modes are solved.  These include low-order
($l = 0,1$) modes, static modes, and near-static modes.  Particularly 
worth noting is our new analytic solution for even-parity static modes 
(Sec.~\ref{sec:static}) and various procedures for coping with near-static 
modes (Secs.~\ref{sec:numericalGeneral} \& \ref{sec:near_static}).  
Sec.~\ref{sec:results} gives our numerical results.  There we compare our 
computed GSF to values given in \cite{AkcaWarbBara13} for a particular orbit 
and provide tables of GSF values, including estimated digits of accuracy, for 
a broader set of orbital parameters (see also App.~\ref{sec:addedGSFtables}).  
We show how the GSF errors vary smoothly as we range over orbital parameter 
space, while the speed of the algorithm changes more abruptly as it copes with 
difficult modes.  We also discuss how flux calculations may be combined with 
the computed oscillatory part of the GSF to obtain sufficient accuracy for 
high eccentricity orbits in long term orbit integrations, a subject we expect 
to return to in a later paper.  Finally, we relegate to App.~\ref{asymp} some 
details on expansions that are used to set accurate boundary 
conditions on mode functions at large $r$ and near the horizon, to 
App.~\ref{static} some details on the expansions from which analytic 
solutions are derived for static modes, and to App.~\ref{sec:littlefs} the 
form of certain force terms used in the mode-sum regularization procedure.

Throughout this paper we set $c = G = 1$ and use metric signature $(-+++)$ and 
sign conventions of Misner, Thorne, and Wheeler \cite{MisnThorWhee73}.  Our 
notation for metric perturbation amplitudes and source terms largely follows 
that of Martel and Poisson \cite{MartPois05} (see also \cite{HoppEvan10}).  In 
particular, while general coordinate indices are denoted by Greek letters 
$\alpha,\beta,\mu,\nu,\ldots$, it is useful to consider a split of the 
four-dimensional manifold into $\mathcal{M}^2\times\mathcal{S}^2$ and adopt 
lowercase Latin letters $a,b,c,\ldots$ for indices associated with 
coordinates $t$ and $r$ and capital Latin letters $A,B,C,\ldots$ for the 
angular coordinates $\theta$ and $\vp$ and associated indices.

\section{Formalism}
\label{sec:formalism}

\subsection{Bound orbits on a Schwarzschild background}
\label{sec:orbits}

We consider in this paper generic bound motion of a point particle of mass 
$\mu$ around a Schwarzschild black hole of mass $M$ under the assumption that
$\mu/M \ll 1$.  We use Schwarzschild coordinates 
$x^{\mu} = (t,r,\theta, \varphi )$, in which the line element takes the 
standard form
\be
ds^2 = -f dt^2 + f^{-1} dr^2
+ r^2 \left( d\theta^2 + \sin^2\theta d\varphi^2 \right) ,
\ee
where $f(r) = 1 - 2M/r$.  

Let the worldline of the particle be
$x_p^{\a}(\tau) =\left[t_p(\tau),r_p(\tau),\th_p(\tau),\varphi_p(\tau)\right]$,
with proper time $\tau$.  In this paper a subscript $p$ indicates a field 
evaluated at the location of the particle.  The four-velocity is 
$u^{\alpha} = dx_p^{\alpha}/d\tau$.  Without loss of generality the motion 
is confined to the equatorial plane, $\theta_p(\tau)=\pi/2$.  At zeroth order 
the motion is geodesic in the static background and the geodesic equations 
yield immediate first integrals.  This allows us to write the four-velocity as
\be
\label{eqn:four_velocity}
u^\a = \l \frac{{\cal{E}}}{f_{p}}, u^r, 0, \frac{{\cal{L}}}{r_p^2} \r,
\ee
where ${\cal{E}}$ and ${\cal{L}}$ are the constant specific energy and 
specific angular momentum, respectively.  Bound orbits have ${\cal{E}} < 1$ 
and require at least ${\cal{L}} > 2 \sqrt{3} M$ for two turning points to 
exist.  The constraint $u^\a u_\a = -1$ yields an expression for the radial 
coordinate velocity
\be
\label{eqn:rpDots}
\dot r_p^2(t) = f_{p}^2 \left( 1 - \frac{U^2_{p}}{{\cal{E}}^2} \right) , 
\ee
where
\be
U^2 \l r,{\cal L}^2 \r \equiv f(r) \l 1 + \frac{{\cal{L}}^2}{r^2} \r ,
\ee
and a dot indicates differentiation with respect to coordinate time.

While eccentric orbits on Schwarzschild can be parameterized by ${\cal E}$ and 
${\cal L}$, alternative pairs of parameters can be chosen.  For example, 
we can use instead the (dimensionless) semi-latus rectum $p$ and the 
eccentricity $e$ (see \cite{CutlKennPois94,BaraSago10}).  A third choice is 
the pericentric and apocentric radii, $r_{\rm min}$ and $r_{\rm max}$.  These 
various parameters are related by the following equations 
\be
\label{eqn:defeandp}
p = \frac{2 r_{\rm max} r_{\rm min}}{M (r_{\rm max} + r_{\rm min})} , 
\q \q
e = \frac{r_{\rm max} - r_{\rm min}}{r_{\rm max} + r_{\rm min}},
\ee
\be
r_{\rm max} = \frac{pM}{1-e}, 
\q \q
r_{\rm min} = \frac{pM}{1+e},
\ee
and
\be
{\cal{E}}^2 = \frac{(p-2)^2-4e^2}{p(p-3-e^2)}, 
\q
{\cal{L}}^2 = \frac{p^2 M^2}{p-3-e^2}.
\ee
To avoid a plunging orbit the inner turning point must lie outside the 
maximum of the effective potential $U^2$, which implies another inequality, 
$p > 6 + 2 e$.  The boundary $p = 6 + 2 e$ of these innermost stable orbits 
is the separatrix indicated in Fig.~\ref{fig:near_static}.

Numerical integration of the trajectory employs an alternate curve parameter, 
$\chi$, in which the radial position on the orbit is given a 
Keplerian-appearing form \cite{Darw59}  
\be
r_p \l \chi \r = \frac{pM}{1+ e \cos \chi} ,
\ee
where $\chi$ differs in general from the true anomaly $\varphi$.  One 
radial libration makes a change $\Delta\chi = 2\pi$.  The orbital equations 
then take the form
\begin{align}
\label{eqn:darwinEqns}
\frac{dt_p}{d \chi} &= \frac{r_p \l \chi \r^2}{M (p - 2 - 2 e \cos \chi)}
 \left[\frac{(p-2)^2 -4 e^2}{p -6 -2 e \cos \chi} \right]^{1/2} ,
\nonumber
\\
\frac{d \varphi_p}{d\chi} 
&= \left[\frac{p}{p - 6 - 2 e \cos \chi}\right]^{1/2} ,
\\
\frac{d\tau_p}{d \chi} &= \frac{M p^{3/2}}{(1 + e \cos \chi)^2} 
\left[ \frac{p - 3 - e^2}{p - 6 - 2 e \cos \chi} \right]^{1/2} ,
\nonumber
\end{align}
with the use of $\chi$ removing singularities in the differential equations 
at the radial turning points (see \cite{CutlKennPois94}).  Integrating the 
first of these equations provides the fundamental frequency and period 
of radial motion,
\be
\label{eq:O_r}
\O_r \equiv   \frac{2 \pi}{T_r},
\q \q 
T_r \equiv \int_{0}^{2 \pi} \l \frac{dt_p}{d\chi} \r d \chi.
\ee
Integrating the second equation determines the azimuthal advance.  The 
average angular frequency $d \varphi_p / dt$ is found by integrating over a 
complete radial oscillation
\be
\label{eq:O_phi}
\O_\varphi \equiv 
\frac{1}{T_r} \int_{0}^{T_r} \l \frac{d \varphi_p}{dt} \r dt .
\ee
In general $\Omega_r \ne \Omega_{\vp}$, except in the Newtonian limit. 

\subsection{First-order metric perturbation equations in Lorenz gauge}
\label{sec:tensors}

The finite mass of the small body induces a first-order perturbation 
$p_{\mu\nu}$ in the background metric $g_{\mu \nu}$: 
${\rm g}_{\mu\nu} = g_{\mu\nu} + p_{\mu\nu}$.  Using the trace reverse 
${\Bar p}_{\mu\nu} = p_{\mu\nu} - \tfrac{1}{2} g_{\mu\nu} p$ 
(with $p = p_{\alpha\beta} \, g^{\alpha\beta}$), linearizing the Einstein 
equations, and imposing the Lorenz gauge condition
\be
\label{eqn:tensor_gauge}
\tensor{\Bar p}{^\mu^\nu_{|\nu}} = 0 ,
\ee
yields the first-order field equations for the MPs
\be
\label{eqn:tensor_field}
^4\Box \bar{p}_{\mu\nu} + 2 \tensor{R}{^\alpha_\mu^\beta_\nu} \, \bar{p}_{\a\b} 
= -16\pi T_{\mu\nu}.
\ee
Here a stroke ${| \mu}$ (or $\nabla_{\mu}$) indicates covariant 
differentiation with respect to $g_{\mu\nu}$ and 
$^4\Box = g^{\mu\nu} \nabla_{\mu} \nabla_{\nu}$.  Additionally, 
$\tensor{R}{^\alpha_\mu_\beta_\nu}$ is the Riemann tensor associated with 
$g_{\mu \nu}$.  Adopting a point particle description, the stress-energy 
tensor in Eqn.~\eqref{eqn:tensor_field} is
\be
T^{\mu \nu} \l x^\a \r
= \mu \int \frac{u^\mu u^\nu}{\sqrt{-g}} \, \d^4 
\left[ x^{\alpha} - x^{\alpha}_p(\tau) \right] d\tau .
\ee

\subsection{Spherical harmonic decomposition}
\label{sec:sph_harm}

Our convention for tensor spherical harmonics and notation for MP amplitudes 
follows that of Martel and Poisson~\cite{MartPois05}, a modification of the 
original notation of Regge and Wheeler \cite{ReggWhee57}.  (An alternative 
notation is found in \cite{BaraLous05,BaraSago07,AkcaWarbBara13}.)  The 
convention we use leaves all tensor harmonics orthogonal and clarifies 
the distinction between even-parity and odd-parity amplitudes.  Odd-parity 
perturbations are expanded in terms of $X_A^{lm}$ and $X_{AB}^{lm}$, while 
even-parity perturbations use $Y^{lm}$, $Y_A^{lm}$, and $Y_{AB}^{lm}$
\begin{align}
\displaystyle
p_{ab} &= \sum_{lm} h^{lm}_{ab}Y^{lm} ,
\nonumber
\\
\label{eq:MP}
p_{aB} &= \sum_{lm} \left(j^{lm}_a Y^{lm}_B + h^{lm}_a X^{lm}_B \right),
\\
\nonumber
p_{AB} &= \sum_{lm} \left[r^2(K^{lm} \O_{AB} Y^{lm}+ G^{lm}Y^{lm}_{AB}) 
+ h^{lm}_2 X^{lm}_{AB} \right] .
\end{align}
The stress-energy tensor is also decomposed and following \cite{MartPois05} 
has even-parity projections
\begin{align}
Q^{ab}_{lm} &= 8\pi \int T^{ab} \bar{Y}^{lm}\, d\Omega,   
\nonumber
\\
Q^a_{lm} &= \frac{8\pi r^2}{\la+1} \int T^{aB} \bar{Y}^{lm}_B\, d\Omega, 
\\ 
Q^\flat_{lm} &= 8\pi r^2 \int T^{AB} \Omega_{AB} \bar{Y}^{lm}\, d\Omega, 
\nonumber
\\ 
Q^\sharp_{lm} &= \frac{8\pi r^4}{\la(\la+1)} \int T^{AB}
\bar{Y}^{lm}_{AB}\, d\Omega ,
\nonumber
\end{align}
and odd-parity projections
\begin{align}
P^a_{lm} &= \frac{8\pi r^2}{\la+1} \int T^{aB} \bar{X}^{lm}_B\, d\Omega, 
\\
P_{lm} &= \frac{4\pi r^4}{\la(\la+1)} \int T^{AB} \bar{X}^{lm}_{AB}\, d\Omega .
\nonumber
\end{align}
The overbar here indicates the complex conjugate and 
$\la \equiv (l+2)(l-1) / 2$.  The sharp ($\sharp$) and flat ($\flat$) 
superscripts merely distinguish two distinct scalar projections.  These 
source terms are given explicitly in Sec.~V of \cite{HoppEvan10}.

\subsection{Lorenz gauge equations for MP amplitudes}
\label{sec:lorenzeqns}

Applying these projections to \eqref{eqn:tensor_field} yields coupled sets 
of field equations in $t$ and $r$ for the MP amplitudes.  Likewise 
\eqref{eqn:tensor_gauge} provides a set of Lorenz gauge conditions 
on the amplitudes.  Lorenz gauge gives each of the ten field equations 
a hyperbolic form, and the principal part of the wave operator in each 
equation can be compactly expressed using the 1+1 dimensional d'Alembertian
\be
\Box \equiv -\pa_t^2 + f \pa_r \l f \pa_r \r = 
-\pa_t^2 + \pa_{r_*}^2 ,
\ee
where $r_*$ is the tortoise coordinate 
\be
r_* = r + 2M \ln \left(\frac{r}{2M} - 1 \right) .
\ee

The seven even-parity and three odd-parity Lorenz gauge field equations 
are well-posed hyperbolic systems, but the Lorenz gauge conditions (three even 
parity and one odd parity) force constraints on the initial conditions.  
These \emph{unconstrained} field equations, along with the Bianchi
identities, ensure that the gauge conditions, if fixed initially, are 
satisfied subsequently.  We present the unconstrained equations first, and 
then introduce modified \emph{constrained} systems.  Equations in this 
subsection are in TD form but can be converted to FD form as discussed in 
Sec.~\ref{sec:fourier}.  In what follows all $l$ and $m$ indices on MP and 
source amplitudes are suppressed for brevity unless otherwise noted. 

\subsubsection{Unconstrained Lorenz gauge field equations}
\label{sec:unconstrained}

The seven even-parity unconstrained Lorenz gauge equations are 
\begin{widetext}
\begin{align}
\begin{split}
\label{eqn:EqnsUnconstrEven}
&\Box h_{tt} 
+ \frac{2 (r-4M) f}{r^2} \pa_r h_{tt}
+ \frac{4 M f}{r^2} \pa_t h_{tr} 
+ \frac{2 M (3 M-2 r) f^2}{r^4} h_{rr} 
\\
& \hspace{40ex} 
+ \frac{4 M f^2}{r^3} K 
+ \frac{2(M^2 -r^2 f) - 2 \la r^2 f}{r^4} h_{tt} 
= -f Q^{rr}-f^2 Q^\flat -f^3 Q^{tt},
\\
&\Box h_{tr} 
+ \frac{2 f^2}{r} \pa_r h_{tr}
+ \frac{2 M f}{r^2} \pa_t h_{rr} 
+ \frac{2 M}{r^2 f} \pa_t h_{tt}
+ \frac{4 (\la +1) f}{r^3} j_t
- \frac{4 (M-r)^2 + 2 \la r^2 f}{r^4} h_{tr}
= 2f Q^{tr},
\\
&\Box h_{rr} 
+ \frac{2f}{r} \pa_r h_{rr}
+\frac{4 M}{r^2 f} \pa_t h_{tr}
+ \frac{2 M (3 M-2 r)}{r^4 f^2} h_{tt} 
+ \frac{4 (r-3M)}{r^3} K  
 \\
& \hspace{34ex}
+ \frac{8 (\la +1) f}{r^3} j_r 
 + \frac{2(r-M)(7M-3r) - 2 \la r^2 f}{r^4} h_{rr} 
= Q^\flat-\frac{1}{f} Q^{rr}-f Q^{tt},
\\
&\Box j_t 
- \frac{2 M f}{r^2} \pa_r j_t
+ \frac{2 M f}{r^2} \pa_t j_r
+ \frac{2 f^2}{r} h_{tr}
- \frac{2 f^2 + 2 \la f}{r^2} j_t
= f^2 Q^t,
\\
&\Box j_r 
+ \frac{2M f}{r^2} \pa_r j_r
+ \frac{2 M}{r^2 f } \pa_t j_t
+ \frac{2 f^2}{r} h_{rr} 
- \frac{2 f}{r} K
+ \frac{2 \la f}{r} G
- \frac{4 f^2 + 2 (\la+1) f}{r^2} j_r 
= -Q^r,
\\
&\Box K 
+ \frac{2 f^2}{r} \pa_r K
- \frac{2 (3 M-r) f^2}{r^3} h_{rr} 
+ \frac{2 M}{r^3} h_{tt}
- \frac{4 (\la +1) f^2}{r^3}  j_r
- \frac{4 f^2 + 2 \la f}{r^2} K
= Q^{rr}-f^2 Q^{tt} ,
\\
&\Box G 
+ \frac{2 f^2}{r} \pa_r G
+ \frac{4 f^2}{r^3} j_r
- \frac{2 \la f}{r^2} G
= -\frac{f}{r^2} Q^\sharp.
\end{split}
\end{align}
The three odd-parity parts of the field satisfy a separate unconstrained 
set of equations in Lorenz gauge
\begin{align}
\begin{split}
\label{eqn:EqnsUnconstrOdd}
&\Box h_t -\frac{2M f}{r^2} \pa_{r} h_t
+ \frac{2M f}{r^2} \pa_{t} h_r
- \frac{2 f^2 + 2 \la f}{r^2} h_t 
= f^2 P^t,\\
&\Box h_r  
+ \frac{2M f}{r^2} \pa_r h_r 
+ \frac{2 M}{r^2 f} \pa_t h_t 
+ \frac{2 \la f}{r^3}h_2 
+ \frac{2(4M-3r)f -2 \la r f}{r^3}h_r = -P^r,\\
&\Box h_2 - \frac{2 f^2}{r} \pa_r h_2
+ \frac{4f^2}{r} h_r 
+ \frac{2 f(r-4M) - 2 \la r f}{r^3} h_2 = -2 f P .
\end{split}
\end{align}

\subsubsection{Lorenz gauge conditions}

The Lorenz gauge conditions \eqref{eqn:tensor_gauge} separate into even- and 
odd-parity equations when expanded in spherical harmonics.  For even parity 
there are three coupled gauge conditions
\begin{align}
\begin{split}
\label{eqn:LGCondEven}
 f \pa_{r} h_{tr} 
&- \frac{f}{2} \pa_t h_{rr} 
- \pa_t K 
- \frac{1}{2f} \pa_t h_{tt} 
 + \frac{2(r-M)}{r^2} h_{tr} 
 - \frac{2(\la + 1)}{r^2} j_t
 = 0 ,\\
 -\frac{1}{f} \pa_t h_{tr} 
&+ \frac{f}{2} \pa_{r} h_{rr}
- \pa_r K 
+ \frac{1}{2f} \pa_{r} h_{tt} 
+ \frac{2(r-M)}{r^2} h_{rr} 
- \frac{2}{r} K
- \frac{2 (\la+1)}{r^2}  j_r
 = 0 , \\
 f \partial_{r} j_r
&-\frac{1}{f} \pa_t j_t
- \frac{f}{2} h_{rr}
+ \frac{1}{2f} h_{tt} 
+ \frac{2(r-M)}{r^2} j_r 
- \la G
 = 0 ,
\end{split}
\end{align}
while in odd parity there is just one condition
\be
\label{eqn:LGCondOdd}
 f\partial_r h_r 
- \frac{1}{f} \pa_t h_t
+ \frac{2(r-M)}{r^2}h_r 
-  \frac{\la}{r^2} h_2 
 = 0 .
\ee

\subsubsection{Fully-constrained field equations}
\label{sec:constrained}

While the unconstrained equations \eqref{eqn:EqnsUnconstrEven} and 
\eqref{eqn:EqnsUnconstrOdd} might be solved numerically, in practice we have 
found it more efficient and accurate to use the gauge conditions
\eqref{eqn:LGCondEven} and \eqref{eqn:LGCondOdd} to produce reduced order
systems of constrained equations.  To do this we rewrite the gauge conditions 
\eqref{eqn:LGCondEven} and \eqref{eqn:LGCondOdd} as expressions for the four 
amplitudes $j_t$, $j_r$, $G$, and $h_2$.  These are used, as necessary, to 
eliminate their appearance in six of the equations in the sets 
\eqref{eqn:EqnsUnconstrEven} and \eqref{eqn:EqnsUnconstrOdd}--specifically 
those equations with wave operators acting on $h_{tt}$, $h_{tr}$, $h_{rr}$, 
$K$, $h_t$, and $h_r$.  These six equations, four even parity and two odd 
parity, once modified only reference these remaining amplitudes.  Once the 
constrained equations are solved, the eliminated fields, $j_t$, $j_r$, $G$, 
and $h_2$, are recovered via the gauge conditions.  

We find the following system of four constrained even-parity equations
\begin{align}
\label{eqn:EqnsConstrEven}
\begin{split}
&\Box h_{tt} 
+\frac{2(r-4M)f}{r^2}\pa_r h_{tt}
+\frac{4M f}{r^2} \pa_t h_{tr} 
+ \frac{2M(3M-2r)f^2}{r^4} h_{rr} \\
&\hspace{40ex}
+ \frac{4Mf^2}{r^3} K 
+ \frac{2(M^2 -r^2 f) - 2 \la r^2 f}{r^4} h_{tt} 
=-f Q^{rr}- f^2 Q^\flat - f^3 Q^{tt}, 
\\
& \Box h_{tr}
+\frac{4f^2}{r} \pa_r h_{tr} 
+ \frac{4M-r}{r^2 f} \pa_t h_{tt} 
+ \frac{(4M-r)f}{r^2} \pa_t h_{rr} 
- \frac{2f}{r} \pa_t K 
+ \frac{4M(M-r) - 2 \la r^2 f }{r^4} h_{tr}
= 2 f Q^{tr} , 
\\
& 
\Box h_{rr}
+\frac{4(r-M)f}{r^2} \pa_r h_{rr}
+ \frac{2}{r}\pa_r h_{tt}
- \frac{4f}{r}\pa_r K
+ \frac{4(3M-r)}{r^2 f} \pa_t h_{tr}
+ \frac{2M(3M-2r)}{r^4 f^2} h_{tt} \\
& 
\hspace{41.5ex}
+ \frac{4(M-r)}{r^3} K 
+ \frac{2(M-r)^2 - 2 \la r^2f}{r^4} h_{rr}
= -\frac{1}{f} Q^{rr}+Q^\flat-f Q^{tt} ,
\\
& \Box K 
+\frac{4f^2}{r} \pa_r K
-\frac{f}{r}\pa_r h_{tt}
- \frac{f^3}{r} \pa_r h_{rr}
+ \frac{2f}{r} \pa_t h_{tr}   
+ \frac{2M}{r^3} h_{tt} 
- \frac{2(r+M)f^2}{r^3} h_{rr} 
- \frac{2 \la f}{r^2} K
= -f^2 Q^{tt} + Q^{rr} ,
\end{split}
\end{align}
and the following system of two constrained odd-parity equations
\begin{align}
\label{eqn:EqnsConstrOdd}
\begin{split} 
& \Box h_t
- \frac{2M f}{r^2} \pa_r h_t
+ \frac{2Mf}{r^2} \pa_t h_r 
- \frac{2 f^2 + 2 \la f}{r^2} h_t 
 = f^2 P^t,
\\
& \Box h_r 
+ \frac{2(r-M)f}{r^2} \pa_r h_r 
- \frac{2(r-3M)}{r^2 f} \pa_t h_t  
- \frac{2 f^2 + 2 \la f}{r^2}  h_r 
= - P^r .
\end{split}
\end{align}
These six equations, supplemented with the gauge conditions 
\eqref{eqn:LGCondEven} and \eqref{eqn:LGCondOdd}, are satisfied by the MPs 
in Lorenz gauge.  However, as discussed in Sec.~\ref{sec:FD_Methods}, to 
find solutions numerically we cast these equations into the FD, reducing 
them to large sets of ordinary differential equations.  Furthermore, in 
certain special cases [i.e., low-order ($l = 0,1$) modes and static 
($\o = 0$) modes] some MP amplitudes cease to be defined or the systems of 
equations reduce further in size, or both.  Sec.~\ref{sec:numerical} discusses 
these special cases, each of which merits unique numerical treatment.  

\subsection{Self-force and mode-sum regularization}
\label{sec:sf}

Once the Lorenz gauge equations in the preceding section are solved using 
causal boundary conditions (i.e., outgoing waves at infinity and downgoing 
waves at the horizon), the MP amplitudes are used to reassemble the retarded 
field $p^{\rm ret}_{\mu\nu}$.  The full retarded field is divergent at the 
location of the point mass, precisely where its action back on the particle's 
motion must be determined.  Regularization is required, and the mode-sum 
regularization (MSR) procedure of Barack and Ori \cite{BaraOri00} is commonly 
used (see e.g., early use \cite{DetwMessWhit03} with a scalar field and for 
the GSF in Lorenz gauge \cite{BaraSago07,Bara09,BaraSago10}).  To discuss MSR 
it is useful to consider the decomposition discovered by Detweiler and 
Whiting \cite{DetwWhit03} that splits the retarded MP within a normal 
neighborhood of the particle \cite{PoisPounVega11} into \emph{regular} ($R$) 
and \emph{singular} ($S$) parts
\be 
p^{\rm ret}_{\mu \nu} = p_{\mu \nu}^R + p_{\mu \nu}^S .
\ee 
The singular part has a divergence that captures the singular behavior of the 
retarded field and satisfies the same inhomogeneous field equations 
\eqref{eqn:tensor_field}, but through design (i.e., appropriate boundary 
conditions) does not contribute at all to the self-force.  The regular part, 
in contrast, is a solution to the homogeneous first-order field equations and 
is entirely responsible for the self-force.  Applying the self-force, the 
corrected motion can be regarded as forced, non-geodesic motion in the 
background spacetime.  With the Detweiler and Whiting split, the motion can 
also be viewed as geodesic in the corrected metric 
$g_{\mu \nu} +  p_{\mu \nu}^R$.  In either viewpoint the self-force becomes 
a term in the equations of motion found from calculating
\be
\label{eqn:sfReg}
F_R^\a = \mu  k^{\a \b \g \d} \bar{p}^R_{\b \g | \d} ,
\ee
which is evaluated at the particle, $x^{\a} = x_p^{\a}(\tau)$.  Here the 
trace-reversed MP is used and the projection operator is
\be
\label{eqn:projector}
k^{\a \b \g \d} \l x_p \r = 
\frac{1}{2} g^{\a \d} u^\b u^\g
-
g^{\a \b} u^\g u^\d
-
\frac{1}{2} u^\a u^\b u^\g u^\d 
+
\frac{1}{4} u^\a g^{\b \g} u^\d
+
\frac{1}{4} g^{\a \d}  g^{\b \g} .
\ee
At this point, $k^{\a \b \g \d}$ is defined only at the particle's location 
(though below we discuss broadening its definition so it can be evaluated off
the worldline).  Its form ensures orthogonality $F_R^{\a} u_{\a} = 0$.  The 
same operator may be applied to $p^{\rm ret}_{\mu \nu}$ and $p^S_{\mu \nu}$ 
to define the retarded and singular self-forces,
\be
\label{eqn:sfRet}
F_{\rm ret}^\a = \mu  k^{\a \b \g \d} \,
\bar{p}^{\rm ret}_{\b \g | \d} ,
\q \q 
F_S^\a = \mu  k^{\a \b \g \d} \,
\bar{p}^S_{\b \g | \d} ,
\ee
both of which diverge at $x^{\a} = x_p^{\a}(\tau)$.  Formally, the regular 
part is formed through the subtraction $F_{R}^\a = F_{\rm ret}^\a - F_{S}^\a$. 
However, since both $F_{\rm ret}^\a$ and $F_{S}^\a$ are infinite at the 
location of interest, a straightforward subtraction is not possible.

The central idea of MSR is to decompose the components of $F_{\rm ret}^\a$ 
and $F_{S}^\a$ into sums over scalar multipole modes $F_{\rm ret}^{\a l'}$ 
and $F_S^{\a l'}$, with every mode being finite at the location of the 
particle.  (We use $l'$ and $m'$ to distinguish from the $l$ and $m$ of our 
tensor spherical harmonic decomposition.)  Then the subtraction can be 
made mode by mode.  There is a subtlety in the decomposition, however, since 
the operator $k^{\a \b \g \d}$ (and therefore the self-force) is only defined 
at this stage at the location of the particle.  To generate a spherical 
harmonic decomposition we must choose a way to extend $k^{\a \b \g \d}$ off 
of the worldline.  Following Ref.~\cite{BaraSago10} we define 
$k^{\a \b \g \d}(x; x_p)$ at field point $x$, when the particle is at $x_p$, 
to have the value given from Eqn.~\eqref{eqn:projector} with $g^{\mu \nu}$
evaluated at $x$ and $u^\a$ evaluated at $x_p$.  Later, in 
Eqn.~\eqref{eqn:GSF_coupling}, when we re-expand our tensor harmonics as 
sums of scalar harmonics, this choice ensures a finite coupling of $l$ modes
for each $l'$.

The mode-sum expansion for $F_{S}^\a$ can written in the form
\be
\label{eqn:FS}
F_{S}^\a =  \sum_{l'=0}^\infty 
\Bigg[ 
\l l' + \frac{1}{2} \r F^\a_{[-1]}
+ 
F^\a_{[0]}
+
\frac{F^\a_{[2]}}{\l  l' - \frac{1}{2} \r \l l' + \frac{3}{2} \r} 
+
\frac{F^\a_{[4]}}{\l  l' - \frac{3}{2} \r \l l' - \frac{1}{2} \r 
\l l' + \frac{3}{2} \r \l l' + \frac{5}{2} \r} 
+ 
\cdots
\Bigg] ,
\ee
where the coefficients $F^\a_{[-1]}, F^\a_{[0]}, F^\a_{[2]}, \ldots$ are the
$l'$-independent \emph{regularization parameters} (RPs), which depend only 
upon position in the eccentric orbit.  (We use the notation of Heffernan 
et al.~\cite{HeffOtteWard12a} for the RPs.)  Then, the mode-sum formula 
\be
\label{eqn:FRegRet}
F_{R}^\a =  \sum_{l'=0}^\infty 
\Bigg[ 
F_{\rm ret}^{\a l'}
-
\l l' + \frac{1}{2} \r F^\a_{[-1]} \\
-
F^\a_{[0]}
-
\frac{F^\a_{[2]}}{\l  l' - \frac{1}{2} \r \l l' + \frac{3}{2} \r}
-
\frac{F^\a_{[4]}}{\l  l' - \frac{3}{2} \r \l l' - \frac{1}{2} \r \l  l' +
\frac{3}{2} \r \l l' + \frac{5}{2} \r} 
- 
\cdots
\Bigg] ,
\ee
determines the regularized self-force.  The first two RPs, $F^\a_{[-1]}$ and
$F^\a_{[0]}$, for the GSF on a Schwarzschild background were originally 
given by Barack et al.~\cite{BaraETC02}.  Indeed, only these first two 
parameters are needed to obtain convergence.  From the structure of the 
$l'$-dependent denominator terms, all of the succeeding terms each converge to 
zero as $l' \rightarrow \infty$.  However, since the series with only 
$F^\a_{[-1]}$ and $F^\a_{[0]}$ converges slowly ($\sim 1/l'_{\rm max}$), 
higher-order RPs are important for hastening convergence when the sum is 
truncated at some finite $l'_{\rm max}$.  Heffernan et 
al.~\cite{HeffOtteWard12a} have calculated the higher-order coefficients 
$F^\a_{[2]}$ and $F^\a_{[4]}$ for the GSF, and their use (along with 
numerically fitting to even higher order) greatly improves convergence.

As described above, MSR requires an expansion of the full retarded self-force 
$F_{\rm ret}^{\a}$ as a sum over scalar spherical harmonic modes 
$F_{\rm ret}^{\a l'}$.  In contrast, our Lorenz gauge calculation yields a set 
of MP amplitudes for each $l$ and $m$ in a tensor spherical harmonic expansion. 
The former can be derived from the latter by re-expanding each tensor 
spherical harmonic in our expression for $F_{\rm ret}^{\a}$ as a sum of 
scalar spherical harmonics.  To that end, we take the definition of 
$k^{\a \b \g \d} (x, x_p)$ given above, along with the tensor spherical 
harmonic expansion of the retarded MP given in Eqn.~\eqref{eq:MP}, and 
substitute in Eqn.~\eqref{eqn:sfRet}.  Taking the limit $r\rightarrow r_p(t)$ 
while maintaining $\theta$ and $\vp$ dependence leaves \cite{BaraSago10}
\begin{align}
\label{eqn:fExansion}
\left[ F_{\rm ret}^\a (t,r_p(t),\theta,\vp ) \right]_{\pm} = 
\frac{\mu}{r_p^2} 
\sum_{l=0}^\infty \sum_{m=-l}^l \Big[
& f_0^{\a lm} \, Y^{lm} 
+ f_1^{\a lm} \, \sin^2\theta \, Y^{lm} 
+ f_2^{\a lm} \, \sin\theta \cos\theta \, Y_{,\theta}^{lm} 
\nonumber
\\
& + f_3^{\a lm} \, \sin^2\theta \, Y_{,\theta\theta}^{lm} 
+ f_4^{\a lm} \, \l \cos\theta \, Y^{lm} - \sin\theta \, Y_{,\theta}^{lm} \r 
\nonumber
\\
&  + f_5^{\a lm} \, \sin\theta \, Y_{,\theta}^{lm} 
+ f_6^{\a lm} \, \sin^3\theta \, Y_{,\theta}^{lm} 
+ f_7^{\a lm} \, \sin^2\theta \, \cos\theta \, Y_{,\theta\theta}^{lm} 
\Big]_{\pm} ,
\end{align}
where a comma indicates a partial derivative.
The vectors $f_0^{\a lm}\ldots f_7^{\a lm}$ are functions of the MP 
amplitudes and their first $t$ and $r$ derivatives.  Our 
tensor harmonic decomposition of the MP  
differs from \cite{BaraSago10} and so we provide the detailed 
form of these functions in Appendix \ref{sec:littlefs}.  The 
MP amplitudes are $\mathcal{O}(\mu)$, which makes the GSF of order
$\mathcal{O}(\mu^2)$.  Each of the functions $f_0^{\a lm}\ldots f_7^{\a lm}$, 
as well as $F_{\rm ret}^\a$, takes on a pair of values ($\pm$) since the 
limit $r\rightarrow r_p(t)$ can be applied from the outside or inside of the 
particle radius $r_p(t)$.  Differing limits on the two sides also appear in 
the RP $F_{[-1]}^\a$ and therefore in $F_S^\a$.  The regularized GSF itself, is 
single-valued though. 

Finally, we obtain $F_{\rm ret}^{\a l'}$ 
by expanding the $\th$-dependent terms in \eqref{eqn:fExansion} as sums of
scalar spherical harmonics. This yields the following expression
\be
\label{eqn:GSF_coupling}
\left[ F_{\rm ret}^{\a l'} \right]_{\pm} = 
\frac{\mu}{r_p^2} \sum_{m=-l'}^{l'} Y^{l'm} \left[
\mathcal{F}_{(-3)}^{\a l'-3,m} 
+ \mathcal{F}_{(-2)}^{\a l'-2,m} 
+ \mathcal{F}_{(-1)}^{\a l'-1,m} 
+ \mathcal{F}_{(0)}^{\a l',m} 
+ \mathcal{F}_{(+1)}^{\a l'+1,m} 
+ \mathcal{F}_{(+2)}^{\a l'+2,m} 
+ \mathcal{F}_{(+3)}^{\a l'+3,m} 
\right]_{\pm} .
\ee
The functions $\mathcal{F}_{(j)}^{\a l,m}$, given in \cite{BaraSago10}, 
are found to each be a linear 
combination of the $f_n^{\a lm}$ of the same $l$ and $m$.  Accordingly, a 
given $l'$ term used in the MSR formula couples only to tensor spherical 
harmonic amplitudes in the range 
$l' - 3 \le l \le l' + 3$.

\subsection{Conservative and dissipative parts of the self-force and
first-order changes in orbital constants}
\label{sec:sf_cons_diss}

The procedure described in the previous subsection takes the retarded field 
and produces the regular ($R$) force (i.e., the self-force).  To make the 
notation clear we can write this retarded self-force as $F_{R,{\rm ret}}^\a$.  
It is also conceivable to calculate the advanced self-force 
$F_{R,{\rm adv}}^\a$, which is obtained by precisely the same procedure 
except in replacing $\bar{p}_{\mu\nu}^{\rm ret}$ with 
$\bar{p}_{\mu\nu}^{\rm adv}$.  The singular field $F_S^\a$ is time symmetric, 
so the RPs are unaffected in swapping `ret' for `adv'.  Hinderer and 
Flanagan \cite{HindFlan08} show that it is convenient to split the retarded 
and advanced self-force into conservative and dissipative parts
\be
F_{R,{\rm ret}}^\a = F_{\rm cons}^\a + F_{\rm diss}^\a ,
\quad \quad
F_{R,{\rm adv}}^\a = F_{\rm cons}^\a - F_{\rm diss}^\a ,
\ee
where
\begin{align}
\label{eqn:consDissSF}
F_{\rm cons}^\a = \frac{1}{2}\l F_{R,{\rm ret}}^\a +F_{R,{\rm adv}}^\a \r , 
\q \q
F_{\rm diss}^\a = \frac{1}{2}\l F_{R,{\rm ret}}^\a -F_{R,{\rm adv}}^\a \r .
\end{align}
See also \cite{mino03}.  Furthermore, because of the symmetry, the 
conservative part actually requires regularization
\be
F_{\rm cons}^\a =  \sum_{l'=0}^\infty 
\left[ 
\frac{1}{2}\l F_{\rm ret}^{\a l'} +F_{\rm adv}^{\a l'} \r - 
F_{S}^{\a l'} \right] ,
\ee
while the dissipative part does not
\be
F_{\rm diss}^\a =  \frac{1}{2} \sum_{l'=0}^\infty 
\l F_{\rm ret}^{\a l'} - F_{\rm adv}^{\a l'} \r .
\ee
Conveniently, for geodesic motion on Schwarzschild the advanced self-force 
can be obtained from the retarded self-force using time reversal and symmetry
\be
F_{R,{\rm adv}}^\a(\tau) = \epsilon_{(\a)} \, F_{R,{\rm ret}}^\a(-\tau) , 
\ee
where $\tau = 0$ corresponds to periastron passage and the Schwarzschild 
components change sign or not according to $\epsilon_{(\a)} = (-1,1,1,-1)$, 
with no implied sum in the equation above.

The self-force produces changes in the orbital constants $\mathcal{E} = -u_t$ 
and $\mathcal{L} = u_\vp$.  Using the first-order equations of motion
\begin{align}
\mu \frac{D u_\a}{D \tau} = g_{\a \b} F_R^\b ,
\end{align}
the $t$ component $F_R^t$ provides a rate of work and the $\vp$ component 
$F_R^\vp$ gives a torque such that
\begin{align}
\label{eqn:eDot_lDot}
\dot{\mathcal{E}} = \frac{f_p}{\mu u^t} F_R^t ,
\q \q
\dot{\mathcal{L}} = \frac{r_p^2}{\mu u^t} F_R^\varphi ,
\end{align}
where dot refers to derivative with respect to Schwarzschild time $t$.  While 
the first-order GSF determines the leading order, adiabatic motion and 
contributes terms to the post-1-adiabatic corrections \cite{HindFlan08}, the 
leading-order adiabatic changes require only the orbit-averaged part of the 
dissipative GSF 
\begin{align}
\label{eqn:adiabatic}
\langle \dot{\mathcal{E}} \rangle = 
\frac{1}{T_r} \int_0^{T_r} \frac{f_p}{\mu u^t} F_{\rm diss}^t \, dt ,
\q \q
\langle \dot{\mathcal{L}} \rangle = 
\frac{1}{T_r} \int_0^{T_r} \frac{r_p^2}{\mu u^t} F_{\rm diss}^\varphi \, dt .
\end{align}
For the geodesic GSF, the first-order rate of work and torque are balanced by 
the energy and angular momentum fluxes (each averaged over the orbital period 
and summed over two-surfaces near infinity and the horizon) calculated from 
the first-order MP (see Sec.~\ref{sec:gsf_fluxes}).

\section{Frequency domain techniques for solving coupled systems}
\label{sec:FD_Methods}

Rather than solve directly the TD Lorenz gauge equations of 
Sec.~\ref{sec:lorenzeqns}, we use FD techniques for their speed and 
accuracy.  Accuracy requirements were discussed in the Introduction and these 
are attained in the FD through solution of ordinary differential equations
(ODEs).  The TD alternative \cite{BaraSago10}, solving 1+1 dimensional partial 
differential equations for each $l, m$, has the compensating advantage of 
allowing the GSF to be applied self-consistently \cite{DienETC12}.  The 
specific equations we solve are the FD version of the fully-constrained 
field equations \eqref{eqn:EqnsConstrEven} and \eqref{eqn:EqnsConstrOdd} and 
the gauge conditions \eqref{eqn:LGCondEven} and \eqref{eqn:LGCondOdd},  
obtained by taking $\pa_t \to -i \o$ and replacing amplitudes, e.g., 
$h_{tt}(t,r) \to \tilde{h}_{tt}(r)$.  Subsequently the solution is returned 
to the TD, whence the GSF can be calculated.  The Fourier synthesis uses the 
method of EHS \cite{BaraOriSago08}, which circumvents the Gibbs phenomenon 
encountered with a distributional source.

Below we set the notation for the Fourier transform, give a matrix notation 
for the coupled sets of FD ODEs, and discuss independent bases of homogeneous 
solutions that appear at leading order asymptotically.  We then discuss the 
use of variation of parameters and how EHS is broadened to encompass systems 
of ODEs.

\subsection{Fourier decomposition}
\label{sec:fourier}

As explained in Sec.~\ref{sec:orbits}, two fundamental frequencies, 
$\Omega_r$ and $\Omega_\varphi$, exist in the eccentric-orbit Schwarzschild 
E/IMRI problem.  In the frame that rotates at the mean azimuthal rate 
($\vp' = \vp - \Omega_{\vp} t$) the MP appears non-sinusoidal but periodic in 
$t$.  It can be represented in a Fourier series in harmonics $n \Omega_r$.  
In the inertial frame, the phase of each multipole with $m \ne 0$ advances 
linearly, giving the Fourier-harmonic modes a spectrum
\be
\o \equiv \omega_{mn} = m\Omega_\vp + n\Omega_r.
\ee
Each MP and source amplitude is replaced by a Fourier series (with a tilde 
denoting a FD amplitude).  For a generic amplitude $X_{lm}$
(not to be confused with the tensor harmonics $X_{A}^{lm}$ and 
$X_{AB}^{lm}$) we have
\begin{align}
\ti{X}_{lmn}(r) = 
\frac{1}{T_r}\int_0^{T_r} X_{lm}(t,r) e^{i\omega_{mn} t} \, dt, 
\hspace{10ex}
X_{lm}(t,r) = 
\sum_{n=-\infty}^{\infty}\ti{X}_{lmn}(r) e^{-i\omega_{mn} t}
\label{eqn:fourier}
\end{align}
Henceforth, not only will indices $l$ and $m$ be suppressed but so will $n$ 
on FD objects (unless otherwise noted).  

\subsection{Matrix notation for coupled ODE systems}
\label{sec:matrixnotation}

It is convenient to place the coupled FD equations in matrix form.  For 
even and odd parities, respectively, the fields appearing in the constrained 
systems are assembled into the vectors 
\be
\label{eqn:EandBvector}
\bscal{\ti E}(r) = 
r \left[
\begin{array}{c}
\ti{h}_{tt} \\ 
f \ti{h}_{tr} \\
f^2 \ti{h}_{rr} \\
\ti{K}
\end{array}
\right],
\q \q
\bscal{\ti B}(r) = 
\left[
\begin{array}{c}
\ti{h}_{t} \\
f \ti{h}_{r}
\end{array}
\right] .
\ee
With this notation the even- and odd-parity FD equations are compactly 
expressed in matrix form
\be
\label{eqn:fieldEqsVector}
\bscal{\ti E}'' + \mathbf{A} \, \bscal{\ti E}' + \mathbf{B} \, \bscal{\ti E} 
= \ti{\bscal{U}} ,
\q \q
\bscal{\ti B}'' + \mathbf{C} \, \bscal{\ti B}' + \mathbf{D} \, \bscal{\ti B} 
= \ti{\bscal{V}} ,
\ee
with prime indicating differentiation with respect to tortoise coordinate 
$r_*$ and where the solution vectors and source vectors have dimension 
$k = 4$ or $k = 2$ for even or odd parity, respectively.  In the general 
case the matrices that couple the amplitudes and their first derivatives are
\begin{align}
\begin{split}
\mathbf{A} &=
\frac{1}{r^2}
\left[ 
\begin{array}{cccc} 
-4M & 0 & 0 & 0 \\
0 & 2(r-4M) & 0 & 0 \\
2rf & 0 & 2(r-4M) & -4rf^2 \\
- r & 0 & -r & 2rf
\end{array} 
\right], \\
\mathbf{B} &=
\l \o^2 - \frac{2 (\la+1) f}{r^2}\r 
\mathbb{I} 
+
\frac{1}{r^4}
\left[
\begin{array}{c@{\hspace{-.5ex}}c@{\hspace{-.5ex}}cc} 
2M(r-M) \;\;
& -4i\o Mr^2 \;\;
& -2M(2r-3M) \;\;
& 4Mrf^2 \;\; \\
\;\; i\o r^2(r-4M) \;\;
& \;\; -2fMr \;\;
& \;\; i\o r^2(r-4M) \;\;
& \;\; 2i\o r^3f^2 \;\; \\
\;\; -2(r-M)^2 \;\;
& \;\; 4i\o r^2 (r-3M) \;\;
& \;\; 2(r^2-3Mr+3M^2) \;\;
& \;\; -4Mrf^2 \;\; \\
\; r^2 
& \; -2i\o r^3
& \; -r^2
& \; 2fMr
\end{array} 
\right],
\end{split}
\end{align}
\begin{align}
\mathbf{C} =
\frac{2}{r^2}\left[
\begin{array}{cc} 
-M & 0 \\
0 & r-3M
\end{array} 
\right] ,
\q \q
\mathbf{D} =
\l \o^2 - \frac{2 f^2 + 2 \la f}{r^2} \r 
\mathbb{I} 
 +
 \frac{2 i \o}{r^2}
\left[ 
\begin{array}{cc}
0
& - M \\
r-3M 
& 0
\end{array}
\right].
\end{align}
where the $\mathbb{I}$'s are relevant-sized identity matrices 
($k \times k = 4\times 4$ or $2\times 2$).  The source vectors are
\begin{equation}
\ti{\bscal{U}} = 
r \left[ 
\begin{array}{c}
-f \ti{Q}^{rr}- f^2 \ti{Q}^\flat - f^3 \ti{Q}^{tt} \\
2 f^2 \ti{Q}^{tr} \\
 -f \ti{Q}^{rr}+f^2\ti{Q}^\flat-f^3 \ti{Q}^{tt} \\
 \ti{Q}^{rr} -f^2 \ti{Q}^{tt}
\end{array}
\right],
\q \q
\ti{\bscal{V}} = 
\left[ 
\begin{array}{c}
 f^2 \tilde{P}^t \\
 -f\tilde{P}^r
\end{array}
\right].
\end{equation}
In certain special cases (low-order modes or static modes) some components 
of the vectors $\bscal{\ti E}$ and $\bscal{\ti B}$ identically vanish, 
effectively reducing the order of the system, with concomitant reduction in 
the source components and elements of $\mathbf{A},\ldots,\mathbf{D}$.  These
special cases are detailed in Sec.~\ref{sec:numerical}.

\subsection{Linearly independent sets of homogeneous solutions}
\label{sec:homog}

The constrained even-parity equations are a set of four, coupled, second-order 
ODEs.  As such they have eight linearly independent homogeneous solutions.  
We divide these into four solutions $\bscal{\ti E}^{+}_i$ (with $i=0,1,2,3$) 
that have causal, running-wave dependence $e^{i\o r_*}$ at $r_* = +\infty$ 
and four solutions $\bscal{\ti E}^{-}_i$ that are downgoing, $e^{-i\o r_*}$, 
at the horizon ($r_* = -\infty$).  For odd parity, where the system is a set 
of two, coupled, second-order ODEs, there are four linearly independent 
homogeneous solutions.  In parallel we denote these by $\bscal{\ti B}^\pm_i$ 
with $i=0,1$.  A complete basis of linearly independent homogeneous solutions 
is of dimension $2k$. 

Upon examining the asymptotic limits of Eqn.~\eqref{eqn:fieldEqsVector} as 
$r_* \to \pm \infty$, we find the following is one possible representation of 
the leading-order behavior of the even-parity homogeneous solutions
\begin{align}
\label{eqn:LeadingOrderEven}
\l \bscal{\ti E}^-_0 \r^\top 
& \sim \l  1, 1, 1, 0 \r e^{-i\o r_*},
&
\l \bscal{\ti E}^+_0 \r^\top & \sim 
\l 1, 0, -1, 0 \r  e^{i\o r_*}, \notag \\
\l \bscal{\ti E}^-_1 \r^\top 
& \sim \l 1, 0, -1, -2(1-4i\o M)^{-1} \r f e^{-i\o r_*},
&
\l \bscal{\ti E}^+_1 \r^\top 
& \sim \l 0, 1, -2, 0 \r e^{i\o r_*},\\
\l \bscal{\ti E}^-_2 \r^\top 
& \sim \l  1, -1, 1, 1 \r f^2 e^{-i\o r_*},
&
\l \bscal{\ti E}^+_2 \r^\top 
& \sim \l 0, 1, -2, 1 \r r^{-1}e^{i\o r_*}, \notag \\
\l \bscal{\ti E}^-_3 \r^\top 
& \sim \l 0,0,0,1 \r e^{-i\o r_*},
&
\l \bscal{\ti E}^+_3 \r^\top & 
\sim \l 0, 0, -2, 1 \r r^{-2} e^{i\o r_*} , \notag 
\end{align}
where $\top$ indicates transpose.  We note that, while these vectors are 
linearly independent, the MP amplitudes (components) do not decouple 
asymptotically.  Likewise the asymptotic limits of the odd-parity equations
allow the following representation of the leading-order behavior of
odd-parity homogeneous solutions
\begin{align}
\label{eqn:LeadingOrderOdd}
\l \bscal{\ti B}^-_0 \r^\top 
& \sim \l  1, 1 \r e^{-i\o r_*},
&
\l \bscal{\ti B}^+_0 \r^\top & \sim 
\l 1, -1 \r  e^{i\o r_*},  \\
\l \bscal{\ti B}^-_1 \r^\top 
& \sim \l 1, -1 \r f e^{-i\o r_*},
&
\l \bscal{\ti B}^+_1 \r^\top 
& \sim \l 0, 1 \r r^{-1} e^{i\o r_*} . \notag
\end{align}
Here again, while the odd-parity vectors are linearly independent, the MP 
amplitudes are still mixed between them asymptotically.  

The limiting behavior for $\bscal{\ti E}^\pm_i$ and $\bscal{\ti B}^\pm_i$ 
displayed in \eqref{eqn:LeadingOrderEven} and \eqref{eqn:LeadingOrderOdd}
is merely one possible choice and we refer to these as the 
\emph{simple bases}.  It is however clearly possible to introduce linear 
transformations on these sets of eight and four homogeneous solutions, and 
we describe in Sec.~\ref{sec:numericalGeneral} clear advantages in doing so 
at least for the even- and odd-parity bases on the near-infinity side.

\subsection{Variation of parameters and extended homogeneous solutions 
for coupled systems}
\label{sec:EHS}

With the assumption that sets of homogeneous solutions $\bscal{\ti E}^\pm_i$ 
and $\bscal{\ti B}^\pm_i$ have been obtained by integrating 
Eqns.~\eqref{eqn:fieldEqsVector} (subject to the boundary conditions of the 
previous section or other equivalently-independent ones), it is straightforward 
to construct solutions to the inhomogeneous equations using variation of 
parameters.  Introducing a set of $2k$ variable coefficients $c_i^{\pm}(r)$ 
that multiply the homogeneous basis elements, the particular solutions are 
assumed to have the forms
\be
\label{eqn:gen_J_form}
\bscal{\ti E} (r) =
 \sum_{i=0}^3 \l \bscal{\ti E}^-_i c^{e,-}_i(r) 
+ \bscal{\ti E}^+_i c^{e,+}_i(r) \r ,
\q \q
\bscal{\ti B} (r) = 
 \sum_{i=0}^1 \l \bscal{\ti B}^-_i c^{o,-}_i(r) 
+ \bscal{\ti B}^+_i c^{o,+}_i(r) \r .
\ee
Variation of parameters then assumes that the first derivative of 
\eqref{eqn:gen_J_form} also depends only on the coefficients $c_i^{\pm}(r)$, 
and not their derivatives, by placing a set of $k$ conditions on 
$\pa_{r_*} c_i^{\pm}(r)$.  Differentiating again and substituting into
Eqns.~\eqref{eqn:fieldEqsVector} yields a second set of $k$ conditions on
$\pa_{r_*} c_i^{\pm}(r)$.  Taken together these conditions form a linear 
system with a $2k \times 2k$ matrix $\mathbf{M}$, formed from the homogeneous 
basis elements and their first derivative, that acts on the vector made up 
of the first derivative of the coefficients $c_i^{\pm}(r)$.  The matrix 
$\mathbf{M}$ is the Wronksian matrix.  In odd parity ($k=2$) these equations 
have the form
\be
\label{eqn:matrixM}
\mathbf{M} 
\left[
\begin{array}{c}
\pa_{r_*} c_0^{o,-} \\
\pa_{r_*} c_1^{o,-} \\
\pa_{r_*} c_0^{o,+} \\
\pa_{r_*} c_1^{o,+} 
\end{array}
\right]
=
\left[
\begin{array}{cccc}
\bscal{\ti B}^-_0 & \bscal{\ti B}^-_1 & \bscal{\ti B}^+_0 & 
\bscal{\ti B}^+_1 \\
\pa_{r_*} \bscal{\ti B}^-_0 & \pa_{r_*} \bscal{\ti B}^-_1 & 
\pa_{r_*} \bscal{\ti B}^+_0 & \pa_{r_*} \bscal{\ti B}^+_1
\end{array}
\right] 
\left[
\begin{array}{c}
\pa_{r_*} c_0^{o,-} \\
\pa_{r_*} c_1^{o,-} \\
\pa_{r_*} c_0^{o,+} \\
\pa_{r_*} c_1^{o,+} 
\end{array}
\right]
=
\left[
\begin{array}{c}
\mathbf{0} \\
\ti{\bscal{V}} 
\end{array}
\right] ,
\ee
where bold entries are $2 \times 1$ column vectors.

The normalization functions are then found by matrix inversion followed by
integration over the source region
\begin{align}
\label{eqn:normLowerCs}
c^{e/o,+}_i(r) = \int_{r_{\text{min}}}^{r} 
\frac{1}{f} \frac{W^{e/o,+}_i}{W^{e/o}} \, dr' , \q \q
c^{e/o,-}_i(r) = -\int_{r}^{r_{\text{max}}} 
\frac{1}{f} \frac{W^{e/o,-}_i}{W^{e/o}} \, dr' .
\end{align}
In these integrals $W^{e/o}$ is the determinant of the Wronskian matrix 
(even or odd parity).  The determinants $W^{e/o,\pm}_i$ are formed by 
replacing the column in the Wronskian corresponding to the $i$th homogeneous 
solution with the column vector $\big( \mathbf{0}, \ti{\bscal{U}}\big)^\top$ 
or $\big( \mathbf{0}, \ti{\bscal{V}}\big)^\top$ (even or odd parity) in 
accordance with Cramer's rule.  Again, for odd parity, the Wronskian 
and one of the modified Wronskians are
\be
\label{eqn:w_det}
W^o=
\left|
\begin{array}{cccc}
\bscal{\ti B}^-_0 & \bscal{\ti B}^-_1 & \bscal{\ti B}^+_0 & 
\bscal{\ti B}^+_1 \\
\pa_{r_*} \bscal{\ti B}^-_0 & \pa_{r_*} \bscal{\ti B}^-_1 & 
\pa_{r_*} \bscal{\ti B}^+_0 & \pa_{r_*} \bscal{\ti B}^+_1
\end{array}
\right|,
\q \q
W^{o,-}_0=
\left|
\begin{array}{cccc}
 \mathbf{0} & \bscal{\ti B}^-_1 & \bscal{\ti B}^+_0 & \bscal{\ti B}^+_1 \\
\ti{\bscal{V}} & \pa_{r_*} \bscal{\ti B}^-_1 & \pa_{r_*} 
\bscal{\ti B}^+_0 & \pa_{r_*} \bscal{\ti B}^+_1
\end{array}
\right|.
\ee
Thus, both $W^o$ and $W^{o,-}_0$ are determinants of $4 \times 4$ matrices.  
In even parity the matrices are $8 \times 8$ and in special cases other 
matrix ranks occur.  In this section we have sketched using Cramer's rule for 
the matrix inversion merely to provide a compact discussion.  In reality we 
use LU decomposition in the code to provide the numerical inversion.

Once the normalization functions $c^{e/o,\pm}_i(r)$ are known, the particular 
solutions \eqref{eqn:gen_J_form} can be computed.  However, since the source 
in the TD problem is distributional, this standard procedure is fraught with 
the appearance of Gibbs behavior in the MP (and GSF) upon returning to
the TD. Its use is now supplanted by the method of EHS, though the EHS method 
uses key parts of the standard-approach machinery. 

Barack, Ori, and Sago \cite{BaraOriSago08} developed the EHS method and 
applied it in computing the scalar field of a charge in eccentric orbit 
about a Schwarzschild black hole.  Subsequently, Hopper and Evans 
\cite{HoppEvan10} employed EHS to compute the MPs of a small mass in 
eccentric orbit on Schwarzschild in the Regge-Wheeler-Zerilli formalism.  EHS 
was also used \cite{Golb09,BaraSago10} to compute the low-order ($l=0,1$) 
modes in Lorenz gauge, which marked its first use for a coupled system.  EHS 
then found use in modeling the scalar self-force on a particle in eccentric 
equatorial orbit on a Kerr black hole \cite{WarbBara11}.  In addition, a 
variant called the method of extended particular solutions was developed 
\cite{HoppEvan13} that is useful for certain problems with non-compact source 
terms.  It was employed to compute the gauge vector that generates the 
odd-parity transformation of the MP from Regge-Wheeler to Lorenz gauge.  

Our application of EHS to general MPs in Lorenz gauge for eccentric 
orbital motion on Schwarzschild was developed contemporaneously with 
Akcay, Warburton, and Barack (see talks at the 2012 Capra meeting
\cite{Bara12,Warb12,EvanOsbuFors12}).  Their code was applied 
\cite{WarbETC12} to long term inspiral and their full method has been 
published \cite{AkcaWarbBara13}.  

EHS uses the matrix inversion and integration involved in computing the 
normalization functions, but extends the integration over the entire 
source region to obtain a set of complex constants.  In practice, the 
integration is done with respect to $\chi$
\be
\label{eq:CapitalCs}
C_i^{e/o,\pm} = \pm \int_{0}^{\pi} 
\frac{1}{f(r_p(\chi))} \frac{W_i^{e/o,\pm}(r_p(\chi))}{W^{e/o}(r_p(\chi))} 
\ \frac{dr_p}{d\chi}
\ d\chi ,
\ee
providing better numerical behavior at the turning points.  These constants 
are used to normalize the basis vectors and to assemble 
specific linear combinations, referred to as FD extended homogeneous 
solutions.  They are smooth functions everywhere outside the horizon ($r>2 M$),
\be
\label{eqn:FD_EHS}
\bscal{\ti E}^\pm (r) = \sum_{i=0}^3 C^{e,\pm}_i \ \bscal{\ti E}^\pm_i ,
\q \q
\bscal{\ti B}^\pm (r) = \sum_{i=0}^1 C^{o,\pm}_i \ \bscal{\ti B}^\pm_i.
\ee
Using these functions, exponentially-convergent Fourier sums then provide the 
TD extended homogeneous solutions
\be
\label{eqn:TD_EHS}
\bscal{E}^\pm (t,r) 
= \sum_{n=-\infty}^\infty \bscal{\ti E}^\pm (r) \, e^{-i \o t},
\q \q
\bscal{B}^\pm (t,r) 
= \sum_{n=-\infty}^\infty \bscal{\ti B}^\pm (r) \, e^{-i \o t},
\ee
which likewise hold for all $r>2 M$ and are smooth in $r$ and $t$.  The 
solutions to Eqn.~\eqref{eqn:EqnsConstrEven} and \eqref{eqn:EqnsConstrOdd} 
then follow by abutting the $+$ and $-$ TD EHS at the location of the 
particle,
\be
\bscal{E}(t,r) 
= \bscal{E}^+ \th \left[ r-r_p (t) \right]
+ \bscal{E}^- \th \left[ r_p (t) - r \right],
\q \q
\bscal{B}(t,r) 
= \bscal{B}^+ \th \left[ r-r_p (t) \right]
+ \bscal{B}^- \th \left[ r_p (t) - r \right].
\ee
In Lorenz gauge all of the MP amplitudes are $C^0$ at $r=r_p(t)$.  The 
discontinuity in the derivative is encoded by the presence of the $\theta$ 
functions.  While the Lorenz gauge MP amplitudes must analytically satisfy
$\bscal{E}^+(t,r_p(t)) = \bscal{E}^-(t,r_p(t))$ and
$\bscal{B}^+(t,r_p(t)) = \bscal{B}^-(t,r_p(t))$, the degree to which this 
equality is satisfied numerically is a measure of convergence.

\section{Numerical algorithm}
\label{sec:numerical}

In this section we provide details on our numerical algorithm.  For a
geodesic given by $p$ and $e$, we seek to compute to sufficient accuracy the 
MP and the GSF, $F_R^\a$, as functions of time around the orbit.  We first 
itemize the principal steps and then follow with detailed discussion on some 
aspects of the procedure. 

\begin{enumerate}

\item
\emph{Orbital parameters:} For a given $p$ and $e$, integrate the orbit 
equations to find the period of radial motion $T_r$, and fundamental 
frequencies $\Omega_r$ and $\Omega_\vp$.  Determine also $\mathcal{E}$, 
$\mathcal{L}$, $r_{\rm min}$, and $r_{\rm max}$ (Sec.~\ref{sec:orbits}).

\item
\emph{Mode characterization:} Fourier-harmonic modes divide into classes 
according to $l,m,n$.  Low-multipole modes $l=0,1$ are handled separately 
from $l \ge 2$ radiative modes.  We further divide modes into static 
($m = n = 0$), near-static ($0 < |\o M|< 10^{-4}$), or general cases.  See 
Table \ref{tbl:roadmap} for overlapping breakdown of modes.

\item
\emph{Linearly independent, causal homogeneous bases:} For every $l,m,n$ mode 
find or compute a complete set of $2k$ independent homogeneous solutions.  
In general, the solution process begins with providing causal initial 
conditions at the boundaries using Taylor series or asymptotic expansions 
(App.~\ref{asymp}) and performing numerical ODE 
integrations (Sec.~\ref{sec:homog}) into the source 
region.  On the horizon side, boundary conditions are set at $r_* = -6 M$ and 
sufficient Taylor expansion terms are included to reach a fractional error of 
$\sim 10^{-15}$.  At large radius, the starting location depends on mode and 
frequency.  Large enough starting radius is taken and short integrations are 
used to confirm the asymptotic expansions have errors of order $10^{-14}$.  
All of the homogeneous solutions are then integrated to $r_* = r_*^{\rm min}$
(i.e. the value of $r_*$ when $r = r_{\rm min}$).
Orthogonality of the initial vectors is carefully considered to minimize 
ill-conditioning of matrix inversion (Sec.~\ref{sec:numericalGeneral}).  For 
near-static $(0<|\o M|< 10^{-4})$ modes we employ special techniques to 
overcome strong ill-conditioning (Secs.~\ref{sec:numericalGeneral}, 
\ref{sec:near_static}).  Static (zero frequency) modes have exact analytic 
homogeneous solutions (Sec.~\ref{sec:static}).  The systems of equations 
change character or reduce in size for low-multipole modes 
(Sec.~\ref{sec:low_multipole}).

\item
\emph{FD extended homogeneous solutions:} For each $l,m,n$ the homogeneous 
solutions are integrated over the source from $r_{\rm min}$ to $r_{\rm max}$ 
to find normalization constants and the linear combinations that represent the 
FD EHS (Sec.~\ref{sec:EHS}).  Again, for near-static $(0<|\o M|< 10^{-4})$ 
modes we employ special techniques to overcome strong ill-conditioning 
(Sec.~\ref{sec:near_static}).  Special consideration occurs again for 
low-multipole modes.

\item
\emph{TD extended homogeneous solutions:} For every $l,m$ construct the TD 
EHS (Sec.~\ref{sec:EHS}) by summing over sufficient positive and negative $n$ 
until the Fourier series on each side converge to a relative error of 
$\sim 10^{-10}$.  Not only can convergence of the EHS on each side of 
$r_p(t)$ be monitored, but each $l,m$ mode should approach becoming $C^0$ 
and the derivative in $r$ at the particle should satisfy a jump condition.

\begin{table}[t]
\begin{center}
\caption{Classification of FD modes as functions of $lm\omega$.  Most modes 
(i.e., general case) are found by solving the complete fully-constrained 
systems \eqref{eqn:EqnsConstrEven} and \eqref{eqn:EqnsConstrOdd} and deriving 
the remaining fields using the gauge conditions \eqref{eqn:LGCondEven} and 
\eqref{eqn:LGCondOdd}.  Special cases include static, near-static, and 
low-multipole ($l = 0,1$) modes.  For static and low-multipole modes the 
system size reduces and some MP amplitudes identically vanish.  Special 
cases are discussed in separate sections as noted. 
\label{tbl:roadmap}}
\begin{tabular*}{7.0in}{@{\extracolsep{\fill}}c|c|c|c|c|c|c|c}
\hline\hline
$l$	&	Parity	&	Frequency	& No. Field Eqns.	
& No. Constrs.	& Variables in Reduced Eqns.	& Variables from Constrs.
&	Section	\\
\hline
\multirow{6}{*}{$l\ge 2\;$}	& \multirow{3}{*}{Even} 	
& General  	& 7 & 3 
& $\ti{h}_{tt}$, $\ti{h}_{tr}$, $\ti{h}_{rr}$, $\ti{K}$ & $\ti{j}_t$, $\ti{j}_r$, $\ti{G}$ & \ref{sec:numericalGeneral} \\
&	& Near-static 	& 7 
& 3 & $\ti{h}_{tt}$, $\ti{h}_{tr}$, $\ti{h}_{rr}$, $\ti{K}$ & $\ti{j}_t$, $\ti{j}_r$, $\ti{G}$
& \ref{sec:near_static} \\
&	& Static 	& 5 & 2 
& $\ti{h}_{tt}$, $\ti{h}_{rr}$, $\ti{K}$ & $\ti{j}_r$, $\ti{G}$  & \ref{sec:static} \\ 
\cline{2-8}
& \multirow{3}{*}{Odd} 	& General 	& 3 & 1 & $\ti{h}_t$, $\ti{h}_r$ & $\ti{h}_2$ & \ref{sec:numericalGeneral} \\
&	& Near-static 	& 3 
& 1 & $\ti{h}_t$, $\ti{h}_r$ & $\ti{h}_2$ 
& \ref{sec:near_static} \\
&	& Static 	& 1 & 0 & $\ti{h}_t$ & - & \ref{sec:static} \\ 
\hline
\multirow{4}{*}{$l=1\;$}	& \multirow{2}{*}{Even} 	
& General  			& 6 
& 3 & $\ti{h}_{tt}$, $\ti{h}_{tr}$, $\ti{h}_{rr}$, $\ti{K}$ & $\ti{j}_t$, $\ti{j}_r$ 
& \ref{sec:numericalGeneral} \\
&	& Near-static 	& 6 
& 3 & $\ti{h}_{tt}$, $\ti{h}_{tr}$, $\ti{h}_{rr}$, $\ti{K}$ & $\ti{j}_t$, $\ti{j}_r$
& \ref{sec:near_static} \\
\cline{2-8}
& \multirow{2}{*}{Odd} 	& General  	& 2 & 1 & $\ti{h}_t$, $\ti{h}_r$ & - & \ref{sec:numericalGeneral} \\
&	& Static	& 1 & 0 & $\ti{h}_t$ & - & \ref{sec:low_multipole} \\ 
\hline
\multirow{2}{*}{$l=0\;$}	& \multirow{2}{*}{Even} 	
& General			& 4 
& 2 & $\ti{h}_{tt}, \ti{h}_{tr}, \ti{h}_{rr}, \ti{K}$ & - 
& \ref{sec:numericalGeneral} \\
&	& Static 		& 3 & 1 & $\ti{h}_{tt}, \ti{h}_{rr}, \ti{K}$ & - & \ref{sec:low_multipole} \\ 
\hline\hline
\end{tabular*}
\end{center}
\end{table}

\item
\emph{Assemble $l'$ contributions to $F_{\rm ret}^{\a l'}$:} Compute force 
terms $f_n^{\a lm}$ (App.~\ref{sec:littlefs}) and linear combinations 
$\mathcal{F}_{(j)}^{\a l,m}$ (Sec.~\ref{sec:sf}) and sum over $m$ for each $l$ 
mode.  Only $m\ge 0$ modes need be computed, since $m<0$ are determined by 
crossing relations on the spherical harmonics.  Assemble the $l'$ part of 
the retarded force by combining $l$ for $l' - 3 \le l \le l' + 3$.

\item
\emph{Apply MSR to obtain GSF:} Sum over $l'$ in the MSR formula until the 
GSF converges to a prescribed tolerance or minimum error (Sec.~\ref{sec:sf}).  
In the process we use available analytically-calculated regularization 
parameters $F^{\a}_{[-1]}$, $F^{\a}_{[0]}$, $F^{\a}_{[2]}$, and 
$F^{\a}_{[4]}$ and least-squares fit for $F^{\a}_{[6]}$ and $F^{\a}_{[8]}$ 
using the last seven $l'$ modes.  We find that the error (by comparing the 
regularized self-force on the two sides of the particle) minimizes for 
$l'_{\text{max}} \simeq 13$ for low eccentricities and several modes lower 
for high eccentricity.  A required $l'_{\text{max}}$ implies that we must 
compute tensor spherical harmonic modes up to 
$l_{\text{max}}=l'_{\text{max}}+3$.

\end{enumerate}

\begin{figure}
{\begin{center}
\includegraphics[scale=1.03]{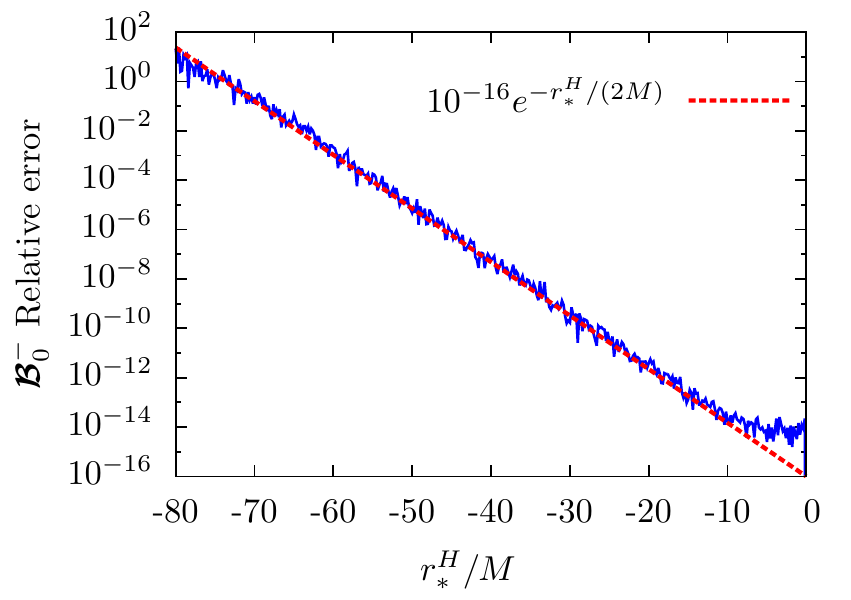}
\includegraphics[scale=1.03]{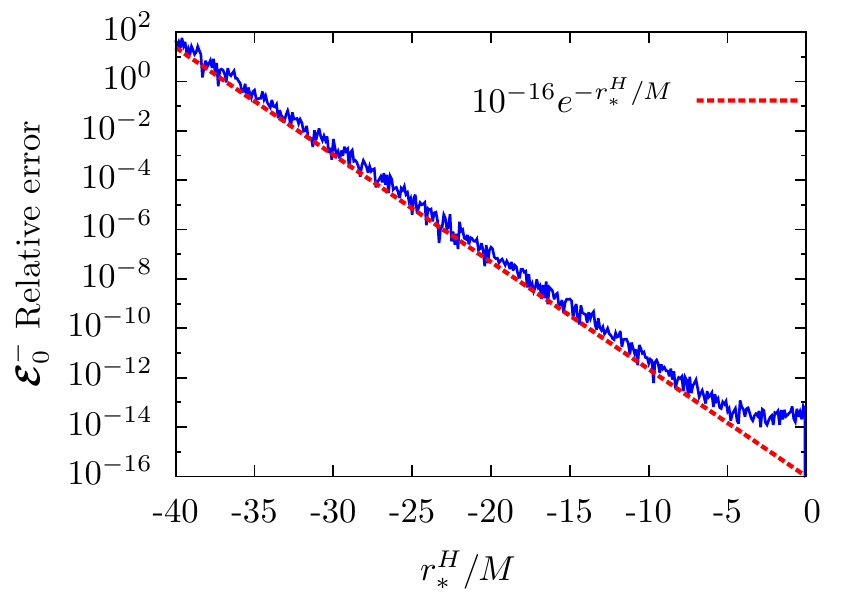}
\end{center}
\caption{\label{fig:hor_growth_1}
Subdominance instability and growth of roundoff errors with starting location.  
We demonstrate the effects of a subdominance instability by comparing results 
of numerical integrations begun at different initial radii $r_{*}^H$ near the 
horizon and ending at $r_*=10M$.  The chosen modes have $l=2$ and $M\o = 1$ 
(odd parity on the left; even parity on the right).  The fiducial, accurate 
solution is obtained from a high-order Taylor expansion, with sufficient 
terms that residuals are at or below roundoff even at a radius of 
$r_{*}^H = 0$.  Using the Taylor expansion at any $-6M < r_{*}^H < 0$ to 
begin an integration that then ends at $r_{*} = 10M$ gives results that are 
consistent with each other.  However, as smaller initial radii are chosen 
($r_{*}^H<-10M$), exponentially greater errors are found in comparing at 
$r_{*} = 10M$ the integrated mode and the fiducial Taylor expansion.  We 
avoid the instability by beginning all integrations at $r_*^H = -6 M$ with
initial conditions from the high-order Taylor expansion.
}  
}
\end{figure}

\subsection{General modes}
\label{sec:numericalGeneral}

We first consider the general case, encompassing all modes with $l\ge 2$ that 
are neither static nor near-static.  The 
expressions \eqref{eqn:LeadingOrderEven} and \eqref{eqn:LeadingOrderOdd} 
provide leading-order behavior for the MP amplitudes as $r_* \to\pm\infty$.  
In practice boundary conditions are set at finite radii and require expansions 
with numerous terms beyond just this leading order.  Appendix \ref{asymp} 
provides details on the asymptotic ($r_* \rightarrow +\infty$) and Taylor 
($r_* \rightarrow -\infty$) expansions that are used to set accurate boundary 
conditions as close to the source region as possible.  Unique numerical
issues are encountered on both the near-horizon and near-infinity sides.

\subsubsection{Boundary conditions near the horizon and subdominance
instability}

On the near-horizon side, using the simple bases of 
\eqref{eqn:LeadingOrderEven} and \eqref{eqn:LeadingOrderOdd} at large
negative $r_*$ is found to generate a subdominance instability.  There is an 
undesired, acausal (up-going) homogeneous solution that can be excited by 
roundoff errors in the numerical boundary condition that grows exponentially 
relative to a desired (subdominant) causal solution.  
Fig.~\ref{fig:hor_growth_1} shows the effect of starting the integration at 
various initial $r_*^H$ and integrating to $r_* = 10M$.  Setting the boundary 
at $r_*^H < -10M$ generates substantial growth of this acausal mode.  We now 
explain briefly why this occurs.  We use odd parity as the example, with 
even parity following a similar analysis.

A complete set of odd-parity independent homogeneous solutions at the event
horizon has leading behavior
\begin{align}
\begin{split}
\l \bscal{\ti B}^-_0 \r^\top &\sim \l  1, 1 \r e^{-i\o r_*} ,
\hspace{18.2ex}
 \l \bscal{\ti B}^-_2 \r^\top \sim \l 1, 1 \r f e^{+i\o r_*} ,
\\
 \l \bscal{\ti B}^-_1 \r^\top &\sim \l 1, -1 \r f e^{-i\o r_*} ,
\hspace{15ex}
\l \bscal{\ti B}^-_3 \r^\top \sim \l  1, -1 \r e^{+i\o r_*} .
\end{split}
\end{align}
$\bscal{\ti B}^-_0$ and $\bscal{\ti B}^-_1$ are the desired causal solutions 
of Eqn.~\eqref{eqn:LeadingOrderOdd}, representing downgoing modes, while 
$\bscal{\ti B}^-_2$ and $\bscal{\ti B}^-_3$ are acausal, representing 
radiation coming up from the black hole.  When we attempt to set boundary 
conditions for $\bscal{\ti B}^-_0$ and $\bscal{\ti B}^-_1$, the inherent 
limitations of our double precision routines produce instead numerical 
superpositions
\begin{align}
\bscal{\ti B}^{-,N}_0  = \bscal{\ti B}^-_0 
+ \a_1 \bscal{\ti B}^-_1 + \a_2 \bscal{\ti B}^-_2 + \a_3 \bscal{\ti B}^-_3 ,
\q \q 
\bscal{\ti B}^{-,N}_1  = \bscal{\ti B}^-_1 
+ \b_0 \bscal{\ti B}^-_0 + \b_2 \bscal{\ti B}^-_2 + \b_3 \bscal{\ti B}^-_3 ,
\end{align}
where all the terms $\a_n \bscal{\ti B}^-_n$ and $\b_n \bscal{\ti B}^-_n$ 
are of order $\sim 10^{-16}$ (roundoff) times the desired dependence.  We 
must be concerned with any of these roundoff terms that are acausal and grow 
relative to the causal terms as we integrate from our starting location, 
$r_*^H$.  Near the horizon $f \sim e^{r_*/2M}$, meaning 
$\a_2 \bscal{\ti B}^-_2$, an acausal contribution to $\bscal{\ti B}^{-,N}_0$,
has precisely this exponential growth relative to $\bscal{\ti B}^-_0$.  This 
prediction is confirmed numerically, as shown in the left panel of 
Fig.~\ref{fig:hor_growth_1}.  On the other hand, $\bscal{\ti B}^-_1$ itself 
grows like $e^{r_*/2M}$, and we see none of the other roundoff terms grow 
relative to it.  As such, this solution does not display a subdominance 
instability.

In the case of even parity, the worst acausal mode has an $f^2$ radial 
dependence.  Accordingly, its relative growth is even worse, i.e. 
$\sim e^{r_*/M}$.  This is shown in the right panel of 
Fig.~\ref{fig:hor_growth_1}.  The figure merely demonstrates the instability.  
In practice, we simply set the boundary condition at $r_*^H=-6M$ using 
Taylor series with sufficient terms to reach roundoff.  The details of those 
Taylor series are found in Appendix \ref{asymp}.  We note finally that it is 
not inconceivable that the instability we discuss here is a result of the 
particular set of MP variables, and therefore the form of the Lorenz gauge 
equations, that we chose to use.

\subsubsection{Boundary conditions at large radius and thin-QR
pre-conditioning}

On the near-infinity side the expansions are asymptotic and require a large 
starting radius $r_*^\infty$, with the radius being roughly inversely related 
to mode frequency $\omega$.  In what follows, we use the odd-parity equations 
as an example.  Even parity follows similar analysis.  After long inward 
integration to $r_*^{\rm min}$ the outer solutions $\bscal{\ti B}^+_i$ can 
be combined with the inner solutions $\bscal{\ti B}^-_i$ to form the 
Wronskian matrix $\mathbf{M}$ [see Eqn.~\eqref{eqn:matrixM}].  
Unfortunately, especially at low frequency, we find the Wronskian matrix to 
be typically ill-conditioned.  Generally one can define a condition number of 
the matrix as $\kappa(\mathbf{M}) = |\lambda_{\rm max}/\lambda_{\rm min}|$, 
where $\lambda_{\rm max}$ and $\lambda_{\rm min}$ are the maximal and minimal 
eigenvalues of $\mathbf{M}$.  Alternatively and conveniently, we may define 
it as $\kappa(\mathbf{M}) = \sigma_{\rm max}/\sigma_{\rm min}$, in terms of 
the singular values $\sigma_i$ of $\mathbf{M}$ in a singular value 
decomposition (SVD).  The condition number is important since one loses 
roughly $\log_{10}(\kappa)$ digits of accuracy in operations like matrix 
inversion \cite{ChenKinc12}.  Starting with the leading-order, near-infinity 
behavior of the simple basis in Eqn.~\eqref{eqn:LeadingOrderOdd} leads to 
condition numbers as large as $\kappa \sim 10^{12}$ in some cases.

Fortunately, it is possible to use a linear transformation on the simple 
basis $\bscal{\ti B}^+_i$ to find a new one $\bscal{\ti B}^+_{i'}$.  
Unfortunately, long integration of the altered set of homogeneous solutions 
to $r_*^{\rm min}$ is required in order to combine them with the inner 
solutions and calculate $\kappa$, making this a hit-or-miss procedure.

We have instead developed a novel means for determining a good linear 
transformation (at $r_*^\infty$) that reduces $\kappa$ by many orders of 
magnitude.  While the method is most effective in handling near-static modes
(discussed below in Sec.~\ref{sec:near_static}), we nevertheless use it for 
all modes and therefore discuss it here.  The technique involves using just 
half the information (outer solutions only) that goes into the Wronskian and 
calculating a ``semi-condition number'' $\rho$.  It begins by picking a basis 
(e.g., the simple one), taking the right half of the matrix $\mathbf{M}$, and 
forming the $4 \times 2$ matrix 
\begin{align}
\mathbf{V} &\equiv
\left[
\begin{array}{cc}
\bscal{\ti B}^+_0 & \bscal{\ti B}^+_1 \\
\pa_{r_*} \bscal{\ti B}^+_0 & \pa_{r_*} \bscal{\ti B}^+_1
\end{array}
\right] .
\end{align}
While $\mathbf{V}$ is a non-square matrix, it has a SVD and yields a set of 
non-negative, real singular values $\sigma_i$.  In our example there are two 
singular values; for even parity there are four.  We call the ratio of the 
largest to smallest, $\rho(\mathbf{V}) = \sigma_{\rm max}/\sigma_{\rm min}$, 
the semi-condition number.  An advantage of $\rho(\mathbf{V})$ is that it can 
be computed immediately once an outer basis is chosen.  However, $\rho$ is not 
the same as the full condition number $\kappa$, which can only be computed 
once the complete set of (inner as well as outer) homogeneous solutions are 
compared.  Empirically, though, we find that $\rho$ is typically large to 
begin with ($\sim 10^7$) and grows by multiple orders of magnitude as the outer 
solutions are integrated inward (see Fig.~\ref{fig:condition_number}), and 
that its value at $r_*^{\rm min}$ tends to be within an order of magnitude 
of $\kappa$.  This strongly suggested that, if $\rho$ could be minimized at 
the starting radius, then $\kappa$ might be greatly reduced in the source 
region.  This guess turned out to be correct. 

\begin{figure}
{\begin{center}
\includegraphics[scale=1.032]{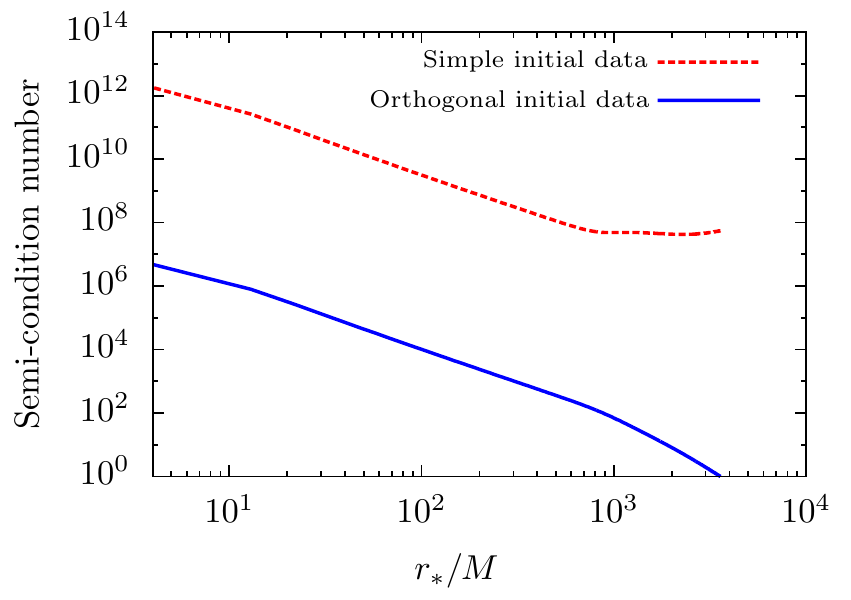}
\includegraphics[scale=1.032]{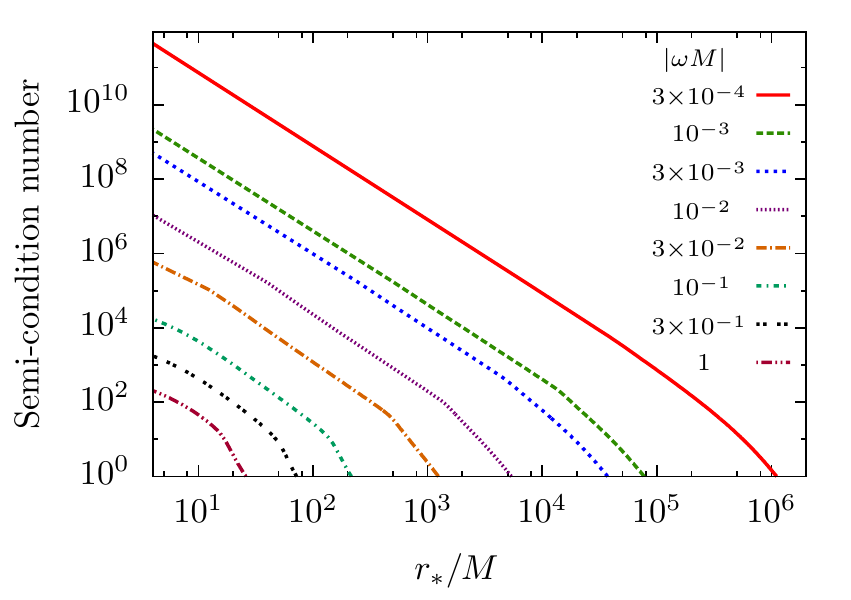}
\end{center}
\caption{\label{fig:condition_number}
Semi-condition number growth of outgoing homogeneous solutions and effect of 
thin-QR pre-conditioning.  The left panel uses the even-parity mode 
$(l,\o)=(5,\,5\times 10^{-3}M^{-1})$ and plots as a function of $r_*$ the 
semi-condition number $\rho$ of the matrix $\mathbf{V}$, which is comprised of 
(the outer solution) half of the Wronskian matrix.  Two initial conditions are 
compared: the simple basis in red (dotted) and the thin-QR pre-conditioned 
basis in blue (solid).  Orthogonalization with the thin-QR pre-conditioner
makes a more than five orders of magnitude improvement.  The right panel uses 
an $l = 16$ even-parity mode and shows the growth of $\rho$ in solutions that 
start with thin-QR orthogonalized initial conditions, as functions of 
frequency.  Once the frequency reaches $|\o M| \le 10^{-4}$, thin-QR
pre-conditioning is no longer sufficient to control the condition number in 
the source region and still allow double precision computations, and we turn 
to added techniques.
}
}
\end{figure}

A linear transformation on the outer boundary conditions can be used to 
mitigate the ill-conditioning [i.e., we are free to choose the starting 
$b$'s in \eqref{eqn:oddInfinity} to begin solving the recurrence relations].
To see how a choice might be made, we start with the simple basis of 
\eqref{eqn:LeadingOrderOdd} to form $\mathbf{V}$ [see also 
Eqn.~\eqref{eqn:oddOutNaive}] and perform a thin-QR decomposition 
\cite{GoluVanl96}.  The matrix is numerically split into a product 
$\mathbf{V} = \mathbf{Q} \mathbf{R}$, where $\mathbf{Q}$ is a $4 \times 2$ 
unitary matrix and $\mathbf{R}$ is a $2 \times 2$ square, upper-triangular 
matrix.  Computed at an initial location $r_*^\infty$, the columns of 
$\mathbf{Q}$ are an alternative, and in this case orthogonal, basis for 
beginning an integration for the homogeneous solutions.  In other words 
$\rho(\mathbf{Q}) = 1$.  We see that the square matrix $\mathbf{R}$ 
multiplies $\mathbf{Q}$ from the right to give $\mathbf{V}$ and 
$\mathbf{R}^{-1}$ multiplies $\mathbf{V}$ from the right to give $\mathbf{Q}$.

In principle, while the columns of $\mathbf{Q}$ (evaluated from $\mathbf{V}$ 
at finite radius $r_{*}^\infty$) do indeed give a new orthogonal basis 
with unit semi-condition number, in practice the use of this basis for 
boundary conditions on the homogeneous solutions (i.e., replacing 
$\mathbf{V} \rightarrow \mathbf{V}' = \mathbf{Q}$) leads to a separate, 
serious numerical problem.  Because $\mathbf{V}$ is ill-conditioned, the 
numerical construction of $\mathbf{Q}$ at finite radius $r_{*}^\infty$ will 
be accompanied by phase and amplitude errors that are well above roundoff, 
some of which will be consistent with undesired acausal modes (see the similar 
discussion in the previous subsection).  In effect, the numerically derived 
new basis could not be obtained (to machine accuracy) from an integration of 
purely outgoing wave solutions at infinity. 

Nevertheless, the thin-QR decomposition provides the route forward.  The 
idea is to use the initial choice for $\mathbf{V}$ at $r_*^\infty$ afforded 
by the simple basis and its related asymptotic expansion.  Then the thin-QR 
decomposition is computed numerically.  With this done, we compute from
$\mathbf{R}$ its inverse numerically.  After that, we use these values of 
$\mathbf{R}^{-1}$ at $r_*^\infty$ to transform the initial conditions for 
solving the recurrence relations, and we solve those again.  The resulting set 
of new asymptotic expansions have built into them proper causal behavior and 
also have $\rho(\mathbf{V}') = 1$.  In effect, $\mathbf{R}^{-1}$ serves as a 
\emph{pre-conditioner} on the linear system.  (Akcay et 
al.~\cite{AkcaWarbBara13} use a different means of pre-conditioning their 
boundary conditions for the outer solutions.)  So we are able to start inward 
integrations with ideal linear independence (by this measure) and obtain 
greatly reduced ill-conditioning (also by this measure) once the source 
region is reached (see Fig.~\ref{fig:condition_number} and six orders of 
magnitude improvement).  Empirically, we then find the full condition number, 
$\kappa$, is also improved by orders of magnitude. 

Since developing this thin-QR pre-conditioning technique, we have thus far 
not been able to find any comparable discussion in the literature.

\subsubsection{Numerical integration}

Having set the boundary conditions, our C code uses the 
Runge-Kutta-Prince-Dormand 7(8)~\cite{PrinDorm81} routine rk8pd of the GNU 
Scientific Library (GSL)~\cite{GSL} to obtain the homogeneous solutions (note that 
GSL documentation incorrectly labels rk8pd a 8(9) method).  We first integrate 
the outer homogeneous solutions from $r_*^\infty$ inward and then through the 
source region to $r_*^{\rm min}$.  We then integrate the inner homogeneous 
solutions from $r_*^H$ to $r_*^{\rm min}$.  Next, we switch to an integration 
over $\chi$ to compute Eqns.~\eqref{eq:CapitalCs} and acquire $C^{e/o,\pm}_i$.
In practice we also find it more efficient to determine the integrands of 
Eqn.~\eqref{eq:CapitalCs} using an LU decomposition of the Wronskian matrix.
Finally, we form the TD EHS as described in Sec.~\ref{sec:EHS}.

A final comment is warranted on the integration over the source region and 
the relative accuracies of various quantities.  In the sweep back over the 
source region, the Wronskian matrix elements are recomputed step-by-step 
alongside the normalization functions $c^{e/o}_i(r)$ within a broadened system
of ODEs.  When the Wronskian matrix is mildly ill-conditioned it becomes 
impractical to enforce the same accuracy criterion on the normalization 
coefficients as the homogeneous solutions that make up the elements of the 
Wronskian.  We instead modify the adaptive step size routine to demand high
accuracy $\sim 10^{-15}$ for the Wronskian elements while ignoring the 
fractional errors in the normalization coefficients unless they exceed 
$\sim 10^{-12}$.  This criterion does not really diminish the achievable 
accuracy in the coefficients, since the condition number of the Wronskian 
may reach or exceed $10^{3}$ near the low frequency limit of our double 
precision code (see Sec.~\ref{sec:near_static} for use of quad precision).  
It does, however, prevent the stepsize from being driven unreasonably small 
and halting the integration.

\subsection{Near-static modes}
\label{sec:near_static}

As mentioned in our step-by-step procedure, near-static modes 
($0 < |\o M|< 10^{-4}$) are a special case subject to separate numerical 
handling.  This problem has also been discussed in \cite{AkcaWarbBara13}.  
The ill-conditioning associated with the outer homogeneous solutions
continues to grow as $\omega\rightarrow 0$, despite the application 
of the orthogonalization technique described in the previous section.  
To compute modes with $10^{-6} \lesssim |\o M| \lesssim 10^{-4}$, we make 
use of three procedures.  Firstly, the thin-QR pre-conditioning discussed in
Sec.~\ref{sec:numericalGeneral}, which is used for all modes, helps to 
minimize the semi-condition number as much as possible.  Secondly, when a 
mode with frequency as low as this is encountered, we switch to the use of 
quad-precision routines to handle integration of the homogeneous solutions 
and source integrations (i.e., steps 3 and 4).  Thirdly, for a given $l,m$, 
we identify the lowest frequency mode $n = n'$ and for it we bypass the source 
integration and instead use the jump conditions to provide its normalization.

The semi-condition number scales roughly as $\rho \sim 10^2 \, (M\o)^{-2}$, 
as can be seen in Fig.~\ref{fig:condition_number}.  Once the condition number 
of the Wronskian matrix reaches $\sim 10^{10}$, too many digits ($\sim 10$) 
are being lost to make double precision calculations viable.  Resorting to 
128-bit floating point arithmetic is a computationally costly but effective 
way of proceeding.  At quad precision, much higher condition numbers 
($\lesssim 10^{22}$) can be tolerated.  Our quad-precision implementation is 
based on modified Numerical Recipes in C \cite{PresETC93} routines.  We 
switch to the Runge-Kutta-Cash-Karp 4(5) method for these calculations.  
While C compiler support for quad precision is available, its use is 
computationally costly on 64-bit hardware.  Fortunately, for broad regions 
of orbital parameter space these modes are few enough that growth in CPU 
time is manageable (see Fig.~\ref{fig:timing}).

\begin{figure}
\begin{center}
\includegraphics[scale=1.02]{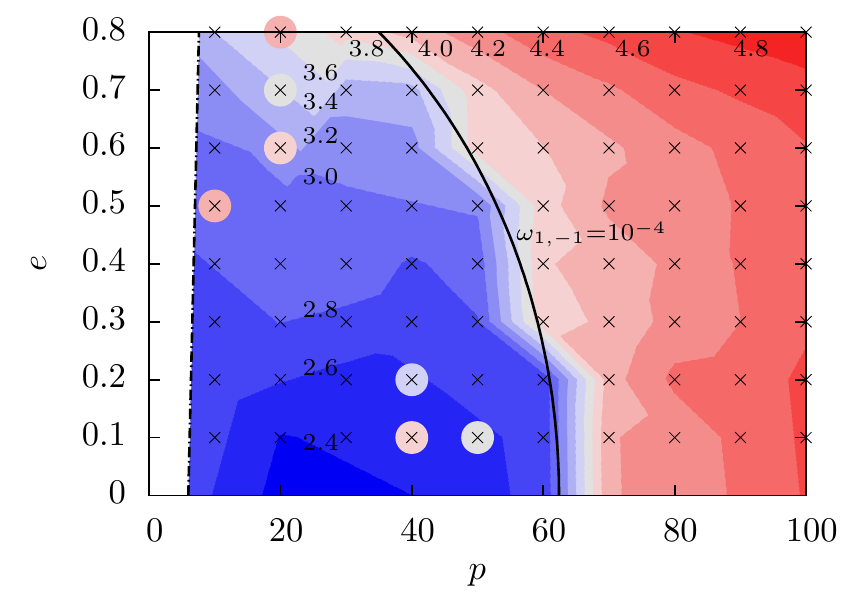}
\includegraphics[scale=1.02]{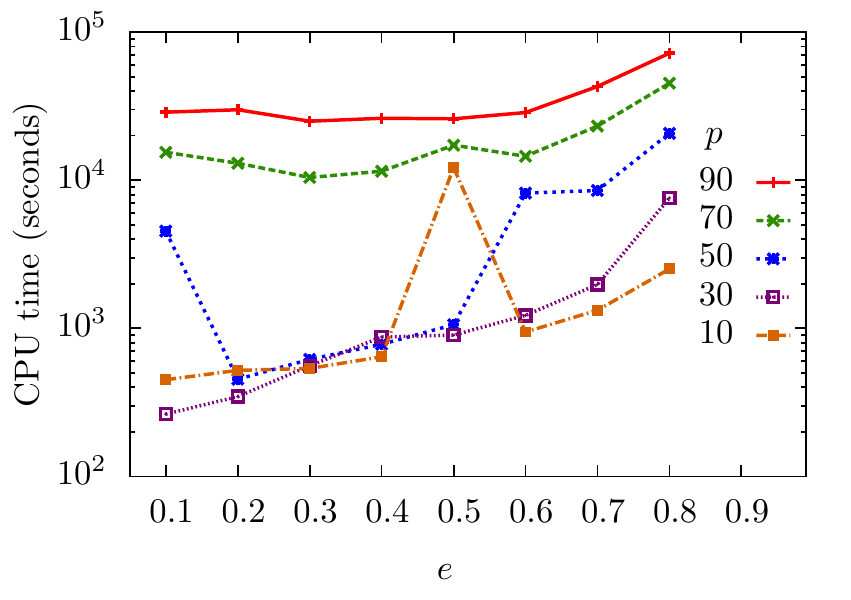}
\end{center}
\caption{Plots of CPU time for GSF calculations as a function of 
orbital parameter space location.  In the left panel labels give the 
$\log_{10}$ of CPU time in seconds for each contour.  The crosses indicate 
where models were computed.  Every orbit on the right of the solid curve 
utilizes quad precision.  Some orbits on the left lie near resonances, as 
indicated by local peaks in the contour plot caused by quad precision 
computing.  Slices of CPU time versus $e$ are shown in the right panel.  
GSF models require single-processor CPU times that range from 4 minutes to 
1 day. \label{fig:timing}} 
\end{figure}

The third element of the procedure focuses on the fact that for a given 
$l,m$ there is always one $n = n'$ that gives the lowest magnitude frequency, 
$\omega'$.  If $\omega'$ is small enough (and there are others like it for 
enough other $l$ and $m$), the quad precision integrations over the source 
might overly dominate the runtime of the code.  This is particularly a 
concern for wide separations and large eccentricities.  Fortunately, for 
each $l$ and $m$ there is a way of bypassing the source integration for this 
one $n'$ mode and obtaining its normalization coefficients more efficiently.  

We use odd parity to illustrate the method.  For a given $l,m$, the jump 
conditions in the TD for the MP amplitudes and their derivatives can be
written in vector form
\begin{align}
\left\llbracket \bscal{B} \right\rrbracket_p (t) 
\equiv
\bscal{B}^+(t,r_p) -\bscal{B}^-(t,r_p)
= 
\left[
\begin{array}{c}
\left\llbracket h_{t} \right\rrbracket_p \\
 \left\llbracket f h_{r} \right\rrbracket_p
\end{array}
\right], \q \q
\left\llbracket \pa_r \bscal{B} \right\rrbracket_p (t) 
\equiv
\pa_r \bscal{B}^+(t,r_p) - \pa_r \bscal{B}^-(t,r_p)
= 
\left[
\begin{array}{c}
\left\llbracket \pa_r h_{t} \right\rrbracket_p \\
\left\llbracket \pa_r \l f h_{r} \r \right\rrbracket_p
\end{array}
\right] .
\end{align}
These jump conditions can be obtained analytically from the field equations 
and the projections of the stress-energy tensor.  They are known to imply 
that the MP is $C^0$ and the radial derivative jump is some function of time, 
$\bscal{J}(t)$.  The jump conditions can be written as the difference
between the TD EHS or using Eqn.~\eqref{eqn:TD_EHS} as the difference of the 
Fourier sums over FD EHS
\begin{align}
\label{eqn:jumpsum}
\left[
\begin{array}{c}
\mathbf{0} \\
\bscal{J}(t)
\end{array}
\right]
=
\left[ 
\begin{array}{c} 
\left\llbracket \bscal{B} \right\rrbracket_p (t)  \\ 
\left\llbracket \pa_r \bscal{B} \right\rrbracket_p (t)  
\end{array} \right] 
=
\sum_{\o} 
\left\{ \left[ 
\begin{array}{c} \bscal{\ti B}^+ (r_p)\\ 
\pa_r \bscal{\ti B}^+ (r_p)
\end{array} \right] 
- \left[ 
\begin{array}{c}
 \bscal{\ti B}^- (r_p)\\ 
 \pa_r \bscal{\ti B}^- (r_p) 
 \end{array} \right] \right\} e^{-i\o t} .
\end{align}
Normally these conditions are used to check the convergence of the Fourier 
sums.  In the case of a near-static mode we first normalize all of the other 
$n \ne n'$ modes in the usual way.  Then the near-static mode is split out
of the sum in \eqref{eqn:jumpsum} and written explicitly in terms of its
individual homogeneous solutions and their normalization coefficients
\begin{align}
&e^{-i \o' t} \left[\begin{array}{cccc}  \bscal{\ti B}^+_0 &  \bscal{\ti B}^+_1 & - \bscal{\ti B}^-_0 & -  \bscal{\ti B}^-_1 \\
\pa_r \bscal{\ti B}^+_0 & \pa_r \bscal{\ti B}^+_1 & - \pa_r\bscal{\ti B}^-_0 & -  \pa_r\bscal{\ti B}^-_1 \end{array} \right] 
\left[\begin{array}{c} C^{o,+}_0 \\ C^{o,+}_1 \\ C^{o,-}_0 \\ C^{o,-}_1 \end{array} \right] = 
\left[ 
\begin{array}{c} 
\mathbf{0} \\
\bscal{J}(t)
\end{array} 
\right] 
- \sum_{\o\ne \o'} 
\left\{ \left[ 
\begin{array}{c} \bscal{\ti B}^+ \\ 
\pa_r \bscal{\ti B}^+ 
\end{array} \right] 
- \left[ 
\begin{array}{c}
 \bscal{\ti B}^- \\ 
 \pa_r \bscal{\ti B}^- 
 \end{array} \right] \right\} e^{-i\o t} .
\end{align}
In this expression, the function $\bscal{J}(t)$ is known analytically and 
all of the terms in the sum on the right have been computed by the standard
procedure.  On the left, the homogeneous solutions for $\omega'$ that make 
up the matrix are computed with quad precision and what remains are the four 
unknowns $C^{o,\pm}_{0/1}$.  This matrix equation is solved at an arbitrary 
time $t$ and in doing so we have obtained the normalization coefficients for
the troublesome mode without integrating over the source region.  It can be
applied for frequencies as small as $|\o|\sim 10^{-6} M^{-1}$. 

An objection might be raised that this ``spends'' the ability to use the jump 
conditions as a convergence check.  But in fact it remains possible to check 
the jumps at any other time within the radial period $T_r$.  Ultimately, the 
techniques presented in this section can be overwhelmed, since as $T_r$ 
becomes large the frequency $\Omega_r$ can become smaller than 
$10^{-4} M^{-1}$, which results in numerous near-static modes per multipole 
(see Fig.~\ref{fig:timing}). 

\subsection{Static modes with $l\ge 2$}
\label{sec:static}

Static modes are another special case and occur when $m=n=0$.  At zero 
frequency, some of the field amplitudes vanish identically, and spur a 
reduction of order in the constrained field equations and gauge equations.  
We discuss odd and even parity in turn.

\subsubsection{Odd-parity static modes}

Analytic homogeneous solutions to the static odd-parity Lorenz gauge 
field equations were first derived by Barack and Lousto \cite{BaraLous05}.
They showed that $\ti{h}_r=\ti{h}_2=0$ and wrote down the inner and outer 
solutions for $\ti{h}_t$ in terms of finite power series.  Here we express 
the solution in slightly different form
\begin{align}
\label{eqn:ht_plusMinus}
\ti{h}_t^- = \frac{r^2}{M} \sum_{k=0}^{l-1} a^{\text{odd}}_k 
\left( \frac{r}{M}\right)^k ,
\q \q
\ti{h}_t^+ = \ti{h}_t^- \ln{f} + \frac{M^2}{r}\sum_{k=0}^{l+2} 
b^{\text{odd}}_k \left(\frac{r}{M}\right)^k .
\end{align}
The determination of the power series coefficients is described in detail 
in Appendix \ref{sec:staticOdd}. 

\subsubsection{Even-parity static modes}

In this paper, we present for the first time analytic solutions for static 
even-parity modes in Lorenz gauge.  (We understand that equivalent analytic 
solutions have been derived recently by others \cite{NolaOtteWarb14} also.)
For static modes in even parity the reduction $\ti{h}_{tr} = \ti{j}_{t} = 0$ 
occurs.  The reduced constrained equations are sixth order and involve 
$\ti{h}_{tt}$, $\ti{h}_{rr}$, and $\ti{K}$.  We had a novel, if circuitous, route to 
discovering these analytic solutions, which we now present step-by-step.

\begin{enumerate}

\item 
\emph{Even-$l$ solution to odd-parity equations:} For static modes $m=n=0$, 
Eqns.~\eqref{eqn:ht_plusMinus} are used with odd $l$ to provide a necessary 
part of the MP.  There is however nothing to bar us from using an even $l$ 
in Eqns.~\eqref{eqn:ht_plusMinus}; these too are solutions to the odd-parity 
Lorenz gauge equations even if they serve no purpose in decomposing the MP.

\item
\emph{Solution to the Regge-Wheeler equation:} Armed with this ``even-$l$ 
solution to the odd-parity Lorenz gauge equations,'' we next form the 
gauge-invariant Cunningham-Price-Moncrief (CPM) \cite{CunnPricMonc78} function 
\be
\label{eqn:psiOdd}
{\ti \Psi}^{\text{odd}} (r) = \frac{r}{\la} 
\l \frac{d \ti{h}_t}{dr} - \frac{2}{r} \ti{h}_t \r .
\ee
Recall that $\la = (l+2)(l-1)/2$.  This master function satisfies the 
homogeneous Regge-Wheeler (RW) equation.  
See also \cite{MartPois05,HoppEvan10}.

\item
\emph{CPM master function to Zerilli master function:} Next use the 
Detweiler-Chandrasekhar transformation \cite{Chan75,ChanDetw75,Chan83} to 
obtain from the CPM function a solution to the homogeneous Zerilli equation 
\be
\label{eqn:psiEven}
{\ti \Psi}^{\text{even}} (r) = \frac{1}{\la(\la+1)} 
\left[ 
\l \la (\la+1) + \frac{9 M^2 f}{3M r + \la r^2} \r {\ti \Psi}^{\rm odd}
+
3 M f \frac{d {\ti \Psi}^{\rm odd}}{dr}
\right] .
\ee

\item 
\emph{MP amplitudes in RW gauge:} Use ${\ti \Psi}^{\rm even}$ to reconstruct 
the non-zero even-parity MP amplitudes in RW gauge.  For purposes of 
presentation, the expressions (see e.g. \cite{HoppEvan10}) simplify greatly 
by using Eqns.~\eqref{eqn:psiOdd} and \eqref{eqn:psiEven} to write the MP 
amplitudes in terms of $\tilde h_t$,
\begin{align}
\label{eqn:evenAsHt}
\ti{K}^{\text{RW}} 
=\frac{2M}{r^2 \la} \tilde{h}_t + \l 1 
+ \frac{2M}{r \la} \r \frac{d\ti{h}_t}{dr} ,
\q \q
\ti{h}_{rr}^{\text{RW}} 
 = - \frac{2M}{r^2 f^2} \tilde{h}_t + \frac{1}{f} \frac{d \tilde{h}_t}{dr}, 
\q \q
\ti{h}_{tt}^{\text{RW}} = 
- \frac{2M}{r^2} \tilde{h}_t + f \frac{d \tilde{h}_t}{dr} ,
\end{align}
where we have used the homogeneous field equations to remove higher 
derivatives of $\tilde{h}_t$.  Given analytic expressions for the even-$l$ 
solutions for $\ti{h}_t$ in step 1, we have obtained even-$l$ static solutions for
the MP in RW gauge.

\item 
\emph{Gauge vector for RW to Lorenz transformation:} We next seek a gauge 
vector to map the even-parity static MPs from RW to Lorenz gauge.
The gauge vector will satisfy the wave equation
\be
\label{eq:gaugeGen}
\nabla_\nu \nabla^\nu \Xi_\mu = \nabla^\nu \Bar p^{\rm RW}_{\mu \nu }.
\ee
The generator $\Xi_\mu$ can be decomposed \cite{HoppEvan13} akin to that 
shown in Sec.~\ref{sec:sph_harm}, and its even-parity part is
\be
\Xi_{a} = \sum_{l,m} 
\left[ {\delta_a}^t \xi^{l m}_t (t,r) 
+ {\delta_a}^r \xi^{l m}_r (t,r) \right]
Y_{l m},
\q \q
\Xi_{A} = \sum_{l,m} \xi^{l m}_{e} (t,r) Y_{A}^{l m} .
\label{eq:xiExpand}
\ee
We insert these into Eqn.~\eqref{eq:gaugeGen} and 
transform to the FD. Then we specialize to the static 
case (where $\tilde{\xi}_t = 0$), and are left with two coupled 
equations (after again dropping $lmn$ indices)
\begin{align}
\label{eqn:xi_field}
\frac{d^2\tilde{\xi}_e}{dr^2}+\frac{2M}{r^2 f} \frac{d\tilde{\xi}_e}{dr}
-\frac{2(\la+1)}{r^2 f}\tilde{\xi}_e+\frac{2}{r}\tilde{\xi}_r &= 0 ,
\\
\label{eqn:xir_field}
\frac{d^2\tilde{\xi}_r}{dr^2}
+\frac{2}{rf}\frac{d\tilde{\xi}_r}{dr}  
-\frac{2(\la+1)+2f}{r^2 f} \tilde{\xi}_r 
+ \frac{4(\la+1)}{r^3 f} \tilde{\xi}_e 
&= 
\frac{2}{rf}\ti{h}_{rr}^{\text{RW}}
- \frac{2}{rf}\ti{K}^{\text{RW}} 
+ \frac{d \ti{h}_{rr}^{\text{RW}}}{dr} 
-\frac{1}{f}\frac{d \ti{K}^{\text{RW}}}{dr} ,
\end{align}
where we have used the homogeneous relation $\ti{h}_{tt}^{\text{RW}}=f^2
\ti{h}_{rr}^{\text{RW}}$.  Solving Eqn.~\eqref{eqn:xi_field} for $\tilde{\xi}_r$ and 
inserting into Eqn.~\eqref{eqn:xir_field} yields a single fourth-order 
equation.  Further, we use Eqn.~\eqref{eqn:evenAsHt} and the $\ti h_t$ field 
equation to write the source term as a function of $\ti h_t$ and its first 
derivative
\begin{align}
\begin{split}
\label{eqn:xiE4th}
\hspace{3ex}
\frac{d^4\tilde{\xi}_e}{dr^4}
+
\frac{4r-2M}{r^2f}\frac{d^3\tilde{\xi}_e}{dr^3} 
-
\frac{4 r (\la + 1) - 4 M}{r^3 f}& \frac{d^2\tilde{\xi}_e}{dr^2} 
+
\frac{8M^2 - 4rM ( \la+2)}{r^5f^2} \frac{d\tilde{\xi}_e}{dr}
+\frac{4 (\la+1)(2M+r \la)}{r^5f^2}\tilde{\xi}_e  
= S_{\xi},
\\
S_{\xi} \equiv &
\frac{8M (\la+f)}{\la f^2 r^4} \ti{h}_t
+
\frac{8 M}{\la f r^3}\frac{d\ti{h}_t}{dr} .
\end{split}
\end{align}
Eqn.~\eqref{eqn:xiE4th} has four independent homogeneous solutions denoted 
by $\tilde{\xi}^{\pm}_{e,H0}$ and $\tilde{\xi}^{\pm}_{e,H1}$ and two independent inhomogeneous 
solutions (since the source has inner and outer instances) denoted by 
$\tilde{\xi}^{\pm}_{e,I}$.  Here the superscript $\pm$ indicates the solution that 
is regular at $r=\infty$ ($+$) or the horizon ($-$).

\item
\emph{Transformation to six independent Lorenz gauge homogeneous solutions:} 
Once the six solutions for the gauge generator have been obtained, we can 
use them to transform the even-parity static MP to Lorenz gauge and derive 
a complete set of homogeneous solutions.  The transformation is 
\cite{MartPois05}
\begin{align}
\label{eqn:push}
\ti{h}_{tt} = \ti{h}_{tt}^{\text{RW}}+\frac{2Mf}{r^2}\tilde{\xi}_r ,
\q \q
\ti{h}_{rr} = \ti{h}_{rr}^{\text{RW}}-\frac{2M}{r^2f}\tilde{\xi}_r-2\frac{d\tilde{\xi}_r}{dr} ,
\q \q
\ti{K} = \ti{K}^{\text{RW}}-\frac{2f}{r}\tilde{\xi}_r+\frac{2(\la+1)}{r^2}\tilde{\xi}_e.
\end{align}
Note that $\tilde{\xi}_r$ is recovered using Eqn.~\eqref{eqn:xi_field}.  We can now 
switch to the vector notation of Sec.~\ref{sec:matrixnotation} and write
\be
\bscal{\ti E} = \bscal{\ti E}^{\rm RW} + \bs{\Delta \tilde{\xi}} ,
\ee
with components
\be
\bscal{\ti E} 
= 
r \left[
\begin{array}{c}
\ti{h}_{tt}  \\ 
0 \\ 
f^2 \ti{h}_{rr} \\ 
\ti{K}
\end{array}
\right],
\q \q
\bs{\Delta \tilde{\xi}}
\equiv
\left[
\begin{array}{c}
\dfrac{2Mf}{r}\tilde{\xi}_{r} \\
0 \\
 -\dfrac{2M f}{r}\tilde{\xi}_{r} -2 r f^2\dfrac{d\tilde{\xi}_{r}}{dr} \\
-2f\tilde{\xi}_{r}+\dfrac{2(\la+1)}{r}\tilde{\xi}_{e}
\end{array}
\right] ,
\ee
and with $\bscal{\ti E}^{\rm RW}$ being obvious.  The zeros in the second 
row follow from $\ti h_{tr}$ vanishing in both Lorenz and RW gauges when 
$\o =0$.  We denote the six Lorenz gauge homogeneous solutions by 
$\bscal{\ti E}_{0}^\pm$, $\bscal{\ti E}_{1}^\pm$, and 
$\bscal{\ti E}_{2}^\pm$ (recall Sec.~\ref{sec:homog}).  The first four 
Lorenz gauge homogeneous solutions derive from the homogeneous solutions to 
Eqn.~\eqref{eqn:xiE4th},
\be
\bscal{\ti E}_{0}^\pm = \bs{\Delta \tilde{\xi}}^\pm_{H0}, \q \q
\bscal{\ti E}_{1}^\pm = \bs{\Delta \tilde{\xi}}^\pm_{H1}. 
\ee
The final two are found by transforming from the RW gauge MP amplitudes of 
step 4 with the inhomogeneous solutions to Eqn.~\eqref{eqn:xiE4th},
\be
\bscal{\ti E}_{2}^\pm =
\bscal{\ti E}^{\rm RW,\pm} + \bs{\Delta \tilde{\xi}}^\pm_{I} .
\ee
The extensive expressions for $\tilde{\xi}^{\pm}_{e,H0}$, 
$\tilde{\xi}^{\pm}_{e,H1}$, and $\tilde{\xi}^{\pm}_{e,I}$
can be found in Appendix \ref{sec:staticEven}.

\end{enumerate}

\subsection{Low-multipole modes}
\label{sec:low_multipole}

The low-multipole ($l=0,1$) components of the MP are as essential to the 
GSF as the radiative modes.  Solutions were first given by Zerilli 
\cite{Zeri70}.  These solutions, specialized to circular orbits, were then 
transformed to Lorenz gauge by Detweiler and Poisson \cite{DetwPois04}.  
Low-multipole mode calculations for circular orbits were considered in 
\cite{BaraLous05,BaraSago07}.  Their solution was extended to eccentric 
orbits in \cite{BaraSago10,AkcaWarbBara13} using the method of EHS.

\subsubsection{Even-parity dipole mode}

In the case of the even-parity dipole mode $l=1$, $m=1$, the amplitude 
$\ti{G}$ is not defined [see Eqn.~\eqref{eq:MP} and note that $Y_{AB}$ 
is not defined for $l<2$].  The fully constrained field equations 
\eqref{eqn:fieldEqsVector} are unaffected however.  The vanishing of $\ti G$ 
does add the subtlety that the individual homogeneous solutions to 
Eqn.~\eqref{eqn:fieldEqsVector} will not, in general, satisfy the Lorenz 
gauge conditions, Eqn.~\eqref{eqn:LGCondEven}.

Numerically, the even-parity dipole mode requires no special treatment.  As 
usual, we use Eqn.~\eqref{eq:CapitalCs} to integrate through the source 
region and find $C_i^{e,\pm}$.  We then find that the solution that results 
from linear superposition of the normalized modes in Eqn.~\eqref{eqn:FD_EHS} 
\emph{does} satisfy the gauge conditions, a byproduct of the source terms 
being consistent with the Bianchi identities.

\subsubsection{Odd-parity dipole mode}

In the case of the odd-parity dipole mode $l=1$, $m=0$, the amplitude
$\ti{h}_2$ is not defined [see Eqn.~\eqref{eq:MP} and note that $X_{AB}$ 
is not defined for $l<2$].  As with the even-parity case, this does not 
affect the fully constrained field equations.  When $\o \ne 0$, this mode 
requires no special treatment.  We find that after normalization and 
superposition, the solution does satisfy the gauge condition.

The static mode, $l=1$, $m=0$, $n=0$, must be handled separately.  In this 
case we use the analytic homogeneous solutions~\cite{DetwPois04}
\begin{align}
\ti{h}_t^{-} = \frac{r^2}{M} ,
\q \q
\ti{h}_t^{+} = \frac{M^2}{r},
\end{align}
and proceed as usual to obtain the FD EHS.

\subsubsection{Monopole mode}

In the case of the monopole mode, $l=m=0$, the amplitudes $\ti{j}_t$, 
$\ti{j}_r$, and $\ti{G}$ are not defined [see Eqn.~\eqref{eq:MP} and note 
that $Y_A$ and $Y_{AB}$ are not defined for $l=0$].  Again, the fully 
constrained field equations are unaffected and no special treatment is 
required to obtain the particular solution as long as $n \ne 0$.  
 
However, the monopole static mode $l=m=n=0$ is exceptional.  The system is 
fourth order and has four independent homogeneous solutions \cite{Bern07}, 
which also satisfy the Lorenz gauge conditions 
\begin{align}
\begin{split}
\hspace{-.5ex}
\bscal{\ti E}^{-}_0 &= 
\frac{1}{f r^3} \left[ 
\begin{array}{c}
-Mf^2(r^2+2 M r+4M^2) \\
0 \\
r^3-Mr^2-2M^2r+12M^3 \\ 
f^2 r(r^2+2 M r+4M^2)
\end{array}
\right],
\q
\bscal{\ti E}^{+}_0 = 
\frac{1}{f^2r} \left[ 
\begin{array}{c}
f^2(3 M -r) \\ 
 0  \\ 
M \\  
 0
 \end{array}
\right],
\q
\bscal{\ti E}^{+}_1 = 
\frac{1}{f^2 r^4} \left[ 
\begin{array}{c}
f^2 M^4 \\
0 \\
M^3(2r-3M) \\
- r f^2 M^3
 \end{array}
\right], \\
\hspace{-.5ex}
\bscal{\ti E}^{+}_2 &= 
\frac{1}{f^2 r^4}
\left[ \begin{array}{c} 
M f^2 \left[r(4M-3r)(M+r)+(8M^3-r^3) \ln{f}+ 8 M^3 
\ln{\l \dfrac{r}{M} \r} \right] \\ 
0 \\
fr(r^3-Mr^2-2M^2r+12M^3)\ln{f}+8M^3(2r-3M)
\ln{\l\dfrac{r}{M} \r} -Mr(r^2-5Mr+12M^2) \\
f^2 r \left[ (r^3-8M^3)\ln{f}-8M^3\ln{\l\dfrac{r}{M}\r}-Mr(r+4M)\right] 
\end{array} \right] .
\end{split}
\end{align}
Recall from Sec.~\ref{sec:static} that $\ti h_{tr}$ vanishes for static modes,
as indicated by the zeros in the second rows of these expressions.

We have made a particular choice with this basis.  The solutions 
$\bscal{\ti E}^+_1$ and $\bscal{\ti E}^+_2$ are the only independent ones 
that are regular at $r=\infty$.  Then, $\bscal{\ti E}^{-}_0$ is the only 
solution that is regular at the horizon and does not perturb the 
mass-energy of the black hole \cite{DolaBara13} (at the horizon).  This leaves 
$\bscal{\ti E}^{+}_0$.  Ordinarily, we would expect two homogeneous solutions 
on the horizon side and two on the infinity side.  But all that is really 
required are four independent solutions and regularity.  This last solution 
is independent and its only irregularity at $r=\infty$ is the well-known 
property of Lorenz gauge that $\ti{h}_{tt}$ approaches a constant as 
$r\rightarrow\infty$~\cite{BaraLous05}.  This behavior leads to a required 
rescaling of the time coordinate \cite{SagoBaraDetw08,BaraSago10,DolaBara13,
AkcaWarbBara13}.  It is precisely what is necessary that the solution perturb 
the mass $M \rightarrow M + \mu \mathcal{E}$ of the spacetime in the region 
exterior to the particle orbit \cite{DetwPois04}.  With this complete set of 
homogeneous solutions, we form the FD EHS.  Rather than using the expression 
in Eqn.~\eqref{eqn:FD_EHS}, for this special case the normalization is
\begin{align}
\bscal{\ti E}^{-} (r) = 
C^{e,-}_0
\bscal{\ti E}^-_0 ,
\q \q
\bscal{\ti E}^+ (r) = 
\sum_{i=0}^2 C^{e,+}_{i}
\bscal{\ti E}^+_i .
\end{align}
Our route to the solution for this mode differs from that of
Akcay et al.~\cite{AkcaWarbBara13} but of course the two approaches are 
ultimately equivalent. 

\section{Additional numerical results}
\label{sec:results}

We give in this section a sampling of added numerical results from computing 
the GSF and discuss the range of applicability of the code.  As mentioned 
in the Introduction, astrophysical EMRI sources are expected to have 
eccentricities as high as $e \simeq 0.8$.  This expectation has motivated 
our effort to develop an efficient and accurate code capable of widely 
spanning $p$ and $e$ space.  

\subsection{GSF results and their accuracies}

We first compare our code to results from \cite{AkcaWarbBara13} for a mildly 
eccentric orbit ($e=0.2$, $p=7.0$).  Table \ref{tbl:e=0.2_p=7.0} shows 
values of the $t$ and $r$ components of both the conservative and dissipative 
parts of the GSF for a set of locations on the orbit.  Our values match 
closely those of Akcay et al.  Our results are presented with the number of 
digits we believe are significant.  Their values were presented with 
uncertainties in the least significant digit, so we have rounded their values 
and present in the table only fully significant digits for comparison.  The 
two codes agree for this orbit to within four to seven digits, but do differ 
in many cases in the least significant figure.  We estimate errors in our 
calculation by examining sensitivity in Fourier convergence and in truncating 
the MSR.  The discrepancy between our two codes likely reflects the 
difficulty in determining \emph{absolute error} when truncating a Fourier sum 
or power series.  In terms of speed, our code generates GSF data rapidly 
($\sim 15$ minutes) for an orbit with an eccentricity as low as this.  CPU 
runtimes can be nearly two orders of magnitude greater for high-eccentricity 
wide-separation orbits (see Fig.~\ref{fig:timing}) where the code begins to 
switch on intermittent use of quad precision.

\begin{table}
\begin{center}
\caption{Comparison of GSF data from two different codes.  We give self-force 
values for an orbit with $p = 7.0$ and $e = 0.2$ and present only significant 
figures for the data from our code (rows without parentheses).  Our results 
are compared to those of Akcay et al.~\cite{AkcaWarbBara13} (parentheses), 
where we have rounded the last digit from values in their table to retain 
only fully significant digits.  Our code took approximately 15 minutes on a 
single core to generate all of the GSF data in this table.  
\label{tbl:e=0.2_p=7.0}} 
\begin{tabular}{c | c | c | c | c}
\hline \hline
$\chi$	&	$(M/\mu)^2 F^t_\text{cons}$	&	
$(M/\mu)^2 F^t_\text{diss}$		& 
$(M/\mu)^2 F^r_\text{cons}$		& 
$(M/\mu)^2 F^r_\text{diss}$	\\
\hline
\multirow{2}{*}{0}	& 0 & $-4.06328\times10^{-3}$ & 
$3.35760\times10^{-2}$ & 0 \\
	&	(0)	&	($-4.063302\times10^{-3}$)	&	
($3.357606\times10^{-2}$)	&	(0)	\\ 
\hline
\multirow{2}{*}{$\pi/4$}	& $8.6473\times10^{-4}$ & 
$-2.15691\times10^{-3}$ &  $2.909881\times10^{-2}$ & 
$4.734956\times10^{-3}$ \\
	&	($8.6472\times10^{-4}$)		&	
($-2.156923\times10^{-3}$)			&	
($2.909881\times10^{-2}$)		&	
($4.734956\times10^{-3}$)	\\ 
\hline
\multirow{2}{*}{$\pi/2$}	& $8.28613\times10^{-4}$ & 
$-2.5168\times10^{-4}$ & $2.125032\times10^{-2}$ & 
$3.204189\times10^{-3}$ \\
	&	($8.28611\times10^{-4}$)	&	
($-2.516803\times10^{-4}$)			&	
($2.125034\times10^{-2}$)	&	($3.204190\times10^{-3}$)\\ 
\hline
\multirow{2}{*}{$3\pi/4$}	& $4.60749\times10^{-4}$ & 
$ -1.1241\times10^{-5}$ &  $1.590147\times10^{-2}$ & 
$9.63378\times10^{-4}$ \\
	&	($4.60750\times10^{-4}$)	&	
($-1.124092\times10^{-5}$)	&	($1.590149\times10^{-2}$)&	
($9.633734\times10^{-4}$)	\\ 
\hline
\multirow{2}{*}{$\pi$}	& 0 & $-3.4613\times10^{-5}$ & 
$1.40888\times10^{-2}$ & 0 \\
	&	(0)		&	($-3.461416\times10^{-5}$)&	
($1.408877\times10^{-2}$)	&	(0)		\\
\hline \hline
\end{tabular}
\end{center} 
\end{table}

We next give in Table \ref{tbl:e=0.1} a set of numerical values for the $t$ 
and $r$ components of the GSF for eccentricity $e = 0.1$ and a range of 
orbital separations.  The full regularized GSF is given at points all 
around one radial libration.  The dissipative and conservative parts can 
be reconstructed through averaging and differencing values across conjugate 
points on the orbit using expressions in Sec.~\ref{sec:sf}.  The $\vp$ 
component of the GSF can be obtained from orthogonality.  We list only 
significant digits.  It is clear that for low eccentricity our code generally 
achieves accuracies of 7 to 10 decimal places.  As we discussed 
in the Introduction, accuracy of 8 or more decimal places is required to keep 
dephasing errors below $\delta\Phi_r \simeq 10^{-2}$ when $\ve = 10^{-6}$.  
The requirement is obviously eased if $\ve = 10^{-5}$.  The results in 
Table \ref{tbl:e=0.1} indicate that our error criterion is attained for 
$e=0.1$.

Remarkably, the accuracy of our code improves as the orbital separation 
increases, as can be seen in Fig.~\ref{fig:contour}.  This trend emerges from 
conflicting aspects of the algorithm.  One aspect, as 
Fig.~\ref{fig:condition_number} shows, is that integration from large $r_*$ 
to the libration region is accompanied by growth in the semi-condition
number of the outgoing homogeneous solutions.  In integrating from 
$r_* \sim 10^2 M$ to $r_* \sim 10 M$, the semi-condition number grows by 
two orders of magnitude.  For larger $p$, many modes will thus have smaller 
$\rho$ in the libration region, leading generally to more accurate GSF
values.  In contrast, larger radius orbits are more likely to yield 
near-static modes (see the $M \o = 10^{-4}$ curve in Fig.~\ref{fig:timing}).  
Yet, as explained in Sec.~\ref{sec:near_static}, when this occurs the
algorithm switches on quad-precision routines for these modes.  We posit that 
the clear benefit of lower semi-condition numbers at large $p$ outweighs 
difficulties induced by added near-static modes, especially as the algorithm 
adapts to the presence of these modes.  The price to be paid is significant
increase in CPU time as larger $p$ orbits are computed.

The situation changes as we consider higher eccentricities.  Table 
\ref{tbl:e=0.5} shows equivalent information for orbits with $e=0.5$.  At 
this eccentricity the GSF values have between 5 and 7 decimal places of 
accuracy.  As before, accuracies improve with wider separations.  In 
App.~\ref{sec:addedGSFtables} we provide two more tables, with $e = 0.3$ 
and $e = 0.7$.  At $e = 0.3$ accuracies are intermediate, with 6 to 9 
decimal places, but at $e = 0.7$ accuracies drop to 3 to 5 significant 
figures.  The trend in accuracy is best displayed semi-quantitatively in 
Fig.~\ref{fig:contour}, where labeled contours trace the iso-surfaces of 
relative error in the GSF.  The general trend of improvement in accuracy 
(in our code) with increasing $p$ is evident, as is the more severe fall-off 
with increasing $e$.  It is worth noting how uniform the trends in accuracy 
are.  This uniformity is in contrast to CPU runtimes seen in
Fig.~\ref{fig:timing}, evidence that the code trades speed for accuracy 
when necessary.  With an error goal of $10^{-7}$ (useful if we consider 
$\ve \gtrsim 10^{-5}$ or are willing to relax to $\delta\Phi_r \simeq
10^{-1}$), our code can directly supply the GSF for long-term orbit 
integrations for $e \lesssim 0.4$-$0.5$ over most of the range of $p$. 

\begin{table}[p]
\begin{center}
\caption{\label{tbl:e=0.1}
GSF results for $e = 0.1$ and a range of $p$.  We present the $t$ and $r$ 
components of the full regularized self-force at a set of points around a 
complete radial libration.  Dissipative and conservative parts can be
obtained by addition or subtraction across conjugate points on the orbit 
according to Eqns.~\eqref{eqn:consDissSF}.  The $\varphi$ component can be 
recovered from the orthogonality relation $F^{\a}u_{\a}=0$.  Results for 
additional eccentricities are found in Table \ref{tbl:e=0.5} and in 
Appendix \ref{sec:addedGSFtables}.
}
\begin{tabular}{c|c|c|c|c|c|c}
\hline \hline
 & $\chi$ & $p=10$ & $p=20$ & $p=30$ & $p=60$ & $p=90$ \\
\hline
\multirow{8}{*}{$F^t$} & 0 & $-2.262915\times 10^{-4}$ & 
$-5.259858\times 10^{-6}$ & $-6.4546731\times 10^{-7}$ & 
$-1.919788169\times 10^{-8}$ & $-2.504370129\times 10^{-9}$ \\
&$\pi/4$ &$ 1.198168\times 10^{-4}$ & $6.729545\times 10^{-5}$ & 
$  2.7706691\times 10^{-5}$ & $5.382433506\times 10^{-6}$ & 
$  2.00678099\times 10^{-6}$ \\
&$\pi/2$ & $2.767753\times 10^{-4}$ & $8.716476\times 10^{-5}$ & 
$3.4907828\times 10^{-5}$ & $6.689040469\times 10^{-6}$ & 
$2.486733999\times 10^{-6}$ \\
&$3\pi/4$ &$ 1.810961\times 10^{-4}$ & $5.416762\times 10^{-5}$ & 
$  2.1556094\times 10^{-5}$ & $4.107340582\times 10^{-6}$ & 
$  1.524147563\times 10^{-6}$ \\
&$\pi$ & $-3.133613\times 10^{-5}$ & $-8.374829\times 10^{-7}$ & 
$-1.0622118\times 10^{-7}$ & $-3.201044762\times 10^{-9}$ & 
$-4.156911071\times 10^{-10}$ \\
&$5\pi/4$ &$ -2.601123\times 10^{-4}$ & $-5.626206\times 10^{-5}$ & 
$  -2.1821319\times 10^{-5}$ & $-4.115331358\times 10^{-6}$ & 
$  -1.525185847\times 10^{-6}$ \\
&$3\pi/2$ & $-4.334752\times 10^{-4}$ & $-9.111977\times 10^{-5}$ & 
$-3.5403737\times 10^{-5}$ & $-6.703942473\times 10^{-6}$ & 
$-2.488674896\times 10^{-6}$ \\
&$7\pi/4$ & $ -4.52304\times 10^{-4}$ & $-7.519925\times 10^{-5}$ & 
$  -2.8683093\times 10^{-5}$ & $-5.411581273\times 10^{-6}$ & 
$  -2.010582879\times 10^{-6}$ \\
\hline
\multirow{8}{*}{$F^r$} & 0 & $1.606774\times 10^{-2}$ & 
$4.972162\times 10^{-3}$ & $2.3630073\times 10^{-3}$ & 
$6.306313760\times 10^{-4}$ & $2.863538695\times 10^{-4}$ \\
&$\pi/4$ &$  1.544991\times 10^{-2}$ & $4.734491\times 10^{-3}$ & 
$  2.2459999\times 10^{-3}$ & $5.984342621\times 10^{-4}$ & 
$  2.71591791\times 10^{-4}$ \\
&$\pi/2$ & $1.360189\times 10^{-2}$ & $4.167538\times 10^{-3}$ & 
$1.9721189\times 10^{-3}$ & $5.238329929\times 10^{-4}$ & 
$2.374664601\times 10^{-4}$ \\
&$3\pi/4$ &$ 1.180704\times 10^{-2}$ & $3.628645\times 10^{-3}$ & 
$  1.7134645\times 10^{-3}$ & $4.538465637\times 10^{-4}$ & 
$  2.055245423\times 10^{-4}$ \\
&$\pi$ & $1.105404\times 10^{-2}$ & $3.413109\times 10^{-3}$ & 
$1.6107881\times 10^{-3}$ & $4.262250069\times 10^{-4}$ & 
$1.929393281\times 10^{-4}$ \\
&$5\pi/4$ &$  1.163054\times 10^{-2}$ & $3.622782\times 10^{-3}$ & 
$  1.712561\times 10^{-3}$ & $4.538069014\times 10^{-4}$ & 
$  2.055180646\times 10^{-4}$ \\
&$3\pi/2$ & $1.322747\times 10^{-2}$ & $4.155439\times 10^{-3}$ & 
$1.9702746\times 10^{-3}$ & $5.237528164\times 10^{-4}$ & 
$2.374533990\times 10^{-4}$ \\
&$7\pi/4$ & $ 1.506292\times 10^{-2}$ & $4.722305\times 10^{-3}$ & 
$  2.244163\times 10^{-3}$ & $5.983551871\times 10^{-4}$ & 
$  2.71578943\times 10^{-4}$ \\
\hline \hline
\end{tabular}
\end{center} 
\end{table}

\begin{table}[p]
\begin{center}
\caption{
\label{tbl:e=0.5}
Same as Table \ref{tbl:e=0.1} but with $e=0.5$.
}
\begin{tabular}{c|c|c|c|c|c|c}
\hline \hline
 & $\chi$ & $p=10$ & $p=20$ & $p=30$ & $p=60$ & $p=90$ \\
\hline
\multirow{8}{*}{$F^t$} & 0 & $-3.6577\times 10^{-3}$ & 
$-7.31834\times 10^{-5}$ & $-8.45794\times 10^{-6}$ & 
$-2.403093\times 10^{-7}$ & $-3.110084\times 10^{-8}$ \\
&$\pi/4$ &$ 2.3230\times 10^{-3}$ & $5.49789\times 10^{-4}$ & 
$2.18176\times 10^{-4}$ & $4.230920\times 10^{-5}$ & 
$1.583867\times 10^{-5}$ \\
&$\pi/2$ & $2.1668\times 10^{-3}$ & $4.64492\times 10^{-4}$ & 
$1.79705\times 10^{-4}$ & $3.379164\times 10^{-5}$ & 
$1.251105\times 10^{-5}$ \\
&$3\pi/4$ &$ 6.5637\times 10^{-4}$ & $1.45139\times 10^{-4}$ & 
$5.53475\times 10^{-5}$ & $1.020907\times 10^{-5}$ & 
$3.752960\times 10^{-6}$ \\
&$\pi$ & $1.0093\times 10^{-6}$ & $1.68029\times 10^{-8}$ & 
$1.87076\times 10^{-9}$ & $5.493613\times 10^{-11}$ & 
$7.802583\times 10^{-12}$ \\
&$5\pi/4$ &$  -6.2100\times 10^{-4}$ & $-1.44416\times 10^{-4}$ & 
$-5.52583\times 10^{-5}$ & $-1.020630\times 10^{-5}$ & 
$-3.752584\times 10^{-6}$ \\
&$3\pi/2$ & $-1.6155\times 10^{-3}$ & $-4.54743\times 10^{-4}$ & 
$-1.78586\times 10^{-4}$ & $-3.375934\times 10^{-5}$ & 
$-1.250680\times 10^{-5}$ \\
&$7\pi/4$ &$ -3.3431\times 10^{-3}$ & $-5.74637\times 10^{-4}$ & 
$  -2.21391\times 10^{-4}$ & $-4.240869\times 10^{-5}$ & 
$  -1.585176\times 10^{-5}$ \\
\hline
\multirow{8}{*}{$F^r$} & 0 & $3.3855\times 10^{-2}$ & 
$9.08159\times 10^{-3}$ & $4.29527\times 10^{-3}$ & 
$1.154902\times 10^{-3}$ & $5.267282\times 10^{-4}$ \\
&$\pi/4$ &$ 3.0193\times 10^{-2}$ & $7.61709\times 10^{-3}$ & 
$3.56654\times 10^{-3}$ & $9.498506\times 10^{-4}$ & 
$4.317999\times 10^{-4}$ \\
&$\pi/2$ & $1.5020\times 10^{-2}$ & $4.30288\times 10^{-3}$ & 
$2.00992\times 10^{-3}$ & $5.283582\times 10^{-4}$ & 
$2.387940\times 10^{-4}$ \\
&$3\pi/4$ &$ 6.3892\times 10^{-3}$ & $1.86095\times 10^{-3}$ & 
$8.61808\times 10^{-4}$ & $2.238814\times 10^{-4}$ & 
$1.007354\times 10^{-4}$ \\
&$\pi$ & $3.9352\times 10^{-3}$ & $1.12769\times 10^{-3}$ & 
$5.19946\times 10^{-4}$ & $1.345001\times 10^{-4}$ & 
$6.043324\times 10^{-5}$ \\
&$5\pi/4$ &$ 6.2762\times 10^{-3}$ & $1.85640\times 10^{-3}$ & 
$8.61052\times 10^{-4}$ & $2.238461\times 10^{-4}$ & 
$1.007295\times 10^{-4}$ \\
&$3\pi/2$ & $1.3115\times 10^{-2}$ & $4.24160\times 10^{-3}$ & 
$2.00050\times 10^{-3}$ & $5.279454\times 10^{-4}$ & 
$2.387264\times 10^{-4}$ \\
&$7\pi/4$ &$ 2.2613\times 10^{-2}$ & $7.40405\times 10^{-3}$ & 
$3.53587\times 10^{-3}$ & $9.485827\times 10^{-4}$ & 
$4.315958\times 10^{-4}$ \\
\hline \hline
\end{tabular}
\end{center} 
\end{table}

For eccentricities above $e = 0.5$ (or in fact above $e = 0.25$ for 
$p \lesssim 10$) computing the full GSF accurately is more problematic, and 
the code, by itself, is not able to meet the goal of $\delta\Phi_r = 0.01$ 
if $\ve = 10^{-6}$.  (For an IMRI, though, with $\ve = 10^{-3}$ we might 
compute inspirals with eccentricities as high as $e \lesssim 0.5$-$0.6$.)  
One recourse would be to switch over much of the computation to 128-bit 
arithmetic, but doing so on 64-bit architecture would be expensive.  So, can 
eccentricities of $\simeq 0.8$ be reached and still maintain the required 
error tolerance?  We believe the answer is yes and propose a hybrid approach.  

The present difficulty stems from asking too much of a single numerical 
method.  Recall that the first-order GSF determines both the adiabatic 
inspiral and its part of the post-1-adiabatic corrections (with additional
correction coming eventually from the orbit-averaged part of the second-order 
GSF).  Hence, we need the code to provide the orbit-averaged part of the 
first-order GSF to a fractional accuracy $\epsilon_0$ that is 
$\mathcal{O}(\ve)$ better than the accuracy $\epsilon_1$ it provides in the 
oscillatory part of the GSF [see the argument centered around 
Eqn.~\eqref{eqn:phase}].  This viewpoint suggests splitting the task, with 
a separate code providing the gravitational wave fluxes that drive the 
inspiral (i.e., post-0-adiabatic) and the Lorenz gauge code providing the 
conservative and oscillatory part of the dissipative GSF (post-1-adiabatic).  
In such a hybrid scheme, the present code must needs only provide the 
oscillatory GSF with relative errors of, say, 
$\epsilon_1 \simeq 10^{-4}$-$10^{-3}$.  The flux code would need to give 
the orbit-averaged force to accuracy of $\epsilon_0 \lesssim 10^{-8}$.  A 
Regge-Wheeler-Zerilli (RWZ) code can achieve this latter accuracy and would 
not add significant computational burden. 

\begin{figure}
\begin{center}
\includegraphics[scale=1.0]{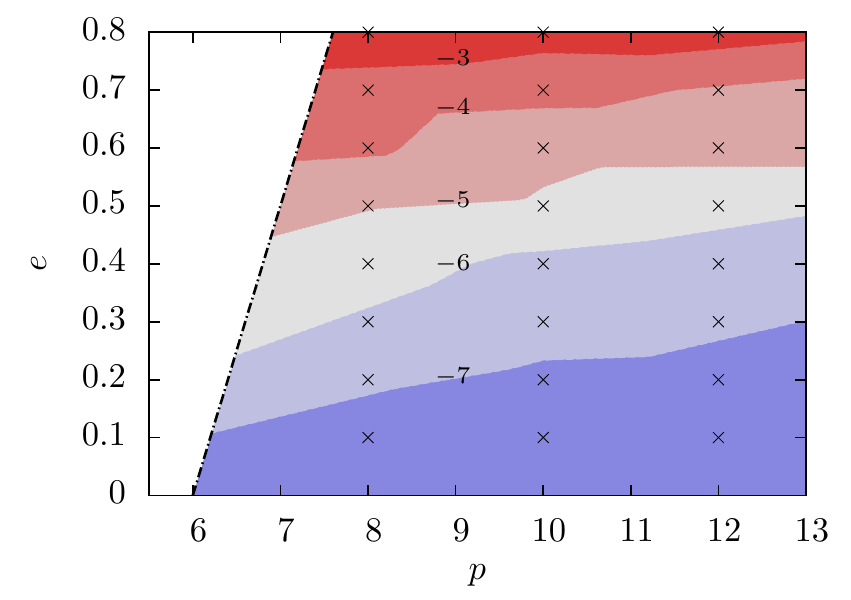}
\includegraphics[scale=1.0]{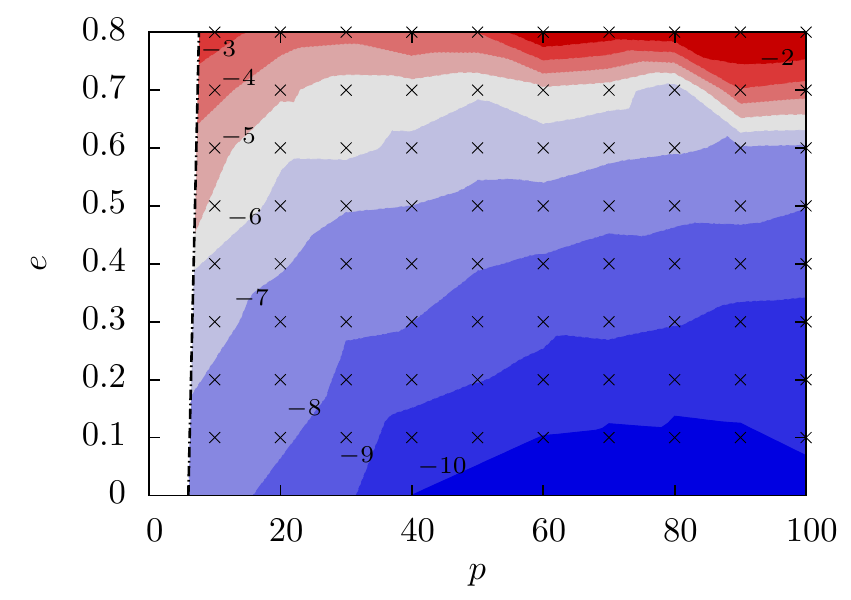}
\end{center}
\caption{Contours of relative errors in the GSF.  A grid of orbital
parameters is chosen (crosses) and the GSF is calculated.  Resulting relative 
errors are used to generate contour levels of relative accuracy.  Numerical 
labels indicate the $\log_{10}$ of the relative error of each contour. 
\label{fig:contour}} 
\end{figure}

\subsection{Improving the GSF with energy and angular momentum fluxes and 
a hybrid approach}
\label{sec:gsf_fluxes}

To assess how this hybrid scheme might work, we first discuss how fluxes 
are extracted from the Lorenz gauge code and compare them to computed local 
rate of change of work and torque.  Energy and angular momentum fluxes can 
be read off if the asymptotic values of the Zerilli-Moncrief (ZM), 
$\Psi_{lm}^{\rm even}$, and CPM, $\Psi_{lm}^{\rm odd}$, master functions 
\cite{MartPois05,HoppEvan10} are available.  When $l+m$ is even we use 
$\Psi_{lm} = \Psi^{\rm even}_{lm}$ and when $l+m$ is odd we use 
$\Psi_{lm} = \Psi^{\rm odd}_{lm}$.  Functions are evaluated at both
asymptotic limits, with $\Psi^+_{lm}$ being the amplitude at $r=\infty$ and 
$\Psi^-_{lm}$ being the one at $r=2M$.  See \cite{HoppEvan10} and their 
Sec.~IV\,B for flux expressions in terms of $\Psi_{lm}^+$ and $\Psi_{lm}^-$.

Expressed in terms of FD amplitudes, the ZM and CPM master functions are 
related to Lorenz gauge amplitudes by
\begin{align}
\label{eqn:master_Lorenz}
{\ti \Psi}^{\rm even}_{lmn} (r) = 
\frac{r}{\la+1} 
\left[ \ti{K} + \frac{f}{\La} \l  f \ti{h}_{rr} - r \pa_r \ti{K} \r \right]
- \frac{2 f}{\La} \ti{j}_r + r \ti{G}, 
\q \q
\ti{\Psi}^{\rm odd}_{lmn} (r) =
\frac{r}{\la} \l \pa_r \ti{h}_t + i \o \ti{h}_r  - \frac{2}{r} \ti{h}_t \r ,
\end{align}
where we define $\La \equiv \la + 3M/r$.  The master functions have
asymptotic running wave behavior 
${\ti \Psi}_{lmn}^\pm \l r_* \to \pm\infty \r = C^\pm_{lmn} e^{\pm i \o r_*}$
and the coefficients can be obtained from the asymptotic behavior of the 
Lorenz gauge amplitudes.  [Note, the $C_{lmn}^\pm$ here are not the same as 
those in Eqn.~\eqref{eq:CapitalCs}.]  Having made these connections to Lorenz 
gauge, we use standard expressions for the fluxes 
\begin{align}
\langle\dot{E}\rangle = 
\sum_{lmn}\frac{\o_{mn}^2}{64\pi}\frac{(l+2)!}{(l-2)!}
\left(|C^{+}_{lmn}|^2 + |C^{-}_{lmn}|^2  \right) ,
\q \q
\langle\dot{L}\rangle = 
\sum_{lmn}\frac{m\,\o_{mn}}{64\pi}\frac{(l+2)!}{(l-2)!}
\left(|C^{+}_{lmn}|^2 + |C^{-}_{lmn}|^2  \right) .
\end{align}

In a geodesic GSF code, the fluxes should match the orbit-averaged rate of 
work and torque that are computed locally at the particle via 
Eqn.~\eqref{eqn:adiabatic}.  The dissipative GSF can be split into sums over
\emph{tensor} spherical harmonic and FD contributions, each of which can be 
taken to be a function of $\chi$
\begin{align}
F^{\a}_\text{diss} (\chi) = \sum_{lmn} F_{ lmn}^{\a,\text{diss}} (\chi).
\end{align}
This decomposition of $F^\a_{\rm diss}$ can be substituted into the integrals 
in Eqn.~\eqref{eqn:adiabatic} to yield the orbit-averaged rates of change of 
energy and angular momentum.  It is possible though to reverse the order of 
sum and integration, and derive individual $l,m$ contributions to the rate of 
work and torque 
\begin{align}
\langle \dot{\mathcal{E}} \rangle = 
\sum_{lm} \langle \dot{\mathcal{E}} \rangle_{lm} = 
\frac{2}{T_r} \sum_{lmn} \left(\int_0^{\pi} \frac{f_p}{\mu u^t} 
\frac{dt}{d\chi} F_{lmn}^{t,\text{diss}} d\chi \right),
\q \q
\langle \dot{\mathcal{L}} \rangle = 
\sum_{lm} \langle \dot{\mathcal{L}} \rangle_{lm} = 
\frac{2}{T_r} \sum_{lmn} \left(\int_0^{\pi} \frac{r_p^2}{\mu u^t} 
\frac{dt}{d\chi} F_{lmn}^{\varphi,\text{diss}} d\chi \right) .
\end{align}
Moreover, the force can be evaluated on either side of the particle and 
should yield the same rates of change (up to numerical errors).  Balance 
between fluxes and local dissipation occurs mode by mode, i.e.,
$\langle\dot{E}\rangle_{lm} = -\mu \langle \dot{\mathcal{E}} \rangle_{lm}$ and
$\langle\dot{L}\rangle_{lm} = -\mu \langle \dot{\mathcal{L}} \rangle_{lm}$.  
Alternatively, we can compare them after summing over all modes.

Table \ref{tbl:oscGSF} compares the balance between fluxes and local 
dissipation for several $p=10$ orbits with different eccentricities.  For 
low eccentricity ($e = 0.1$) we see a high degree of fidelity between the 
local dissipation, computed on both sides of the particle, and the fluxes 
derived from the Lorenz gauge fields.  The comparison continues to hold but 
the accuracy drops markedly as orbits with $e=0.5$ and $e=0.7$ are considered.
We also then show the results of computing the fluxes with a RWZ code 
\cite{HoppEvan10} and a Teukolsky code \cite{FujiHikiTago09}.  Much smaller 
fractional errors, $\simeq 10^{-10}$-$10^{-9}$, are typically obtained, a 
result due at least in part to computing more $l,m$ modes.

\begin{table}
\caption[Comparisons of adiabatic contribution for $p=10$]
{\label{tbl:oscGSF} 
Comparisons between energy and angular momentum fluxes and locally computed 
dissipation.  Several orbits with $p=10$ and differing eccentricities are 
considered.  Local changes in energy and angular momentum (computed with 
the GSF on both sides of the particle) are compared to total fluxes radiated 
to infinity and down the horizon.  One set of fluxes is calculated using the 
present GSF code by extracting asymptotic values of the Lorenz gauge 
amplitudes.  These results are then compared to published values that were 
computed using RWZ and Teukolsky codes. The changes in energy
are measured in units of $M^2/\mu^2$ while the changes in angular
momentum are measured in units of $M/\mu^2$.}
\begin{center}
\begin{tabular}{cc|l|l|l}
\hline \hline
&& $e=0.1$ & $e=0.5$ & $e=0.7$ \\ 
\hline 
$-\mu \langle \dot{\mathcal{E}}^+ \rangle$ & This paper &
$6.3190584052\times 10^{-5}$ & $9.2871\times 10^{-5}$& $9.49\times 10^{-5}$\\
$-\mu \langle \dot{\mathcal{E}}^- \rangle$ &  This paper & 
$6.3190584053\times 10^{-5}$ &$9.2871\times 10^{-5}$ & $9.49\times 10^{-5}$\\
$\langle\dot{E}\rangle $ &  This paper & 
$6.319058405374\times 10^{-5}$ &$9.287477\times 10^{-5}$ & 
$9.5052\times 10^{-5}$\\
$\langle\dot{E}\rangle $ & Hopper and Evans & 
$6.319058405375\times 10^{-5}$ & $9.287480002\times 10^{-5}$ & 
$9.505332849\times 10^{-5}$\\
$\langle\dot{E}\rangle $ & Fujita et al. & $6.3190584054\times 10^{-5}$ & 
$9.287480001\times 10^{-5}$ & $9.505332847\times 10^{-5}$ \\
\hline
$-\mu \langle \dot{\mathcal{L}}^+ \rangle$ &  This paper&
$1.9531904845\times 10^{-3}$ & $1.9765\times 10^{-3}$ & $1.63\times 10^{-3}$ \\
$-\mu \langle \dot{\mathcal{L}}^- \rangle$ &  This paper&
$1.9531904845\times 10^{-3}$ &$1.9765\times 10^{-3}$& $1.63\times 10^{-3}$ \\
$\langle\dot{L}\rangle $ &  This paper &
$1.953190484551\times 10^{-3}$&$1.976807\times10^{-3}$&$1.6348\times 10^{-3}$\\
$\langle\dot{L}\rangle $ & Hopper and Evans&
$1.953190484552\times 10^{-3}$ &$1.976807667\times 10^{-3}$ & 
$1.634854630\times 10^{-3}$ \\ 
\hline \hline
\end{tabular}
\end{center} 
\vspace{-3ex}
\end{table}

A hybrid method would make use of the substantially smaller relative error 
$\epsilon_0 \simeq 10^{-10}$-$10^{-9}$ of a RWZ code to provide the 
orbit-averaged first-order GSF.  A question arises, however, as to what 
exactly orbit-averaged means.  Pound and Poisson \cite{PounPois08b} discuss 
various secular and radiative approximations.  As they point out, an average 
$\langle F_R^{\alpha} \rangle_{\chi}$ over $\chi$ is not the same as, for 
example, the average $\langle F_R^{\alpha} \rangle_{t}$ over $t$.  A hybrid 
method would use a very specific average.  A glance at \eqref{eqn:adiabatic} 
shows that the net fluxes will be balanced by integrals over 
\emph{proper time} $\tau$ of the relevant \emph{covariant} components of 
the dissipative part, $F_{\alpha}^{\rm diss}$, of the GSF.  These averages 
are then related to fluxes by
\begin{align}
\begin{split}
\langle\dot{E}\rangle_{\rm flux} 
&= - \mu \langle \dot{\mathcal{E}} \rangle_{\rm diss} 
= \frac{1}{T_r} \int_0^{\mathcal{T}_r} F_t^{\rm diss} \, d\tau 
= \frac{\mathcal{T}_r}{T_r} \langle F_t^{\rm diss} \rangle_{\tau} ,
\\
\langle\dot{L}\rangle_{\rm flux} 
&= - \mu \langle \dot{\mathcal{L}} \rangle_{\rm diss}
= - \frac{1}{T_r} \int_0^{\mathcal{T}_r} F_{\vp}^{\rm diss} \, d\tau 
= - \frac{\mathcal{T}_r}{T_r} \langle F_{\vp}^{\rm diss} \rangle_{\tau} ,
\end{split}
\end{align}
where $\mathcal{T}_r$ is the lapse of proper time in one radial orbit.  If 
we assume that the fluxes are computed with a RWZ code, we can infer from 
them an orbit-averaged dissipative force
\begin{align}
\langle F_t^{\rm diss} \rangle_{\rm RWZ}
= \frac{T_r}{\mathcal{T}_r} \langle\dot{E}\rangle_{\rm RWZ} ,
\q\q
\langle F_{\vp}^{\rm diss} \rangle_{\rm RWZ}
= - \frac{T_r}{\mathcal{T}_r} \langle\dot{L}\rangle_{\rm RWZ} ,
\end{align}
with vanishing $r$ component.  The process of constructing the hybrid force 
involves first taking the GSF from the Lorenz gauge code and constructing 
the oscillatory part 
\begin{align}
F_{\alpha}^{\rm osc} = F_{\alpha}^{\rm cons}
+ F_{\alpha}^{\rm diss}
- \langle F_{\alpha}^{\rm diss} \rangle_{\tau} ,
\end{align}
by computing the $\tau$-average of the full force (the conservative part has 
zero mean) and subtracting it off.  The hybrid GSF is then the sum of the 
dissipative term from a RWZ code and the oscillatory part from the Lorenz 
gauge code
\begin{align}
F_{\alpha}^{\rm hybrid} 
= \langle F_{\alpha}^{\rm diss} \rangle_{\rm RWZ}
+ F_{\alpha}^{\rm osc} .
\end{align}

If the Lorenz gauge code and the RWZ code had comparable accuracies, this 
construction would have little value.  But circumstances are different if 
the RWZ code can provide the average force, which drives secular changes, 
with relative errors as small as $\epsilon_0 \simeq 10^{-10}$-$10^{-9}$, 
while the Lorenz code supplies the oscillatory part of the GSF with relative 
errors of $\epsilon_1 \sim 10^{-10}$-$10^{-3}$ (depending on eccentricity).
Substantially tighter tolerance, and hence smaller $\epsilon_0$, is required 
on the former because the secular changes drive a large accumulation in 
the orbital phase $\Phi_r \simeq 1/\ve$ in a long-term evolution.  The 
oscillatory part contributes to $\kappa_1$ and its fractional errors 
$\epsilon_1$ need only be $\lesssim 10^{-3} \lesssim \delta\Phi_r$, 
consistent with the criterion outlined in the Introduction.

\section{Conclusions and future work}
\label{sec:conclusions}

We have described in this paper the key elements in our development of a FD 
method to compute the gravitational self-force in Lorenz gauge.  With this 
method we have extended the region in $p$ and $e$ of orbital parameter space 
within which accurate GSF results can be obtained.  The GSF can be calculated 
out to $p \simeq 100$ and up to $e \simeq 0.5$ (with this code alone).  New 
features in our approach include: (1) use of fully constrained Lorenz gauge 
equations for both odd and even parity; (2) discovery of analytic solutions 
for arbitrary-$l$ even-parity static modes; (3) development of a thin-QR 
pre-conditioning technique for orthogonalizing outer homogeneous solutions 
and reducing condition number; (4) adaptive use of quad-precision arithmetic 
to maintain accuracy of near-static modes; (5) an application of the jump 
conditions to avoid source integration for the lowest frequency mode; and 
(6) outlining a proposal for a novel hybrid approach to combine the Lorenz 
gauge code with a RWZ code to allow GSF calculation up to $e \simeq 0.8$.

This last proposal is an important idea to explore next and should be done 
in the context of using our code with a separate osculating orbits code to 
revisit long-term orbit evolutions \cite{WarbETC12}.  Our existing Lorenz 
gauge code, with minor tightening of tolerances, should be able to push to 
inspirals of orbits that start with $e \simeq 0.5$.  By including parallel 
computation of radiative modes with an existing, separate RWZ code, we 
should be able to reach initial orbits with $e \simeq 0.8$, near the peak 
in the expected EMRI distribution.

An ambitious downstream effort would involve finding some way to include the 
orbit-averaged second-order GSF (i.e., second-order fluxes).  Preliminary 
work is underway \cite{Poun12b} with applications to circular orbits on a 
Schwarzschild background \cite{WarbWard14,Warb14a}.  If it proves possible 
to find and implement such a scheme, we would be able to compute inspirals 
accurately enough for matched filtering and detector applications (within 
the restrictions of a Schwarzschild background and no spin in the secondary 
body).

A more immediate next application might involve the inclusion of spin in 
the small body and calculating not just the regularized perturbation of 
the spin precession for circular orbits \cite{DolaETC14a} but for eccentric 
orbits also.  More generally, the code might be used as a laboratory to 
explore other self-interaction effects, like tidal moments 
\cite{DolaETC14b}, with attention to their behavior in eccentric orbits.
We anticipate also using the code to explore overlap with a newly developed 
MST code that uses analytic function expansions to find the GSF for 
eccentric orbits \cite{ForsEvanHopp14}.

\acknowledgments

We thank Sarp Akcay, Leor Barack, Sam Dolan, Niels Warburton, Steve Detweiler, 
and Eric Poisson for helpful discussions.  E.F.~and T.O.~acknowledge support 
from the North Carolina Space Grant's Graduate Research Assistantship 
Program.  T.O.~also acknowledges support from the Tom and Karen Sox Summer 
Research Fellowship, and E.F.~acknowledges support from the Royster Society
of Fellows at the University of North Carolina-Chapel Hill. C.R.E.~is 
grateful for the hospitality of the Kavli Institute for Theoretical Physics 
at UCSB (which is supported in part by the National Science Foundation under 
Grant No. NSF PHY11-25915) and the Albert Einstein Institute in Golm, Germany, 
where part of this work was completed.  C.R.E.~also acknowledges support 
from the Bahnson Fund at the University of North Carolina-Chapel Hill.

\appendix

\section{Asymptotic boundary conditions}
\label{asymp}

We give here the recurrence relations for asymptotic and Taylor expansions 
that provide boundary conditions for mode integrations.  Expansions about 
$r_* = \pm\infty$ for homogeneous Lorenz gauge solutions were first given 
by Akcay \cite{Akca11} but with a different initial basis and for a larger, 
partially-constrained even-parity system.  The fully constrained even-parity 
system we use makes the generic recurrence relations valid for $l=0,1$ modes 
when $\o\ne 0$.  Throughout this section we use $\s = M \o$ for brevity.

\subsection{Near-horizon even-parity Taylor expansions}

The even-parity homogeneous solutions can be expanded around $r=2M$ in a 
Taylor series in powers of $f(r)$
\begin{align}
\bscal{\ti E}^- =
r \left[\begin{array}{c}
\ti{h}_{tt} \\
f \ti{h}_{tr} \\
f^{2} \ti{h}_{rr} \\
\ti{K}
\end{array}\right] = 
M e^{-i\o r_*} 
\sum_{k=0}^{\infty}
\left[\begin{array}{c}
a^{(tt)}_k \\
a^{(tr)}_k \\
 a^{(rr)}_k \\
a^{(K)}_k
\end{array}\right] f^k.
\end{align}
Recurrence relations for the coefficients can be found via the method of 
Frobenius
\begin{align}
&\left[
\begin{array}{cccc}
1+8i\s+2k(k-2-4i\s) \hspace{-1ex} & -8i\s & -1 & 0 \\
-2i\s & \hspace{-1ex} k(k-2)-4i\s(k-1) \hspace{-1ex} & -2i\s & 0 \\
-1 & -8i\s & \hspace{-1ex} 1+8i\s+2k(k-2-4i\s) \hspace{-1ex}  & 0 \\
-k+1+2i\s & -4i\s & -k-1+2i\s & \hspace{-1ex} k(k-4i\s)
\end{array}
\right]
\left[
\begin{array}{cccc}
a^{(tt)}_k \\
a^{(tr)}_k \\
a^{(rr)}_k \\
a^{(K)}_k
\end{array}
\right]
=
\left[
\begin{array}{cccc}
A^{(tt)}_k \\
A^{(tr)}_k \\
A^{(rr)}_k \\
A^{(K)}_k
\end{array}
\right] .
\end{align}
The RHS contains only lower order coefficients in the expansion
\begin{align}
\begin{split}
\hspace{-2.5ex}
A^{(tt)}_k \equiv & [-4 i (k-4) \s +2 k (5 k-34)+3 l (l+1)+113]2 a^{(tt)}_{k-3} -2
   [2 k (5 k-6 i \s -26)+3 l (l+1)+36 i \s
   +65] a^{(tt)}_{k-2}  \\
 & \hspace{-2ex} + [2 k (5 k-12 i \s -18)+2 l (l+1)+48 i
   \s +29] a^{(tt)}_{k-1} +[2 (42-5 k) k-2 l (l+1)-173] a^{(tt)}_{k-4}-8 i \s
    a^{(tr)}_{k-3}  \\
 & \hspace{-2ex} +24 i \s  a^{(tr)}_{k-2}-24 i \s 
   a^{(tr)}_{k-1}+(2 (k-5)^2-1) a^{(tt)}_{k-5}+3 a^{(rr)}_{k-5}-11
   a^{(rr)}_{k-4}+14 a^{(rr)}_{k-3}-6 a^{(rr)}_{k-2}-a^{(rr)}_{k-1} 
   -4 a^{(K)}_{k-6}  \\
 & \hspace{-2ex} +16 a^{(K)}_{k-5}-24 a^{(K)}_{k-4}
   +16 a^{(K)}_{k-3}-4 a^{(K)}_{k-2} ,
\\
\hspace{-2.5ex} A^{(tr)}_k \equiv & [5 k^2-4 i (3 k-7) \s -20 k+l^2+l+16] a^{(tr)}_{k-1} + [2 k (5 k-2 i \s -40)+3 (l^2+l+52)+20 i \s ] a^{(tr)}_{k-3}  \\
 & \hspace{-2ex} +  [2 k (-5 k+6 i \s +30)-3 l (l+1)-44 i \s -84] a^{(tr)}_{k-2}+[-5 (k-10) k-l (l+1)-124] a^{(tr)}_{k-4}-4 i \s  a^{(tt)}_{k-3}  \\
  & \hspace{-2ex} +10 i \s  a^{(tt)}_{k-2}-8 i \s  a^{(tt)}_{k-1}-4 i \s  a^{(rr)}_{k-3} +10 i \s a^{(rr)}_{k-2}-8 i \s  a^{(rr)}_{k-1}-4 i \s  a^{(K)}_{k-4}+8 i \s  a^{(K)}_{k-3}-4 i \s  a^{(K)}_{k-2}+(k-6)^2 a^{(tr)}_{k-5} ,
\\
\hspace{-2.5ex} A^{(rr)}_k \equiv & 2 [-4 i (k-5) \s +10 (k-8) k+3 l (l+1)+155] a^{(rr)}_{k-3} - 2 [2 k (5 k-6 i \s -30)+3 l (l+1)+44 i \s +83]  a^{(rr)}_{k-2}  \\
 & \hspace{-2ex} + [2 k (5 k-12 i \s -20)+2 l (l+1)+56 i \s +33] a^{(rr)}_{k-1} + [-10 (k-10) k-2 l (l+1)-249] a^{(rr)}_{k-4}  \\
 & \hspace{-2ex} + (-24 k+8 i \s +74) a^{(tt)}_{k-3}+2 (8 k-8 i \s -17) a^{(tt)}_{k-2}+(-4 k+8 i \s +3) a^{(tt)}_{k-1}-24 i \s a^{(tr)}_{k-3}+56 i \s  a^{(tr)}_{k-2}  \\
 & \hspace{-2ex} -40 i \s  a^{(tr)}_{k-1}+8 (6 k-2 i \s -21) a^{(K)}_{k-4}+(-32 k+32 i \s +80) a^{(K)}_{k-3}+4 (2 k-4 i \s -3) a^{(K)}_{k-2}+(19-4 k) a^{(tt)}_{k-5}  \\
 & \hspace{-2ex} +(16 k-63) a^{(tt)}_{k-4}+(2 (k-12) k+73) a^{(rr)}_{k-5}+(8 k-44) a^{(K)}_{k-6}+(144-32 k) a^{(K)}_{k-5} ,
\\
\hspace{-2.5ex} A^{(K)}_k \equiv & [5 k^2-2 k (5+6 i \s )+l^2+l+16 i \s +4] a^{(K)}_{k-1} + [2 k (-5 k+6 i \s +20)-3 (l^2+l+12)-32 i \s ]  a^{(K)}_{k-2}  \\
& \hspace{-2ex} + [2 k (5 k-2 i \s -30)+3 l (l+1)+16 i \s +84]  a^{(K)}_{k-3} + [-5 (k-8) k-l (l+1)-76] a^{(K)}_{k-4} \\
& \hspace{-2ex} +(6 k-2 i \s -15) a^{(tt)}_{k-2} 
 +(-4 k+4 i \s +7) a^{(tt)}_{k-1}+4 i \s  a^{(tr)}_{k-2}-8 i \s a^{(tr)}_{k-1} + (6 k-2 i \s -9) a^{(rr)}_{k-2} \\
 & \hspace{-2ex} +(4 i \s -4 k +1) a^{(rr)}_{k-1}+(k-4) a^{(tt)}_{k-4}  +(13-4 k) a^{(tt)}_{k-3}+(k-4) a^{(rr)}_{k-4}+(11-4 k) a^{(rr)}_{k-3}+(k-6) (k-4) a^{(K)}_{k-5} .
\end{split}
\end{align}
In these recurrence relations, a coefficient vanishes anytime a negative 
index appears.  
Because the matrix is 
singular when $k\le 2$ the first few terms are evaluated separately 
\begin{align}
\begin{split}
a^{(tr)}_0 &= a^{(tt)}_0 , 
\q \q
a^{(rr)}_0 = a^{(tt)}_0 ,
\q \q
a^{(tr)}_1 = -\frac{l(l+1)+1}{1+4i\s}a^{(tt)}_0 ,
\q \q
a^{(rr)}_1 = -\frac{2(l(l+1)+1)}{1+4i\s}a^{(tt)}_0 - a^{(tt)}_1 ,
\\
a^{(K)}_1 &= -\frac{4(l(l+1)+1)}{1+16\s^2}a^{(tt)}_0 +\left( \frac{l(l+1)}{1-4i\s} -1 \right) a^{(K)}_0  - \frac{2}{1-4i\s}a^{(tt)}_1 ,
\\
a^{(tr)}_2 &= \frac{l^2(l+1)^2-8i\s(l(l+1)+1)+16\s^2}{4i\s(1+16\s^2)} a^{(tt)}_0 - \frac{1}{1-4i\s} a^{(K)}_0 + \frac{l(l+1)-1}{1-4i\s}a^{(tt)}_1 - a^{(tt)}_2 ,
\\
a^{(rr)}_2 &= -\frac{2(l(l+1)+1)(l(l+1)-3)}{1+16\s^2} a^{(tt)}_0 +2\left( 1+\frac{1}{1-4i\s} \right) a^{(K)}_0 -\frac{2(l(l+1)-1)}{1-4i\s} a^{(tt)}_1 + a^{(tt)}_2 ,
\\
a^{(K)}_2 &= \frac{-4 (l(l+1)-11) (l(l+1)+1) \s-i (3 l (l (3 l (l+2)+1)-2)-16)+16 i \s^2}{8 \s (1+8 \s (2 \s+i))+4i} a^{(tt)}_{0} \\
&\hspace{25ex}-\frac{(l(l+1)+4 i \s)^2-36 i \s+12}{4 (8 \s^2+6 i \s-1)}a^{(K)}_{0}  +\frac{2 l (l+1) (1-i \s)+4 i \s-1}{8 \s^2+6 i \s-1}a^{(tt)}_{1} +a^{(tt)}_{2} .
\end{split}
\end{align}
The freely chosen coefficients $a^{(tt)}_0$,  $a^{(tt)}_1$, $a^{(tt)}_2$, 
and $a^{(K)}_0$ control the boundary conditions.  For example, at leading 
order we can choose the simple basis
\begin{align}
\begin{split}
\left( a^{(tt)}_0 , a^{(tt)}_1 , a^{(tt)}_2 , a^{(K)}_0 \right) =
\left(1, 0, 0, 0 \right)
\q\q &\rightarrow \q\q \l \bscal{\ti E}^-_0 \r^\top \sim \l  1, 1, 1, 0 \r e^{-i\o r_*}
\\
\left(a^{(tt)}_0, a^{(tt)}_1, a^{(tt)}_2, a^{(K)}_0 \right) =
\left(0, 1, 0, 0 \right)
\q\q &\rightarrow \q\q \l \bscal{\ti E}^-_1 \r^\top 
\sim \l 1, 0, -1, -2(1-4i\s)^{-1} \r f e^{-i\o r_*} ,
\\
\left(a^{(tt)}_0, a^{(tt)}_1, a^{(tt)}_2, a^{(K)}_0 \right) =
\left(0, 0, 1, 0 \right)
\q\q &\rightarrow \q\q \l \bscal{\ti E}^-_2 \r^\top \sim \l  1, -1, 1, 1 \r f^2e^{-i\o r_*} ,
\\
\left( a^{(tt)}_0, a^{(tt)}_1, a^{(tt)}_2, a^{(K)}_0 \right) =
\left(0, 0, 0, 1 \right)
\q\q &\rightarrow \q\q \l \bscal{\ti E}^-_3 \r^\top 
\sim \l 0,0,0,1 \r e^{-i\o r_*} .
\end{split}
\end{align}

\subsection{Near-horizon odd-parity Taylor expansions}

The odd-parity homogeneous solutions can also be expanded around $r=2M$ in 
powers of $f(r)$
\begin{align}
\bscal{\ti B}^- =
\left[\begin{array}{c}
\ti{h}_{t} \\
f \ti{h}_{r}
\end{array}\right] = 
M e^{-i\o r_*} 
\sum_{k=0}^{\infty}
\left[\begin{array}{c}
a^{(t)}_k \\
a^{(r)}_k
\end{array}\right] f^k.
\end{align}
Recurrence relations for the coefficients are again found via the method of 
Frobenius,
\begin{align}
&\left[
\begin{array}{cccc}
k(k-1-4i\s)+2i\s & -2i\s \\
-2i\s & k(k-1-4i\s)+2i\s
\end{array}
\right]
\left[
\begin{array}{cccc}
a^{(t)}_k \\
a^{(r)}_k
\end{array}
\right]
=
\left[
\begin{array}{cccc}
A^{(t)}_k \\
A^{(r)}_k
\end{array}
\right] .
\end{align}
Once again these result in a linear system to be solved and the RHS has only 
lower order coefficients
\begin{align}
\begin{split}
A^{(t)}_k \equiv & (4 k^2-22 k+l^2+l+24) a^{(t)}_{k-3}-2 [k (3 k-2 i \s-12)+l^2+l+5 i \s+9] a^{(t)}_{k-2}  \\
 & \hspace{12ex} + [2 k (2 k-4 i \s-5)+l^2+l+12 i \s+4] a^{(t)}_{k-1} +2 i \s a^{(r)}_{k-2}-4 i \s a^{(r)}_{k-1}-(k-5) (k-2) a^{(t)}_{k-4} ,
\\
 A^{(r)}_k \equiv & -2 [k (3 k-2 i \s-15)+l^2+l+7 i \s+15] a^{(r)}_{k-2}+ [4 k (k-2 i \s-3)+l^2+l+16 i \s+6] a^{(r)}_{k-1}  \\
 & \hspace{12ex}  +[4 (k-7) k+l^2+l+42] a^{(r)}_{k-3}+6 i \s a^{(t)}_{k-2}-8 i \s a^{(t)}_{k-1}-(k-6) (k-3) a^{(r)}_{k-4} .
\end{split}
\end{align}
Any negative-index coefficients vanish.  This linear system is singular 
for $k\le 1$ and starting conditions for the recursion are calculated 
separately 
\begin{align}
a^{(r)}_0 = a^{(t)}_0 ,
\q \q
a^{(r)}_1 = -\frac{(l+2)(l-1)}{2i\s}a^{(t)}_0 - a^{(t)}_1 .
\end{align}
The freely chosen coefficients $a^{(t)}_0$ and $a^{(t)}_1$ control the 
boundary conditions.  We can choose a simple basis, which at leading order 
has the form
\begin{align}
\begin{split}
\left( a^{(t)}_0 , a^{(t)}_1\right) =
\left(1, 0\right)
\q\q &\rightarrow \q\q \l \bscal{\ti B}^-_0 \r^\top \sim \l  1, 1 \r e^{-i\o r_*},
\\
\left(a^{(t)}_0, a^{(t)}_1 \right) =
\left(0, 1\right)
\q\q &\rightarrow \q\q \l \bscal{\ti B}^-_1 \r^\top 
\sim \l 1, -1 \r fe^{-i\o r_*} .
\end{split}
\end{align}
In practical applications, we evaluate these expansions at $r_*=-6M$ 
and add terms in the series until the relative 
size of the last term drops below machine precision.

\subsection{Near-infinity even-parity asymptotic expansions}

The even-parity homogeneous solutions can be expanded about $r = \infty$ as
\begin{align}
\label{eqn:evenInfinity}
\bscal{\ti E}^+ =
r \left[\begin{array}{c}
\ti{h}_{tt} \\
f \ti{h}_{tr} \\
f^{2} \ti{h}_{rr} \\
\ti{K}
\end{array}\right] = 
M e^{i\o r_*} 
\sum_{k=0}^{k_{\text{max}}}
\left[\begin{array}{c}
b^{(tt)}_k \\
b^{(tr)}_k \\
b^{(rr)}_k \\
b^{(K)}_k
\end{array}\right] \l \frac{M}{r}\r^k.
\end{align}
Recurrence relations for the coefficients are a linear system of equations
\begin{align}
&\qquad\qquad\q \left[
\begin{array}{cccc}
-2 i \s k & 0 & 0 & 0 \\
-i\s & 2i\s(k-1) & -i\s & -2i\s \\
2i\s & 4i\s & -2i\s(k-1) & -4i\s \\
-i\s & -2i\s & -i\s & -2i\s(k-1)
\end{array}
\right]
\left[
\begin{array}{cccc}
b^{(tt)}_k \\
b^{(tr)}_k \\
b^{(rr)}_k \\
b^{(K)}_k
\end{array}
\right]
=
\left[
\begin{array}{cccc}
B^{(tt)}_k \\
B^{(tr)}_k \\
B^{(rr)}_k \\
B^{(K)}_k
\end{array}
\right] .
\end{align}
As with the horizon-side expansions, the RHS groups all of the lower order 
coefficients
\begin{align}
\begin{split}
B^{(tt)}_k \equiv & [-k^2-4 i (k-2) \s+k+l^2+l] b^{(tt)}_{k-1} -2 [(7-2 k) k+l^2+l-5] b^{(tt)}_{k-2}+4 i \s b^{(tr)}_{k-1}  \\
& \hspace{10ex} -2 [2 (k-6) k+17] b^{(tt)}_{k-3} -6 b^{(rr)}_{k-3}+4 b^{(rr)}_{k-2}-16 b^{(K)}_{k-4}+16 b^{(K)}_{k-3}-4 b^{(K)}_{k-2} ,
\\
 B^{(tr)}_k \equiv & 2 [(11-2 k) k+l^2+l-15] b^{(tr)}_{k-2}+[k (k+4 i \s-3)-l (l+1)-12 i \s+2] b^{(tr)}_{k-1}  \\
& \hspace{10ex} -4 i \s b^{(tt)}_{k-1}- 4 i \s b^{(rr)}_{k-1}+8 i \s b^{(K)}_{k-2}-8 i \s b^{(K)}_{k-1}+4 (k-4)^2 b^{(tr)}_{k-3} ,
\\
 B^{(rr)}_k \equiv & [-k^2-4 i (k-3) \s +3 k+l^2+l-4] b^{(rr)}_{k-1}-2
   [(11-2 k) k+l^2+l-17] b^{(rr)}_{k-2}  \\
& \hspace{10ex} +2 (k+2 i \s) b^{(tt)}_{k-1}+12 i \s b^{(tr)}_{k-1}+4 (6 k+4 i \s-11)
   b^{(K)}_{k-2}-4 (k+4 i \s -1) b^{(K)}_{k-1}+(8 k-22)
   b^{(tt)}_{k-3}  \\
& \hspace{10ex} +(12-8 k) b^{(tt)}_{k-2} +(-4 (k-8) k-66) b^{(rr)}_{k-3}+16 (2
   k-7) b^{(K)}_{k-4}+(128-48 k) b^{(K)}_{k-3},
\\
B^{(K)}_k \equiv & [-k (k+4 i \s-3)+l^2+l+8 i \s-2] b^{(K)}_{k-1} -2 [(9-2 k) k+l^2+l-9] b^{(K)}_{k-2}  \\
& \hspace{10ex} +2 (k-2) b^{(tt)}_{k-2}-k b^{(tt)}_{k-1}+2 (k-2) b^{(rr)}_{k-2}+(2-k) b^{(rr)}_{k-1}-4 (k-4) (k-2) b^{(K)}_{k-3} .
\end{split}
\end{align}
All appearances of a negative index imply a vanishing coefficient.  The 
linear system is singular here when $k\le 2$ and starting coefficients are 
obtained from the reduced equations 
\begin{align}
\begin{split}
b^{(rr)}_0 &= -b^{(tt)}_0-2 b^{(tr)}_0 ,
\q \q
b^{(K)}_0 = 0 ,
\q \q
b^{(tt)}_1 = -\frac{l(l+1)+4i\s}{2i\s} b^{(tt)}_0 - 2 b^{(tr)}_0 ,
\\
b^{(rr)}_1 &= \frac{l(l+1)+4(1+i\s)}{2i\s}b^{(tt)}_0 + 2\left( 1+\frac{1}{i\s} \right)b^{(tr)}_0 - 2 b^{(tr)}_1 ,
\q \q
b^{(K)}_1 = -\frac{1}{i\s}b^{(tt)}_0 + \frac{(l+2)(l-1)}{2i\s}b^{(tr)}_0 + b^{(tr)}_1 ,
\\
b^{(tt)}_2 &= -\frac{l(l+1)((l+2)(l-1)+8i \s)+4 i \s}{8\s^2}b^{(tt)}_0 + \frac{l(l+1)+2}{2i \s} b^{(tr)}_0 - b^{(tr)}_1 ,
\\
b^{(tr)}_2 &= -\frac{1}{i\s}b^{(tt)}_0 + \frac{l(l+1)(l+2)(l-1)+4i\s(l(l+1)+3)}{8\s^2} b^{(tr)}_0 - \left(1 +\frac{l(l+1)}{2i\s} \right) b^{(tr)}_1 ,
\\
b^{(rr)}_2 &= \frac{l(l+1)((l+2)(l-1)+8i\s)+20i\s}{8\s^2} b^{(tt)}_0 + \frac{l(l+1)((l+2)(l-1)-2i\s)+8i\s}{4\s^2} b^{(tr)}_0 + 3 b^{(tr)}_1 - 2 b^{(K)}_2 .
\end{split}
\end{align}
The freely chosen coefficients $b^{(tt)}_0$, $b^{(tr)}_0$, $b^{(tr)}_1$, 
and $b^{(K)}_2$ control the boundary conditions and a simple choice for the 
basis gives the following lowest-order form
\begin{align}
\begin{split}
\left( b^{(tt)}_0 , b^{(tr)}_0 , b^{(tr)}_1 , b^{(K)}_2 \right) =
\left(1, 0, 0, 0 \right)
\q\q &\rightarrow \q\q \l \bscal{\ti E}^+_0 \r^\top \sim \l 1, 0, -1, 0 \r e^{i\o r_*} ,
\\
\left( b^{(tt)}_0 , b^{(tr)}_0 , b^{(tr)}_1 , b^{(K)}_2 \right) =
\left(0, 1, 0, 0 \right)
\q\q &\rightarrow \q\q \l \bscal{\ti E}^+_1 \r^\top 
\sim \l 0, 1, -2, 0 \r e^{i\o r_*} ,
\\
\left( b^{(tt)}_0 , b^{(tr)}_0 , b^{(tr)}_1 , b^{(K)}_2 \right) =
\left(0, 0, 1, 0 \right)
\q\q &\rightarrow \q\q \l \bscal{\ti E}^+_2 \r^\top \sim \l 0, 1, -2, 1 \r r^{-1} e^{i\o r_*} ,
\\
\left( b^{(tt)}_0 , b^{(tr)}_0 , b^{(tr)}_1 , b^{(K)}_2 \right) =
\left(0, 0, 0, 1 \right)
\q\q &\rightarrow \q\q \l \bscal{\ti E}^+_3 \r^\top 
\sim \l 0, 0, -2, 1 \r r^{-2} e^{i\o r_*} .
\end{split}
\end{align}
Note though, that as described in Sec.~\ref{sec:numericalGeneral}, we take 
this simple basis and apply a linear transformation called thin-QR
pre-conditioning. 

\subsection{Near-infinity odd-parity asymptotic expansions}

The odd-parity homogeneous solutions can be expanded about $r = \infty$ as
\begin{align}
\label{eqn:oddInfinity}
\bscal{\ti B}^+
=
\left[\begin{array}{c}
\ti{h}_{t} \\
f \ti{h}_{r}
\end{array}\right] &= 
M e^{i\o r_*} 
\sum_{k=0}^{k_{\text{max}}}
\left[\begin{array}{c}
b^{(t)}_k \\
b^{(r)}_k
\end{array}\right] \l \frac{M}{r} \r^k .
\end{align}
Again, the recurrence relations are found to satisfy a linear system
\begin{align}
\left[
\begin{array}{cccc}
-2ik\s & 0 \\
-2i\s & 2i\s(k-1)
\end{array}
\right]
\left[
\begin{array}{cccc}
b^{(t)}_k \\
b^{(r)}_k
\end{array}
\right]
=
\left[
\begin{array}{cccc}
B^{(t)}_k \\
B^{(r)}_k
\end{array}
\right] ,
\end{align}
where again the RHS contains all lower order coefficients
\begin{align}
\begin{split}
\hspace{-2ex} B^{(t)}_k \equiv & [l(l+1)-k (k+4 i \s-1)+6 i \s] b^{(t)}_{k-1}-2 [l(l+1) -2 (k-3) k-2] b^{(t)}_{k-2}+2 i \s b^{(r)}_{k-1}-4 (k-4) (k-1) b^{(t)}_{k-3} ,
\\
\hspace{-2ex} B^{(r)}_k \equiv & [k (k+4 i \s-3)-l (l+1)-10 i \s+2] b^{(r)}_{k-1} -2 [2 (k-5) k-l(l+1)+10] b^{(r)}_{k-2} \\
& \hspace{70ex} -6 i \s b^{(t)}_{k-1}+4 (k-5) (k-2) b^{(r)}_{k-3} ,
\end{split}
\end{align}
and any negative index that appears implies a vanishing coefficient.  This 
linear system is singular for $k\le 1$ and starting conditions are evaluated 
individually 
\begin{align}
b^{(r)}_0 = -b^{(t)}_0 ,
\q \q
b^{(t)}_1 = -\frac{l(l+1)}{2i\s}b^{(t)}_0 .
\end{align}
The freely chosen coefficients $b^{(t)}_0$ and $b^{(r)}_1$ determine the 
boundary conditions and a simple choice for the basis yields the following 
lowest-order form 
\begin{align}
\begin{split}
\label{eqn:oddOutNaive}
\left( b^{(t)}_0 , b^{(r)}_1\right) =
\left(1, 0\right)
\q\q 
&\rightarrow \q\q \l \bscal{\ti B}^+_0 \r^\top \sim \l  1, -1 \r e^{i\o r_*}
\\
\left(b^{(t)}_0, b^{(r)}_1 \right) =
\left(0, 1\right)
\q\q &\rightarrow \q\q \l \bscal{\ti B}^+_1 \r^\top 
\sim \l 0, 1 \r r^{-1} e^{i\o r_*} .
\end{split}
\end{align}
As with even parity, the method described in Sec.~\ref{sec:numericalGeneral} 
transforms this simple basis to a more orthogonal one using thin-QR
pre-conditioning. 

With these asymptotic series care must be exercised with the number of 
terms and the starting radius $r_*^\infty$.  The test for convergence is 
whether a numerical integration through a distance $\sim \o^{-1}$ starting 
with an initial evaluation of the asymptotic expansion agrees with a second 
evaluation of the expansion at the end point of the trial.  If the test 
fails, we increase $r_*^\infty$ by some factor (say $\sim 1.5$) and repeat.

\section{\label{static} Homogeneous static modes}

Here we provide the details of the power series used to construct exact 
analytic homogeneous solutions for static modes when $l\ge 2$, as were 
discussed in Sec.~\ref{sec:static}.  Throughout this section we set 
$\rho=r/M$.  Regularity at $\rho=2$ and $\rho=\infty$ governs our choice 
for inner and outer solutions. 

\subsection{Odd-parity}
\label{sec:staticOdd}

In Sec.~\ref{sec:static} we gave expressions for $\ti{h}_t^-$ and $\ti{h}_t^+$
as finite sums.  The coefficients in those sums are
\begin{align} 
a^{\text{odd}}_k &= \frac{3(-1)^k 2^{1-k}(l+k+1)!}{l(l+1)k!(k+3)!(l-k-1)!} ,
\\
b^{\text{odd}}_0 &= \frac{96}{l^2(l+1)^2(l+2)(l-1)} ,
\q
b^{\text{odd}}_1 = \frac{24}{l^2(l+1)^2} ,
\q
b^{\text{odd}}_2 = \frac{6}{l(l+1)} ,
\q
b^{\text{odd}}_3 = \frac{1}{l(l+1)}+\frac{11}{6}-2 H_l , \nonumber
\\
b^{\text{odd}}_k &=  \frac{1}{-4k(k-3)}
\Big[ 4(k-3)a^{\text{odd}}_{k-4}+(12-8k)a^{\text{odd}}_{k-3} 
\\
& \hspace{30ex}
+(12-7k+k^2-l(l+1)) b^{\text{odd}}_{k-2} + 2 (10k-2k^2-10+l(l+1)) b^{\text{odd}}_{k-1} \Big], \nonumber
\end{align}
where $H_k$ is the $k^{\rm th}$ harmonic number defined as
\begin{align}
\label{eqn:harmNum}
H_k &\equiv \left\{ \begin{array}{lr} \displaystyle 0&,\;k=0 \\ \sum_{j=1}^k \; j^{-1} &,\;k\ge 1 \end{array} \right. .
\end{align}
We have found the expression for $\tilde{h}_t^+$ in 
Eqn.~\eqref{eqn:ht_plusMinus} to be impractical to use numerically for
large $r$ because of a large number of cancellations between the two sums. 
We instead re-expand the solution as an infinite series
\begin{align}
\label{eqn:ht_plus_inf}
\ti{h}_t^+ = \frac{M}{\rho^l}
\sum_{k=0}^{\infty} \frac{d^{\text{odd}}_k}{\rho^k},
\q \q
d^{\text{odd}}_k = 
\frac{2^{2l+k+1}(l+k+1)!(l+k-2)!\G(l+3/2)}{k!(l+1)(l-2)!(2l+k+1)!\sqrt{\pi}} ,
\end{align}
which agrees with the expression in Eqn.~\eqref{eqn:ht_plusMinus} up to a 
constant factor.  This is a convergent Taylor series if $\rho>2$.

\subsection{Even-parity}
\label{sec:staticEven}

As summarized in Sec.~\ref{sec:static}, we find the even-parity static modes 
through a series of steps.  We give here the complete expressions for the 
gauge variables $\tilde{\xi}^{\pm}_{e,H0}$, $\tilde{\xi}^{\pm}_{e,H1}$ and 
$\tilde{\xi}^{\pm}_{e,I}$ that are defined in that section.  We construct 
power series expansions and we seek series that are exact solutions with 
finite numbers of terms.  This condition imposes constraints on otherwise 
freely chosen coefficients.  The variables $\tilde{\xi}^{\pm}_{e,H0}$ are 
found from the finite sums
\begin{align}
\tilde{\xi}^{-}_{H0} = M^2 \sum_{k=0}^l a^{H0}_k \rho^k ,
\q \q
\tilde{\xi}^{+}_{H0} = 
\tilde{\xi}^{-}_{H0} \ln{f} + M^2 \sum_{k=0}^{l-1} b^{H0}_k \rho^k ,
\end{align}
where the coefficients $a^{H0}_k$ and $b^{H0}_k$ are given by the closed-form 
expressions
\be
a^{H0}_k = \frac{(-1)^k(l+k)!}{2^{k}(k!)^2(l-k)!} ,
\q \q
b^{H0}_k = 2 a^{H0}_k(H_k-H_l).
\ee
In practice we find the above expression for $\tilde{\xi}^{+}_{H0}$ to be
impractical to use numerically due to a large number of cancellations 
between the two sums.  Instead we use the following equivalent Taylor series, 
which converges for all $\rho>2$
\begin{align}
\label{eqn:xi_plus0_inf}
\tilde{\xi}^{+}_{H0} = 
\frac{M^2}{\rho^{l+1}}\sum_{k=0}^{\infty} \frac{d^{H0}_k}{\rho^k} ,
\q \q
d^{H0}_k = \frac{(-1)^{l+1}2^{l+k+1}[(l+k)!]^2}{k!(2l+k+1)!}.
\end{align}
The variables $\tilde{\xi}^{\pm}_{e,H1}$ are given by
\begin{align}
\label{eqn:xiH1}
\begin{split}
\tilde{\xi}^{-}_{H1} &= \tilde{\xi}^{-}_{H0} \ln{\rho} + M^2 \sum_{k=0}^{l+2} a^{H1}_k \rho^k  ,
\\
\tilde{\xi}^{+}_{H1} &= \tilde{\xi}^{-}_{H0}\left[ {\rm Li}_2(f)
-\frac{1}{2}\ln\left(\frac{\rho}{4}\right)\ln{f}
-\frac{\pi^2}{6}-\frac{a^{H1}_{l+2}+a^{H1}_{l+1}
+b^{H1}_l}{a^{H0}_l}\right]
+\frac{1}{2}\tilde{\xi}^{+}_{H0} \ln{\rho} 
- \frac{1}{2} \tilde{\xi}^{-}_{H1}\ln{f}
+M^2 \sum_{k=0}^{l+1} b^{H1}_k \rho^k ,
\end{split}
\end{align}
where we have introduced the dilogarithm function
${\rm Li}_2(f) \equiv -\int_0^f x^{-1}\ln(1-x) dx$, and
the coefficients follow from the recurrences
\begin{align}
\begin{split}
a^{H1}_0 &= 0,
\q \q
a^{H1}_1 = l(l+1)+\frac{1}{2},
\q \q
a^{H1}_2 = \frac{1}{16} (2 - l (-2 + l (1 + 3 l (2 + l)))) ,
\\
a^{H1}_3 &= -\frac{1}{8} - \frac{1}{12} l + \frac{539}{864} l^2 + \frac{91}{288} l^3 - \frac{7}{216} l^4 + \frac{11}{288} l^5 +  \frac{11}{864} l^6 ,
\\
8(-3&+k)(-2+k)k^2 a^{H1}_k = \\
& \hspace{3ex} 2(-7+2k)(11+(-7+k)k-l(1+l))a^{H0}_{k-3} \\
& \hspace{3ex} -2(-99k^2+12k^3-4k(-65+2l(1+l))+3(-72+7l(1+l)))a^{H0}_{k-2} \\
& \hspace{3ex} +4(-68+2k(65+6(-6+k)k)+7l-4kl+(7-4k)l^2)a^{H0}_{k-1} \\
& \hspace{3ex} -8k(12+k(-15+4k))a^{H0}_{k} +(-5+k-l)(-3+k-l)(-4+k+l)(-2+k+l)a^{H1}_{k-3} \\
& \hspace{3ex} -2(128-33k^3+3k^4+(-5+l)l(1+l)(6+l)-2k^2(-65+2l(1+l))+3k(-72+7l(1+l)))a^{H1}_{k-2} \\
& \hspace{3ex}+4(24+(-4+k)k(17+3(-4+k)k)-7l+(7-2k)kl+(-7+(7-2k)k)l^2)a^{H1}_{k-1},
\end{split}
\end{align}
\begin{align}
\begin{split}
b^{H1}_0 &= 0 ,
\q \q
b^{H1}_1 = -\frac{1}{2}\left(1+H_l(2\,l(l+1)+1) \right),
\\
b^{H1}_2 &= \frac{1}{16}\left(\frac{6+l-27 l^2-16 l^3}{l(l+1)}+H_l (-2 + l (-2 + l (1 + 3\, l (2 + l)))) \right),
\\
b^{H1}_3 &= \frac{216 + l (216 + l (-656 + l (-1229 + l (-1279 + l (-463 + 27\, l)))))}{1728\, l (l + 1)} 
\\&\q\q\q\q\q\q\q\q\q\q\q\q\q\q+ \frac{H_l}{864}(-180 + l (-120 + l (613 + l (495 + l (28 - 11 \,l (3 + l)))))),
\\
8 (-3 &+ k) (-2 + k) k^2 b^{H1}_k = \\
& \hspace{3ex} -4(-3+k)a^{H0}_{k-2}+2(-7+4k)a^{H0}_{k-1}-4(-3+k)(8+(-6+k)k-l(1+l))a^{H1}_{k-2} \\
& \hspace{3ex} +2(-51k^2+8k^3+7(-8+l+l^2)+k(99-4l(1+l)))a^{H1}_{k-1}-4k(12+k(-15+4k))a^{H1}_k \\
& \hspace{3ex} +(-7+2k)(11+(-7+k)k-l(1+l))b^{H0}_{k-3} \\
& \hspace{3ex} +(99k^2-12k^3+4k(-65+2l(1+l))-3(-72+7l(1+l)))b^{H0}_{k-2} \\
& \hspace{3ex} +2(-68+2k(65+6(-6+k)k)+7l-4kl+(7-4k)l^2)b^{H0}_{k-1} -4k(12+k(-15+4k))b^{H0}_k \\
& \hspace{3ex} +(-5+k-l)(-3+k-l)(-4+k+l)(-2+k+l)b^{H1}_{k-3} \\
& \hspace{3ex} -2(128-33k^3+3k^4+(-5+l)l(1+l)(6+l)-2k^2(-65+2l(1+l))+3k(-72+7l(1+l)))b^{H1}_{k-2} \\
& \hspace{3ex} +4(24+(-4+k)k(17+3(-4+k)k)-7l+(7-2k)kl+(-7+(7-2k)k)l^2)b^{H1}_{k-1} .
\end{split}
\end{align}
As with $\tilde{\xi}^{+}_{H0}$, we find the expression for 
$\tilde{\xi}^{+}_{H1}$ to be impractical for numerical use at large radius 
and replace it with a convergent Taylor series [this expression for  
$\tilde{\xi}^{+}_{H1}$ is equivalent to that in Eqn.~\eqref{eqn:xiH1} 
up to a linear combination with $\tilde{\xi}^{+}_{H0}$]
\begin{align}
\begin{split}
\label{eqn:xi_plus1_inf}
&\tilde{\xi}^{+}_{H1} =  \frac{1}{2} \tilde{\xi}^{+}_{H0} \ln{\rho} + \frac{M^2}{\rho^{l-1}}\sum_{k=0}^{\infty} \frac{d^{H1}_k}{\rho^k} ,
\\
& d^{H1}_0 =\frac{3 + 8 l + 4 l^2}{4 l} d^{H0}_0,
\q \q
d^{H1}_1 =\frac{l^2+l-1}{l}d^{H1}_0,
\q \q
d^{H1}_2 = 0,
\\
&(-2 + k) k (-1 + k + 2 l) (1 + k + 2 l)d^{H1}_k = \\
& \hspace{2ex} 2 (-3 + k + l) (3 + 4 k^2 + l (-9 + 4 l) + k (-9 + 8 l)) d^{H0}_{k-4}  \\
& \hspace{2ex}- (-2 + 8 k^3 + 3 k^2 (-9 + 8 l) + l (27 - 26 l + 4 l^2) + k (22 - 58 l + 20 l^2)) d^{H0}_{k-3}  \\
&\hspace{2ex}+ (-1 + 2 k + 2 l) (-1 - 2 l + k (-1 + k + 2 l)) d^{H0}_{k-2} - 4 (-3 + k + l)^2 (-1 + k + l) (k + l) d^{H1}_{k-2}  \\
& \hspace{2ex}+2((-2+k)k(1+k(-5+2k))-6l+k(27+k(-29+8k))l+2(-2+k)(-3+5k)l^2+2(-3+2 k) l^3) d^{H1}_{k-1} .
\end{split}
\end{align}

The remaining unknown gauge variables are $\tilde{\xi}^{\pm}_{e,I}$, which 
satisfy the inhomogeneous ODE Eqn.~\eqref{eqn:xiE4th}.  In order to find 
expressions for them we must first write the source term of that equation 
as a power series.  The source term that is regular at the horizon is 
\begin{align}
\label{eqn:S_minus}
S_{\xi}^- = \frac{1}{ M^2\rho^3 f^2}\sum_{k=0}^l y^-_k  \rho^k ,
\q \q
y^-_k = \begin{cases} 
- \dfrac{96}{(l+2)(l-1)}  &\; k=0 \\
 \dfrac{8 a^{\text{odd}}_{k-1}(2(k+1)+l(l+1))-32(k+3)a^{\text{odd}}_k}{(l+2)(l-1)} &\; 0<k<l \\
 \dfrac{8(l+1)}{l-1}a^{\text{odd}}_{l-1} &\; k=l \end{cases}.
\end{align}
The corresponding term that is regular at infinity is
\begin{align}
\begin{split}
\label{eqn:S_plus}
S_\xi^+ &= S_\xi^- \ln{f}+\frac{1}{M^2 \rho^5 f^2}\sum_{k=0}^{l+2} y^+_k  \rho^k ,
\\
y^+_k &= \begin{cases} \dfrac{8(l(l+1)+2(k-1))b^{\text{odd}}_k-32(k+1)b^{\text{odd}}_{k+1}}{(l+2)(l-1)} &\; k=0,1 \\
\dfrac{8(l(l+1)+2(k-1))b^{\text{odd}}_k-32(k+1)b^{\text{odd}}_{k+1}+32 a^{\text{odd}}_{k-2}}{(l+2)(l-1)} &\; 1<k< l+2 \\\
\dfrac{8(l+1)}{l-1}b^{\text{odd}}_{l+2}&\; k=l+2 \end{cases} .
\end{split}
\end{align}
With these in hand, we can write power series for $\tilde{\xi}^{\pm}_{e,I}$. 
\begin{align}
\label{eqn:xiE_plusMinus}
\tilde{\xi}^{-}_{e,I} = M^2 \rho^2 \sum_{k=0}^l a^{I}_k  \rho^k ,
\q \q
\tilde{\xi}^{+}_{e,I} = \tilde{\xi}^{-}_{e,I} \ln{f} 
+\b \tilde{\xi}^{-}_{H0} \ln{f} + M^2 \sum_{k=0}^{l+1} b^{I}_k  \rho^k ,
\end{align}
where $\b\equiv -3072(l(l+1)-7)/[l^4(l+1)^4(l+7)(l+2)(l-1)]$.  The 
coefficients follow from the recurrences
\begin{align}
\begin{split}
&a^I_0 = -\frac{12}{\,(l+2)(l+1)l(l-1)} ,
\q \q
a^I_1 = -\frac{2}{l(l+1)} ,
\\
&4k(k-1)  (k+2)^2 a^I_k = y^-_{k-1}-(k-l-2)(k-l)(k+l-1)(k+l+1)a^I_{k-2} 
\\
&\hspace{35ex} -2(l+l^2-2-k^3-2k^4+k(1+l+l^2)+2k^2(2+l+l^2))a^I_{k-1} ,
\end{split}
\end{align}
\begin{align}
\begin{split}
&b^I_0 = \sum_{k=0}^l \frac{2^{k+2}}{k+2}a^I_k+\b \sum_{k=1}^l \frac{2^k}{k} a^{H0}_k ,
\q \q
b^I_1 = \frac{1}{4}(-3l(l+1)\b + 2\b a^{H0}_1 - 2l(l+1) b^I_0) + \frac{1}{8}y^+_0 ,
\\
&b^I_2 = \b \Big[ \frac{3}{32} (2 - 3 l - 2 l^2 + 2 l^3 + l^4)+ \frac{1}{2l(l+1)}\Big( -\frac{2 - 3 l - 2 l^2 + 2 l^3 + l^4}{8} a^{H0}_1 + \frac{2 + 5 l + 5 l^2}{2} a^{H0}_2 + 9 a^{H0}_3 \Big) \Big]  \\
&\q +\frac{l(l+1)(l+2)(l-1)}{16} b^I_0+\frac{1}{2l(l+1)}\Big(\frac{2 + 5 l + 5 l^2}{2} a^I_0+9a^I_1 -\frac{2 - 3 l - 2 l^2 + 2 l^3 + l^4}{32} y^+_0-\frac{1}{8} y^+_1+\frac{1}{4}y^+_2 \Big) ,
\\
&b^{I}_3 = 
\sum_{k=2}^l \frac{2^{k-1}}{k-1}a^I_k+\b \sum_{k=4}^l \frac{2^{k-3}}{k-3} a^{H0}_k ,
\\
& 8k^2(k-2) (k-3) b^I_k = \\
& \hspace{4ex} 2 y^+_{k-1} - y^+_{k-2}+ 8 (-3 + k) [8 + (-6 + k) k - l (1 + l)] a^I_{k-4}  \\
& \hspace{4ex} +  4 [51 k^2 - 8 k^3 - 7 (-8 + l + l^2) + k (-99 + 4 l (1 + l))] a^I_{k-3} + 
 8 k [-4 + 4 (-2 + k)^2 + k] a^I_{k-2}  \\
& \hspace{4ex} - 2 [128 - 33 k^3 + 3 k^4 + (-5 + l) l (1 + l) (6 + l) - 
    2 k^2 (-65 + 2 l (1 + l)) + 3 k (-72 + 7 l (1 + l))] b^I_{k-2}  \\
& \hspace{4ex}  + 4 [24 + (-4 + k) k (17 + 3 (-4 + k) k) - 7 l + (7 - 2 k) k l + (-7 + (7 - 2 k) k) l^2] b^I_{k-1}  \\
& \hspace{4ex} + (-5 + k - l) (-3 + k - l) (-4 + k + l) (-2 + k + l) b^I_{k-3}
 +  8 \b(-3 + k) [8 + (-6 + k) k - l (1 + l)] a^{H0}_{k-2}  \\
& \hspace{4ex}+  4\b [51 k^2 - 8 k^3 - 7 (-8 + l + l^2) + k (-99 + 4 l (1 + l))] a^{H0}_{k-1} + 8\b k [12 + k (-15 + 4 k)] a^{H0}_{k} .
\end{split}
\end{align}
We have found the expressions for $S_\xi^+$ and $\tilde{\xi}_{e,I}^+$ in 
Eqns.~\eqref{eqn:S_plus} and \eqref{eqn:xiE_plusMinus} to also be impractical 
for numerical use at large $r$.  Again, we replace them with infinite 
series.  For the source term we have
\begin{align}
\label{eqn:S_plus_inf}
S_\xi^+ &= \frac{1}{M^2 \rho^{l+4}f^2} \sum_{k=0}^{\infty} \frac{v_k}{\rho^k} ,
\q \q
v_k = \frac{2^{2l+k+4}l(l+k) [(l+k-1)!]^2 \G(l+3/2)}{(l+2)k! (l-1)!(2l+k+1)!\sqrt{\pi}} ,
\end{align}
while for $\tilde{\xi}_{e,I}^+$ we use
\begin{align}
\begin{split}
\label{eqn:xi_plus2_inf}
\tilde{\xi}_{e,I}^{+} &= \frac{M^2}{\rho^{l-1}}\sum_{k=0}^{\infty} d^{I}_k \frac{1}{\rho^k} ,
\\
d^{I}_0 &= \frac{2-l^2}{4l(l+1)} v_0+\frac{1}{4l} v_1 ,
\q \q
d^{I}_1 = -\frac{l^3-3l+2}{4l^2} v_0 + \frac{l(l+1)-1}{4l^2} v_1 ,
\q \q
d^{I}_2 = 0 ,
\\
 k(k&-2)  (-1 + k + 2 l) (1 + k + 2 l) d^{I}_k = \\
& \hspace{0ex} v_{k-1} - 4 (-3 + k + l)^2 (-1 + k + l) (k + l) d^{I}_{k-2}  \\
& \hspace{0ex} + 2 [(k-2) k (1 + k (2 k-5)) + (-1 + k) (6 + k (-21 + 8 k)) l + 2 (-2 + k) (-3 + 5 k) l^2 + 2 (-3 + 2 k) l^3] d^{I}_{k-1} ,
\end{split}
\end{align}
which agrees with \eqref{eqn:xiE_plusMinus} up to a constant factor and 
linear combination with $\tilde{\xi}^{+}_{H0}$.  It is important when 
constructing the ``plus-side'' solutions to use either
Eqns.~\eqref{eqn:ht_plusMinus}, \eqref{eqn:xiH1}, and 
\eqref{eqn:xiE_plusMinus} or Eqns.~\eqref{eqn:ht_plus_inf}, 
\eqref{eqn:xi_plus1_inf}, and \eqref{eqn:xi_plus2_inf}.  Mixing these sets 
of equations will introduce an inconsistency.

\section{Explicit form of the force terms $f_n^{\a}$}
\label{sec:littlefs}

Here we give the explicit form of the various force terms $f_n^{\a}$ defined 
in Sec.~\ref{sec:sf}.  Only the $t$ and $r$ components are necessary.  The 
$\theta$ component vanishes and the $\vp$ component can be derived from the 
other two.  These functions depend upon the position on the orbit, the 
constants of motion, and the MP amplitudes and their first derivatives.
There is implied dependence on $l$ and $m$.
\begin{align} 
\begin{split}
\hspace{-2ex} f^t_0 &=
\left[
\frac{i m \mathcal{E} \mathcal{L} \left(3 f_p+U_p^2-2\mathcal{E}^2\right)}{4 f_p^3}
-\frac{M \mathcal{E}^2 \left(3 f_p+U_p^2-4 \mathcal{E}^2\right)\dot{r}_p}{2 f_p^5}
\right] h_{{tt}}
+
\frac{r_p^2 \l f_p \left(U_p^2+\mathcal{E}^2\right)-f_p^2+\mathcal{E}^2 \left(U_p^2-2 \mathcal{E}^2\right) \r}{4 f_p^4} \pa_t h_{{tt}}
\\
& \hspace{2ex}  
+
\frac{r_p^2 \mathcal{E}^2 \left(3 f_p+U_p^2-2 \mathcal{E}^2\right) \dot{r}_p}{4 f_p^4} \pa_r h_{{tt}} 
 +
\left[
\frac{i m \mathcal{E} \mathcal{L} \left(f_p+U_p^2-2 \mathcal{E}^2\right)}{4 f_p}
+ \frac{M \mathcal{E}^2 \left(f_p+U_p^2\right) \dot{r}_p }{2 f_p^3}
\right] h_{{rr}} 
 \\
& \hspace{2ex} 
+\frac{r_p^2 \left(f_p+\mathcal{E}^2\right) \left(f_p+U_p^2-2 \mathcal{E}^2\right)}{4f_p^2} \pa_t h_{{rr}} 
+\frac{r_p^2 \mathcal{E}^2 \left(f_p+U_p^2-2 \mathcal{E}^2\right) \dot{r}_p}{4 f_p^2} \pa_r h_{{rr}} \\
 & \hspace{2ex} 
 + 
\left[-\frac{M \left(f_p U_p^2+\mathcal{E}^2 U_p^2-2 \mathcal{E}^4\right)}{f_p^3}
+\frac{i m \mathcal{E} \mathcal{L} \left(f_p-\mathcal{E}^2\right) \dot{r}_p}{f_p^3}\right] h_{{tr}} 
-
\frac{r_p^2 \mathcal{E}^4 \dot{r}_p}{f_p^4} \pa_t h_{{tr}}
+
\frac{r_p^2 \left(f_p-\mathcal{E}^2\right) \left(\mathcal{E}^2-U_p^2\right)}{f_p^2} \pa_r h_{{tr}}  
\\
& \hspace{2ex} +
\left[
-\frac{m^2 \mathcal{L}^2 \left(f_p-\mathcal{E}^2\right)}{r_p^2 f_p^2}
+
\frac{i m \mathcal{E}\mathcal{L} \left(\mathcal{E}^2 f_p-2 f_p^2+\mathcal{E}^2\right) \dot{r}_p}{r_p f_p^4}
\right] j_t
-
\frac{i m \mathcal{E}^3 \mathcal{L} }{f_p^3} \pa_t j_t 
+ 
\frac{i m \mathcal{E} \mathcal{L} \left(f_p-\mathcal{E}^2\right) \dot{r}_p}{f_p^3} \pa_r j_t
 \\
& \hspace{2ex} 
+\left[
\frac{i m \mathcal{E} \mathcal{L} \left(-5 f_p U_p^2+4 \mathcal{E}^2 f_p+U_p^2\right)}{2 r_p f_p^2}
+\frac{m^2 \mathcal{E}^2 \mathcal{L}^2 \dot{r}_p}{r_p^2 f_p^2}
\right] j_r
-
\frac{i m \mathcal{E} \mathcal{L} \left(f_p+\mathcal{E}^2\right) \dot{r}_p}{f_p^3} \pa_t j_r
-\frac{i m \mathcal{E} \mathcal{L}  \left(\mathcal{E}^2-U_p^2\right)}{f_p} \pa_r j_r
\\
& \hspace{2ex} 
+
\frac{i m \mathcal{E} \mathcal{L} \left(f_p-U_p^2\right)}{2 f_p^2} K
+
\frac{r_p^2 \left(f_p-U_p^2\right) \left(f_p+\mathcal{E}^2\right)}{2 f_p^3} \pa_t K 
+
\frac{r_p^2 \mathcal{E}^2  \left(f_p-U_p^2\right) \dot{r}_p}{2 f_p^3} \pa_r K
 \\
& \hspace{2ex} 
+ 
\left[
\frac{i m \left(m^2+4\right) \mathcal{E} \mathcal{L}^3}{4 r_p^2 f_p}
-
\frac{m^2 \mathcal{E}^2 \mathcal{L}^2 \dot{r}_p}{r_p f_p^2}
 \right] G
+
\frac{m^2 \mathcal{L}^2 \left(f_p+\mathcal{E}^2\right)}{4 f_p^2} \pa_t G 
+
\frac{m^2 \mathcal{E}^2 \mathcal{L}^2 \dot{r}_p}{4 f_p^2} \pa_r G ,
\end{split}
\\
\begin{split}
\hspace{-2ex} f_1^t & =
\left[
-\frac{i m \mathcal{E}  \mathcal{L}^3}{4 r_p^2 f_p^2}
+ 
\frac{M \mathcal{E}^2 \mathcal{L}^2 \dot{r}_p}{2 r_p^2 f_p^4}
\right] h_{{tt}} 
-
\frac{\mathcal{L}^2 \left(f_p+\mathcal{E}^2\right)}{4 f_p^3} \pa_t h_{{tt}}
-
\frac{\mathcal{E}^2 \mathcal{L}^2 \dot{r}_p }{4 f_p^3} \pa_r h_{{tt}} 
\\
& \hspace{2ex}
+ 
\left[
\frac{i m \mathcal{E} \mathcal{L}^3}{4 r_p^2}
+ 
\frac{\mathcal{E}^2 \mathcal{L}^2 \left(1-5 f_p\right) \dot{r}_p}{4 r_p f_p^2}  
\right] h_{{rr}}
+ 
\frac{\mathcal{L}^2 \left(f_p+\mathcal{E}^2\right)}{4 f_p}  \pa_t h_{{rr}}
+
\frac{\mathcal{E}^2 \mathcal{L}^2 \dot{r}_p}{4 f_p} \pa_r h_{{rr}} 
 \\
 & \hspace{2ex}
 +
 \frac{\mathcal{L}^2 \left(f_p-\mathcal{E}^2\right)}{r_p f_p} h_{{tr}}
-
\frac{i m \mathcal{E} \mathcal{L}^3}{r_p^3} j_r
+
\frac{\mathcal{E}^2 \mathcal{L}^2 \dot{r}_p}{r_p f_p^2} K
+ 
\left[
-\frac{i m \mathcal{E} \mathcal{L}^3}{r_p^2 f_p}
+ \frac{l (l+1) \mathcal{E}^2 \mathcal{L}^2 \dot{r}_p}{2 r_p f_p^2}
\right] G ,
\end{split}
\end{align}

\begin{align}
\hspace{-2ex} f_2^t & =
\frac{\mathcal{L}^2 \left(f_p-\mathcal{E}^2\right)}{r_p^2 f_p^2} j_t
-
\frac{\mathcal{E}^2 \mathcal{L}^2 \dot{r}_p}{r_p^2 f_p^2} j_r
+ 
\left[ 
-
\frac{5 i m \mathcal{E} \mathcal{L}^3}{4 r_p^2 f_p}
+
\frac{\mathcal{E}^2 \mathcal{L}^2 \dot{r}_p}{r_p f_p^2}
\right] G 
-
\frac{\mathcal{L}^2 \left(f_p+\mathcal{E}^2\right)}{4 f_p^2} \pa_t G
-
\frac{\mathcal{E}^2 \mathcal{L}^2 \dot{r}_p}{4 f_p^2} \pa_r G , 
\\
\hspace{-2ex} f_3^t 
& =
\frac{i m \mathcal{E} \mathcal{L}^3}{4 r_p^2 f_p} G
+
\frac{\mathcal{L}^2 \left(f_p+\mathcal{E}^2\right)}{4 f_p^2} \pa_t G 
+
\frac{\mathcal{E}^2 \mathcal{L}^2\dot{r}_p}{4 f_p^2} \pa_r G,
\\
\hspace{-2ex} f_4^t & =
\frac{i m \mathcal{L}^2 \left(\mathcal{E}^2 - f_p\right)}{r_p^2 f_p^2} h_t 
+
\frac{i m \mathcal{E}^2 \mathcal{L}^2 \dot{r}_p}{r_p^2 f_p^2} h_r 
-
\left[
\frac{m^2 \mathcal{E} \mathcal{L}^3}{r_p^4 f_p}
+ 
\frac{2 i m \mathcal{E}^2 \mathcal{L}^2 \dot{r}_p}{r_p^3 f_p^2}
\right] h_2 +
\frac{i m \mathcal{L}^2 \left(f_p+\mathcal{E}^2\right)}{2 r_p^2 f_p^2} \pa_t h_2
+
\frac{i m \mathcal{E}^2 \mathcal{L}^2 \dot{r}_p}{2 r_p^2 f_p^2} \pa_r h_2, \\
\begin{split}
\hspace{-2ex} f_5^t & =
 \frac{\mathcal{E} \mathcal{L} \left(\mathcal{E}^2 f_p-2 f_p^2+\mathcal{E}^2\right) \dot{r}_p}{r_p f_p^4} h_t 
- 
\frac{\mathcal{E}^3 \mathcal{L}}{f_p^3} \pa_t h_t
+
\frac{\mathcal{E} \mathcal{L} \l f_p-\mathcal{E}^2\r \dot{r}_p}{f_p^3} \pa_r h_t 
+
\frac{\mathcal{E} \mathcal{L} \left(-5 f_p U_p^2+4 \mathcal{E}^2 f_p+U_p^2\right)}{2 r_p f_p^2} h_r
\\
& 
\hspace{2ex}
-
\frac{\mathcal{E} \mathcal{L} \left(f_p+\mathcal{E}^2\right) \dot{r}_p}{f_p^3} 
\pa_t h_r
+
\frac{\mathcal{E} \mathcal{L} \left(U_p^2-\mathcal{E}^2\right)}{f_p} \pa_r h_r
-
\frac{\left(m^2-1\right) \mathcal{E} \mathcal{L}^3}{2 r_p^4 f_p} h_2,
 \end{split}
 \\
\hspace{-2ex} f_6^t & =
-\frac{\mathcal{E} \mathcal{L}^3}{r_p^3} h_r
-\frac{\mathcal{E} \mathcal{L}^3}{2 r_p^4 f_p} h_2 ,
\\
\hspace{-2ex} f_7^t & =
-\frac{\mathcal{E} \mathcal{L}^3}{2 r_p^4 f_p} h_2 ,
\end{align}

\begin{align}
\hspace{-2ex} f_0^r &= 
\left[
\frac{ M \left(\mathcal{E}^2 f_p - f_p^2 -  \l 5 \mathcal{E}^2 - U_p^2 \r U_p^2 +
4 \mathcal{E}^4\right)}{2 f_p^3}
-
\frac{i m \mathcal{E} \mathcal{L} \left(f_p-U_p^2+2 \mathcal{E}^2\right) \dot{r}_p}{4 f_p^3}\right] h_{{tt}} 
-
\frac{r_p^2 \mathcal{E}^2 \left(f_p-U_p^2+2\mathcal{E}^2\right) \dot{r}_p }{4 f_p^4} \pa_t h_{{tt}}
\notag \\
& \hspace{2ex} + 
\frac{r_p^2 \l \mathcal{E}^2 f_p+f_p^2+ \l 3 \mathcal{E}^2 - U_p^2 \r U_p^2-2 \mathcal{E}^4 \r }{4 f_p^2} \pa_r h_{{tt}} 
 \notag \\
& \hspace{2ex}
+ \left[ 
-\frac{M \left(-f_p \left(2 U_p^2+\mathcal{E}^2\right)+f_p^2 + \left(U_p^2-\mathcal{E}^2\right) U_p^2\right) }{2 f_p}
- 
\frac{i m \mathcal{E} \mathcal{L} \left(3 f_p-U_p^2+2 \mathcal{E}^2\right) \dot{r}_p}{4 f_p}
   \right] h_{{rr}} 
\notag \\
& \hspace{2ex} 
-\frac{r_p^2 \mathcal{E}^2 \left(3 f_p-U_p^2+2 \mathcal{E}^2\right) \dot{r}_p}{4 f_p^2} 
\pa_t h_{{rr}} 
+ 
\frac{r_p^2 \l 2 f_p U_p^2-\mathcal{E}^2 f_p-f_p^2+ \l 3 \mathcal{E}^2 - U_p^2 \r U_p^2 -2 \mathcal{E}^4 \r}{4} \pa_r h_{{rr}}
\notag \\
& \hspace{2ex}
+
\left[
\frac{i m \mathcal{E} \mathcal{L} \left(U_p^2 - \mathcal{E}^2 - f_p \right)}{f_p}
+ 
\frac{M \mathcal{E}^2 \left(2 f_p-U_p^2+2 \mathcal{E}^2\right) \dot{r}_p}{f_p^3}
\right] h_{{tr}} 
-
\frac{r_p^2 \mathcal{E}^2 \left(f_p-U_p^2+\mathcal{E}^2\right)}{f_p^2} \pa_t h_{{tr}}
\\
& \hspace{2ex}
+
\frac{r_p^2 \mathcal{E}^2 \left(U_p^2-\mathcal{E}^2\right) \dot{r}_p}{f_p^2} \pa_r h_{{tr}} 
+ 
\left[ 
\frac{i m \mathcal{E} \mathcal{L} \left(f_p+1\right) \left(\mathcal{E}^2-U_p^2\right)}{r_p f_p^2}
+
\frac{m^2 \mathcal{E}^2 \mathcal{L}^2 \dot{r}_p}{r_p^2 f_p^2}
\right] j_t
-
\frac{i m \mathcal{E}^3 \mathcal{L} \dot{r}_p}{f_p^3} \pa_t j_t 
\notag \\
& \hspace{2ex} 
+ 
\frac{i m \mathcal{E} \mathcal{L} \left(f_p+U_p^2-\mathcal{E}^2\right)}{f_p} \pa_r j_t
+ 
\left[ 
\frac{m^2 \mathcal{L}^2 \left(f_p-U_p^2+\mathcal{E}^2\right)}{r_p^2}
+
\frac{i m \mathcal{E} \mathcal{L}  \left(-5 f_p U_p^2+4 \mathcal{E}^2 f_p+4 f_p^2+U_p^2\right)\dot{r}_p }{2 r_p f_p^2}
\right] j_r 
\notag \\
& \hspace{2ex} 
-
\frac{i m \mathcal{E} \mathcal{L} \left(f_p-U_p^2+\mathcal{E}^2\right)}{f_p} \pa_t j_r
- 
\frac{i m \mathcal{E} \mathcal{L} \left(\mathcal{E}^2-U_p^2\right) \dot{r}_p}{f_p}
\pa_r j_r
+
\frac{i m \mathcal{E}\mathcal{L} \left(f_p-U_p^2\right) \dot{r}_p}{2 f_p^2} K
+
\frac{r_p^2 \mathcal{E}^2 \left(f_p-U_p^2\right) \dot{r}_p}{2 f_p^3} \pa_t K
\notag \\
& \hspace{2ex}
-
\frac{r_p^2 \left(f_p-U_p^2\right) \left(f_p+U_p^2-\mathcal{E}^2\right)}{2 f_p} \pa_r K
+
\left[
-\frac{m^2 \mathcal{L}^2
\left(f_p-U_p^2+\mathcal{E}^2\right)}{r_p}
+
\frac{i m \left(m^2+4\right) \mathcal{E} \mathcal{L}^3 \dot{r}_p}{4 r_p^2 f_p}
\right] G
+ 
\frac{m^2 \mathcal{E}^2 \mathcal{L}^2 \dot{r}_p}{4 f_p^2} \pa_t G
\notag \\
& \hspace{2ex}
-
\frac{m^2 \mathcal{L}^2 \left(f_p+U_p^2-\mathcal{E}^2\right)}{4} \pa_r G ,
\notag \\
\hspace{-2ex} f_1^r &=
\left[-\frac{M \mathcal{L}^2 \left(f_p+U_p^2-\mathcal{E}^2\right)}{2 r_p^2f_p^2}
-
\frac{i m \mathcal{E} \mathcal{L}^3 \dot{r}_p}{4 r_p^2 f_p^2}\right] h_{{tt}}
-
\frac{\mathcal{E}^2 \mathcal{L}^2 \dot{r}_p}{4 f_p^3} \pa_t h_{{tt}}
+
\frac{\mathcal{L}^2 \left(f_p+U_p^2-\mathcal{E}^2\right)}{4 f_p} \pa_r h_{{tt}}
\notag \\
& \hspace{2ex}
+ 
\left[
\frac{\mathcal{L}^2 \left(5 f_p  \l U_p^2 - \mathcal{E}^2 \r - 3 f_p^2 - f_p - U_p^2 +\mathcal{E}^2\right)}{4 r_p}
+
\frac{i m \mathcal{E} \mathcal{L}^3 \dot{r}_p}{4 r_p^2}\right] h_{{rr}}
+
\frac{\mathcal{E}^2 \mathcal{L}^2 \dot{r}_p}{4 f_p} \pa_t h_{{rr}}
-
\frac{\mathcal{L}^2 f_p \left(f_p+U_p^2-\mathcal{E}^2\right)}{4} \pa_r h_{{rr}}
\hspace{3ex}
\notag \\
& \hspace{2ex} 
-
\frac{\mathcal{E}^2 \mathcal{L}^2 \dot{r}_p}{r_p f_p} h_{{tr}} 
-
\frac{i m \mathcal{E} \mathcal{L}^3 \dot{r}_p}{r_p^3} j_r 
+
\frac{\mathcal{L}^2 \left(f_p-U_p^2+\mathcal{E}^2\right)}{r_p} K 
+ 
\left[\frac{l (l+1) \mathcal{L}^2 \left(f_p-U_p^2+\mathcal{E}^2\right)}{2 r_p}-\frac{i m \mathcal{E} \mathcal{L}^3 \dot{r}_p}{r_p^2 f_p}\right] G , 
\end{align}

\begin{align}
\begin{split}
f_2^r &=
-
\frac{\mathcal{E}^2 \mathcal{L}^2 \dot{r}_p}{r_p^2 f_p^2} j_t
-
\frac{\mathcal{L}^2 \left(f_p-U_p^2+\mathcal{E}^2\right)}{r_p^2} j_r 
+ 
\left[
\frac{\mathcal{L}^2 \left(f_p-U_p^2+\mathcal{E}^2\right)}{r_p} 
-
\frac{5 i m \mathcal{E} \mathcal{L}^3 \dot{r}_p}{4 r_p^2 f_p} \right] G 
-
\frac{\mathcal{E}^2 \mathcal{L}^2 \dot{r}_p}{4 f_p^2} \pa_t G 
\\
&  \hspace{2ex} 
+
\frac{\mathcal{L}^2 \left(f_p+U_p^2-\mathcal{E}^2\right)}{4} \pa_r G ,
\end{split} \\
f_3^r &= 
\frac{i m \mathcal{E} \mathcal{L}^3 \dot{r}_p}{4 r_p^2 f_p} G
+
\frac{\mathcal{E}^2 \mathcal{L}^2 \dot{r}_p}{4 f_p^2} \pa_t G
-
\frac{\mathcal{L}^2 \left(f_p+U_p^2-\mathcal{E}^2\right)}{4} \pa_r G , \\
\begin{split}
\hspace{-2ex} f_4^r &= 
\frac{i m \mathcal{E}^2 \mathcal{L}^2 \dot{r}_p}{r_p^2 f_p^2} h_t 
+
\frac{i m \mathcal{L}^2 \left(f_p-U_p^2+\mathcal{E}^2\right)}{r_p^2} h_r 
+ 
\left[\frac{2 i m \mathcal{L}^2 \left(U_p^2-\mathcal{E}^2\right)}{r_p^3}
-
\frac{m^2 \mathcal{E} \mathcal{L}^3 \dot{r}_p}{r_p^4 f_p} 
\right] h_2 
+
\frac{i m \mathcal{E}^2 \mathcal{L}^2 \dot{r}_p}{2 r_p^2 f_p^2} \pa_t h_2
\\
&  \hspace{2ex} 
-
\frac{i m \mathcal{L}^2  \left(f_p+U_p^2-\mathcal{E}^2\right)}{2 r_p^2} \pa_r h_2 ,
\end{split}
\\
\hspace{-2ex} f_5^r &=
\frac{\mathcal{E} \mathcal{L} \left(f_p+1\right) \l\mathcal{E}^2-U_p^2\r}{r_p f_p^2} h_t 
-
\frac{\mathcal{E}^3 \mathcal{L} \dot{r}_p }{f_p^3} \pa_t h_t
+
\frac{\mathcal{E} \mathcal{L} \left(f_p+U_p^2-\mathcal{E}^2\right)}{f_p} \pa_r h_t  
-
\frac{\mathcal{E} \mathcal{L} 
\left(5 f_p U_p^2-4 \mathcal{E}^2 f_p-4 f_p^2-U_p^2\right) \dot{r}_p}{2 r_p f_p^2}  h_r
\notag \\
 & \hspace{2ex}
-
\frac{\mathcal{E} \mathcal{L} \left(f_p-U_p^2+\mathcal{E}^2\right)}{f_p} \pa_t h_r
+
\frac{\mathcal{E} \mathcal{L} \l U_p^2-\mathcal{E}^2\r \dot{r}_p}{f_p} \pa_r h_r 
-
\frac{\left(m^2-1\right) \mathcal{E} \mathcal{L}^3 \dot{r}_p}{2 r_p^4 f_p} h_2,
\\
\hspace{-2ex} f_6^r &=
-\frac{\mathcal{E} \mathcal{L}^3 \dot{r}_p}{r_p^3} h_r
-\frac{\mathcal{E} \mathcal{L}^3 \dot{r}_p}{2 r_p^4 f_p} h_2, 
\\
\hspace{-2ex} f_7^r &=
-\frac{\mathcal{E} \mathcal{L}^3 \dot{r}_p}{2 r_p^4 f_p} h_2.
\end{align}

\section{Additional self-force values}
\label{sec:addedGSFtables}

The following two tables (Tables \ref{tbl:e=0.3} and \ref{tbl:e=0.7}) provide GSF data that compliments that presented in Tables \ref{tbl:e=0.1} and \ref{tbl:e=0.5}.

\begin{table}[H]
\begin{center}
\caption{\label{tbl:e=0.3} Same as Table \ref{tbl:e=0.1} with $e=0.3$.}
\begin{tabular}{c|c|c|c|c|c|c}
\hline \hline
 & $\chi$ & $p=10$ & $p=20$ & $p=30$ & $p=60$ & $p=90$ \\
\hline
\multirow{8}{*}{$F^t$} & 0 & $-1.02425\times 10^{-3}$ & 
$-2.195889\times 10^{-5}$ & $-2.619956\times 10^{-6}$ & 
$-7.6424003\times 10^{-8}$ & $-9.94261928\times 10^{-9}$ \\
&$\pi/4$ &$ 7.93725\times 10^{-4}$ & $2.611011\times 10^{-4}$ & 
$  1.056357\times 10^{-4}$ & $2.0516060\times 10^{-5}$ & 
$  7.66603327\times 10^{-6}$ \\
&$\pi/2$ & $1.12072\times 10^{-3}$ & $2.704049\times 10^{-4}$ & 
$1.061464\times 10^{-4}$ & $2.0147891\times 10^{-5}$ & 
$7.47715975\times 10^{-6}$ \\
&$3\pi/4$ &$ 5.23325\times 10^{-4}$ & $1.237182\times 10^{-4}$ & 
$  4.795782\times 10^{-5}$ & $8.9766821\times 10^{-6}$ & 
$  3.31514100\times 10^{-6}$ \\
&$\pi$ & $2.83617\times 10^{-7}$ & $-3.146284\times 10^{-8}$ & 
$-4.916581\times 10^{-9}$ & $-1.5203069\times 10^{-10}$ & 
$-1.85477251\times 10^{-11}$ \\
&$5\pi/4$ &$  -5.01242\times 10^{-4}$ & $-1.234054\times 10^{-4}$ & 
$  -4.792378\times 10^{-5}$ & $-8.9756562\times 10^{-6}$ & 
$  -3.31499843\times 10^{-6}$ \\
&$3\pi/2$ & $-1.05385\times 10^{-3}$ & $-2.698687\times 10^{-4}$ & 
$-1.061119\times 10^{-4}$ & $-2.0147281\times 10^{-5}$ & 
$-7.47706856\times 10^{-6}$ \\
&$7\pi/4$ &$  -1.52944\times 10^{-3}$ & $-2.782552\times 10^{-4}$ & 
$  -1.077562\times 10^{-4}$ & $-2.0579481\times 10^{-5}$ & 
$  -7.67431383\times 10^{-6}$ \\
\hline
\multirow{8}{*}{$F^r$} & 0 & $2.30316\times 10^{-2}$ & 
$6.836866\times 10^{-3}$ & $3.254304\times 10^{-3}$ & 
$8.7346506\times 10^{-4}$ & $3.97631238\times 10^{-4}$ \\
&$\pi/4$ &$ 2.10318\times 10^{-2}$ & $6.042486\times 10^{-3}$ & 
$  2.859673\times 10^{-3}$ & $7.6346290\times 10^{-4}$ & 
$  3.46940757\times 10^{-4}$ \\
&$\pi/2$ & $1.41875\times 10^{-2}$ & $4.215713\times 10^{-3}$ & 
$1.985131\times 10^{-3}$ & $5.2535511\times 10^{-4}$ & 
$2.37910729\times 10^{-4}$ \\
&$3\pi/4$ &$ 8.91644\times 10^{-3}$ & $2.676966\times 10^{-3}$ & 
$  1.253239\times 10^{-3}$ & $3.2907620\times 10^{-4}$ & 
$  1.48590757\times 10^{-4}$ \\
&$\pi$ & $7.11090\times 10^{-3}$ & $2.131279\times 10^{-3}$ & 
$9.953187\times 10^{-4}$ & $2.6059554\times 10^{-4}$ & 
$1.17547270\times 10^{-4}$ \\
&$5\pi/4$ &$ 8.70369\times 10^{-3}$ & $2.669408\times 10^{-3}$ & 
$  1.252041\times 10^{-3}$ & $3.2902233\times 10^{-4}$ & 
$  1.48581907\times 10^{-4}$ \\
&$3\pi/2$ & $1.30565\times 10^{-2}$ & $4.179240\times 10^{-3}$ & 
$1.979557\times 10^{-3}$ & $5.2511220\times 10^{-4}$ & 
$2.37871148\times 10^{-4}$ \\
&$7\pi/4$ &$ 1.86761\times 10^{-2}$ & $5.972195\times 10^{-3}$ & 
$  2.849306\times 10^{-3}$ & $7.6302517\times 10^{-4}$ & 
$  3.46870034\times 10^{-4}$ \\
\hline \hline
\end{tabular}
\end{center} 
\end{table}

\begin{table}[H]
\begin{center}
\caption{
\label{tbl:e=0.7}
Same as Table \ref{tbl:e=0.1} with $e=0.7$.}
\begin{tabular}{c|c|c|c|c|c|c}
\hline \hline
 & $\chi$ & $p=10$ & $p=20$ & $p=30$ & $p=60$ & $p=90$ \\
\hline
\multirow{8}{*}{$F^t$} & 0 & $-1.12\times 10^{-2}$ & 
$-2.101\times 10^{-4}$ & $-2.3435\times 10^{-5}$ & 
$-6.4473\times 10^{-7}$ & $-1.97\times 10^{-7}$ \\
&$\pi/4$ & $ 6.33\times 10^{-3}$ & $9.829\times 10^{-4}$ & 
$3.7436\times 10^{-4}$ & $7.1857\times 10^{-5}$ & 
$2.68\times 10^{-5}$ \\
&$\pi/2$ & $3.53\times 10^{-3}$ & $6.765\times 10^{-4}$ & 
$2.5715\times 10^{-4}$ & $4.7747\times 10^{-5}$ & 
$1.76\times 10^{-5}$ \\
&$3\pi/4$ &$ 6.28\times 10^{-4}$ & $1.300\times 10^{-4}$ & 
$4.8732\times 10^{-5} $ & $8.8491\times 10^{-6}$ & 
$3.24\times 10^{-6}$ \\
&$\pi$ & $9.81\times 10^{-8}$ & $1.933\times 10^{-9}$ & 
$2.2050\times 10^{-10}$ & $6.4536\times 10^{-12}$ & $-3.01\times 10^{-11}$ \\
&$5\pi/4$ &$ -6.10\times 10^{-4}$ & $-1.296\times 10^{-4}$ & 
$-4.8678\times 10^{-5}$ & $-8.8474\times 10^{-6}$ & 
$-3.24\times 10^{-6}$ \\
&$3\pi/2$ & $-2.15\times 10^{-3}$ & $-6.524\times 10^{-4}$ & 
$-2.5435\times 10^{-4}$ & $-4.7665\times 10^{-5}$ & $-1.76\times 10^{-5}$ \\
&$7\pi/4$ &$ -6.48\times 10^{-3}$ & $-9.957\times 10^{-4}$ & 
$-3.7692\times 10^{-4}$ & $-7.1956\times 10^{-5}$ & $-2.68\times 10^{-5}$ \\
\hline
\multirow{8}{*}{$F^r$} & 0 & $5.24\times 10^{-2}$ & 
$1.185\times 10^{-2}$ & $5.5084\times 10^{-3}$ & 
$1.4758\times 10^{-3}$ & $6.74\times 10^{-4}$ \\
&$\pi/4$ &$ 4.66\times 10^{-2}$ & $9.581\times 10^{-3}$ & 
$4.3877\times 10^{-3}$ & $1.1591\times 10^{-3}$ & 
$5.27\times 10^{-4}$ \\
&$\pi/2$ & $1.62\times 10^{-2}$ & $4.435\times 10^{-3}$ & 
$2.0476\times 10^{-3}$ & $5.3292\times 10^{-4}$ & $2.40\times 10^{-4}$ \\
&$3\pi/4$ &$  4.22\times 10^{-3}$ & $1.180\times 10^{-3}$ & 
$5.3937\times 10^{-4} $ & $1.3833\times 10^{-4}$ & 
$6.20\times 10^{-5}$ \\
&$\pi$ & $1.55\times 10^{-3}$ & $4.224\times 10^{-4}$ & 
$1.9202\times 10^{-4}$ & $4.9022\times 10^{-5}$ & $2.19\times 10^{-5}$ \\
&$5\pi/4$ &$ 4.19\times 10^{-3}$ & $1.179\times 10^{-3}$ & 
$5.3905\times 10^{-4}$ & $1.3832\times 10^{-4}$ & $6.20\times 10^{-5}$ \\
&$3\pi/2$ & $1.35\times 10^{-2}$ & $4.348\times 10^{-3}$ & 
$2.0342\times 10^{-3}$ & $5.3232\times 10^{-4}$ & $2.40\times 10^{-4}$ \\
&$7\pi/4$ &$ 2.68\times 10^{-2}$ & $9.061\times 10^{-3}$ & $4.3148\times
   10^{-3}$ & $1.1561\times 10^{-3}$ & $5.26\times 10^{-4}$ \\
\hline \hline
\end{tabular}
\end{center} 
\end{table}

\end{widetext}

\bibliography{Lorenz}

\end{document}